%% file: sig_asia_arxiv.tex
\documentclass[acmtog]{acmart}

\usepackage{booktabs} 

\usepackage[ruled]{algorithm2e} 
\usepackage{cleveref}

\newcommand{\RNum}[1]{\uppercase\expandafter{\romannumeral #1\relax}}
\SetAlFnt{\small}
\SetAlCapFnt{\small}
\SetAlCapNameFnt{\small}
\SetAlCapHSkip{0pt}
\usepackage{float}
\usepackage{enumitem}
\usepackage{makecell}
\usepackage{float}

\acmJournal{TOG}

\input{preamble}

\newcommand\rev[1]{\textcolor{black}{#1}}
\newenvironment{revsection}{%
    \color{black} 
}{%
}

\begin{document}
\title{Learnable Persistent Wrinkle Formation in Cloth Simulation}

\author{Deshan Gong}
\affiliation{%
 \institution{The University of Hong Kong}
 \city{Hong Kong}
 \country{China}}
\email{deshan@hku.hk}
\author{Ningtao Mao}
\affiliation{%
 \institution{University of Leeds}
 \city{Leeds}
 \country{United Kingdom}
}
\email{n.mao@leeds.ac.uk}
\author{Xiaoyuan Yang}
\affiliation{%
\institution{University of Leeds}
\city{Leeds}
\country{United Kingdom}}
\email{sc21xy@leeds.ac.uk}
\author{Xinyu Lu}
\affiliation{%
 \institution{The University of Hong Kong}
 \city{Hong Kong}
 \country{China}
}
\email{lxy819469559@gmail.com}
\author{He Wang\textsuperscript{*}}
\affiliation{%
 \institution{University College London}
 \city{London}
 \country{United Kingdom}}
\thanks{Co-corresponding author: he\_wang@ucl.ac.uk}
\author{Taku Komura\textsuperscript{*}}
\affiliation{%
\institution{The University of Hong Kong}
 \department{School of Computing}
 \city{Hong Kong}
 \country{China}
}
\thanks{Co-corresponding author: taku@cs.hku.hk}

\renewcommand\shortauthors{Gong, et al.}
    
\begin{abstract}

The mechanical memory of fabrics often leads to persistent wrinkles, which reflect key physical properties and habitual wear patterns. Simulating these wrinkles accurately is essential for visual plausibility in digital garments, yet no dedicated approach exists for inferring the parameters that govern their formation due to the lack of precise datasets and estimation methods. We introduce \textbf{Fabric-101}, an \textit{inclusive}, \textit{accurate}, and \textit{extendable} fabric dataset comprising over 101 common fabrics following textile standards. Unlike existing datasets, it captures three physically distinct deformation components (i.e., self-recoverable (elastic), recoverable (friction-driven), and unrecoverable (plastic)), from cyclic loading-unloading measurements. Building on this data, we propose a differentiable cloth simulator combining an elasto-plastic model \rev{with friction}, designed to capture recoverable and unrecoverable wrinkle formation. Our simulator is differentiable and uses adjoint method to learn fabric physical parameters from the measured hysteresis curves, learning fabric-specific wrinkle behaviors. Through extensive experiments, we demonstrate that our model reproduces persistent wrinkles that are visually and physically similar to real fabrics across diverse materials and motions. \rev{Dataset and code are available in \url{https://github.com/GongDeshan/Fabric_101_for_Wrinkles}.}

\end{abstract}

%
%


%
%


\begin{teaserfigure}
        \centering
        \includegraphics[width=\textwidth]{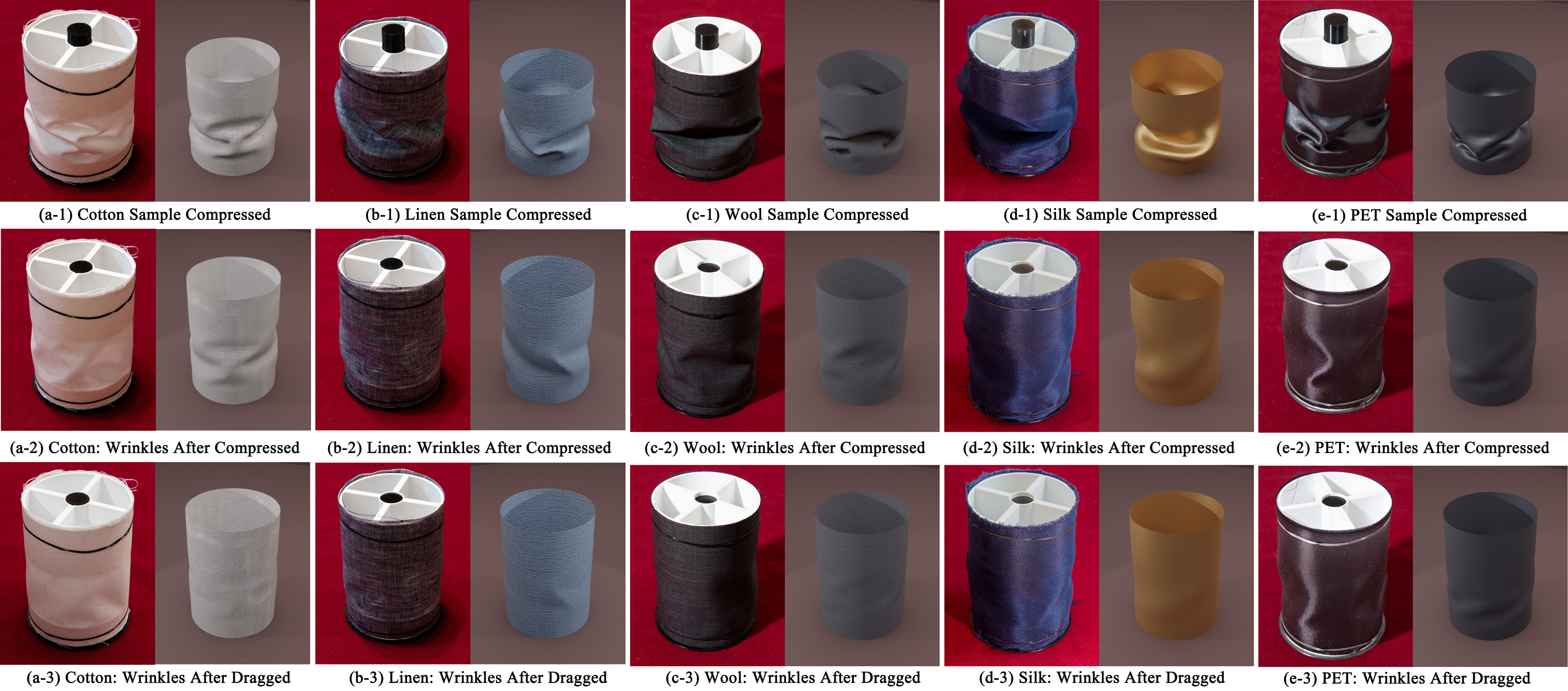}
        \caption{Our differentiable simulator reproduces persistent wrinkles on cylindrical fabric samples of different materials (a to e-1) caused by axial compression (the same as our tester). By combining internal friction and plasticity, it \rev{plausibly simulates} both recoverable and unrecoverable wrinkles. Recoverable wrinkles flatten when the sample is dragged upward (a to e-2), while unrecoverable wrinkles persist (a to e-3).}
        \label{fig:teaser}
\end{teaserfigure}

\maketitle

\input{siga_body_arxiv}
\input{siga_app_arxiv}

\end{document}

%% file: siga_body_arxiv.tex
\section{Introduction}

Cloth simulation in computer graphics strives for ever-greater visual realism and is widely used in 3D animation, fashion design, and robotics~\cite{stuyck2022cloth, moll2009automated}. A key challenge lies in simulating persistent wrinkles—long-lasting deformations that real fabrics retain after folding or manipulation~\cite{brenner1964mechanical}. Their absence significantly undermines visual persuasiveness \cite{gong2025cloth}, yet a noticeable gap remains between simulated results and real-world observations. This gap stems from two interdependent issues: no comprehensive dataset, and no matching physical model.

First, no existing dataset is simultaneously \textit{inclusive}, \textit{accurate}, \textit{extendable}, and open-source while measuring the physical properties governing persistent wrinkles. Most datasets focus narrowly on cotton and synthetics~\cite{wang2010example, gong2024bayesian}, omitting diverse fibers, weaves, and post-processing treatments. Measurements are often acquired with custom apparatus in coarsely controlled settings~\cite{wang2010example,miguel2012data}, and even when standard equipment is used, data is rarely released openly~\cite{ngo2004nonlinear}, hindering reproducibility and extension. To address this, we introduce Fabric-101, a benchmark dataset of over 101 distinct fabrics, \rev{designed to be \textit{inclusive} across diverse fibers, weaves, and finishes.} \rev{For \textit{accuracy} and \textit{extendability}}, all samples are tested in a professional textile lab under controlled conditions using commercially-available and industry-standard equipment.

Second, there is no physical model tailored to the phenomena \rev{observed in our measurements, which reveal} fabric deformation comprises three components: self-recoverable (elasticity), recoverable (reversible upon unloading), and unrecoverable (permanent)~\cite{mao2014towards}, corresponding to rest-shape recovery, transient wrinkles, and persistent wrinkles~\cite{gong2025cloth}. \textbf{However, no existing model can accurately transfer these observations to simulated fabrics.} To bridge this gap, we propose a differentiable simulator that combines elasto-plasticity with friction to model these components jointly. It fits hysteresis data and reproduces realistic persistent wrinkles across diverse fabrics. Our contributions are:
\begin{enumerate}[leftmargin=*]
\item \textbf{Fabric-101}: the first open-source fabric dataset for persistent wrinkle simulation, covering over 101 fabrics with diverse fibers, weaves, and treatments.
\item A \textbf{differentiable elasto-plastic-friction simulator} that learns self-recoverable, recoverable, and unrecoverable deformations from fabric hysteresis measurements.
\end{enumerate}

\section{Related Works}

\paragraph{Cloth Simulation and Wrinkles}

Cloth animation usually models fabrics as (hyper-)elastic materials~\cite{terzopoulos1987elastically,baraff2023large,sperl2020homogenized}, but real fabrics exhibit non-elastic behavior and form persistent wrinkles~\cite{benusiglio2012anatomy}. Wrinkles are typically categorized as dynamic (transient) or static (permanent)~\cite{larboulette2004real}. While dynamic wrinkles have received extensive attention~\cite{bridson2003simulation,lahner2018deepwrinkles}, simulators for static persistent wrinkles are less common. Existing approaches simulate persistent wrinkles via variable rest postures~\cite{kim2011persistent}, hardening plasticity~\cite{narain2013folding}, or internal friction~\cite{miguel2013modeling,wong2013modelling}. However, internal friction and plasticity coexist in real fabrics~\cite{olofsson1969rheology} and jointly contribute to persistent wrinkle formation~\cite{gong2025cloth}. To realistically reproduce persistent wrinkles, our work integrates both internal friction and an elasto-plastic model.

\paragraph{Cloth Dataset}

Fabric datasets for physical simulation have evolved over the past two decades. \cite{miguel2012data} proposed a data-driven estimation framework, though their dataset was not publicly released. \cite{sperl2022estimation} compiled an open database of knitted fabrics with diverse knit patterns and yarn compositions, enabling inverse modeling of yarn-level mechanics. The robotics community also contributed benchmarks: \cite{coltraro2025tracking} introduced a motion capture dataset of four fabric types under dynamic scenarios, while \cite{rodriguez2023will} released depth images with mechanical ground truth. More recently, \cite{dominguez2024practical} assembled a large dataset of 1,565 fabrics but used proprietary, undisclosed testing equipment and \rev{only released the estimated parameters, without providing the raw measurement data used to obtain them.} Despite these advances, existing datasets remain limited to narrow material ranges (e.g., predominantly cotton or knitted) or specific applications. Moreover, they mostly rely on flat specimen testing, which does not capture cylindrical deformation modes common in garment wear. This gap motivates our Fabric-101 dataset: a benchmark that combines professional measurements of a broad spectrum of real-world fabrics.

\paragraph{Cloth Physical Parameters Estimation}

Parameter estimation methods fall into \textit{implicit} and \textit{explicit} categories. Implicit methods train numerical models to map observations to parameters without modeling physics \cite{bhat2003estimating,yang2017learning}. Explicit methods model cloth physics and optimize parameters to fit ground truth data, requiring less training data but more accurate physical modeling \rev{\cite{clyde2017modeling, wang2011data}}. Recently, differentiable cloth simulators have enabled gradient-based optimization \rev{\cite{liang2019differentiable,du2021diffpd,li2022diffcloth}}, but their accuracy depends on the fidelity of the underlying physics. \rev{\cite{liang2019differentiable,li2022diffcloth} model cloth as an elastic material, which cannot account for the hysteresis in deformation and persistent wrinkles.} \rev{\cite{gong2022fine, sperl2022estimation} can estimate cloth parameters at the yarn-level, \cite{zhang2024estimating} learns fabric yarn-level mechanics through homogenized model \cite{sperl2020homogenized}, \cite{gong2024bayesian} infers fabric material heterogeneity by Bayesian inference.} However, no existing differentiable simulator models persistent wrinkles by combining internal friction and plasticity—a gap our work addresses. \rev{In contrast to prior work that relies on manual parameter tuning~\cite{gong2025cloth}, our model learns parameters directly from measured data, enabling automatic reproduction of real fabric wrinkles.}

\section{A Novel Dataset: Fabric-101}

Unlike existing datasets, which are mainly limited to cotton or knits, our Fabric-101 is an inclusive dataset of over 101 commercially available fabrics spanning diverse fibers, weaves, structures, and post-processing treatments: all key factors determining wrinkle formation. The fabrics are tested by a standard tester that can distinguish recoverable and unrecoverable wrinkles via energy analysis~\cite{mao2014towards, wang2016discrimination, gong2025cloth}, and captures wrinkles in garment-like geometry by testing cylindrical samples~\cite{aatcc128, shaikhzadeh2009investigation}. Full fabric list is in the Supplementary Material (SM).

\begin{figure}[tb]
    \centering
    \includegraphics[width=\linewidth]{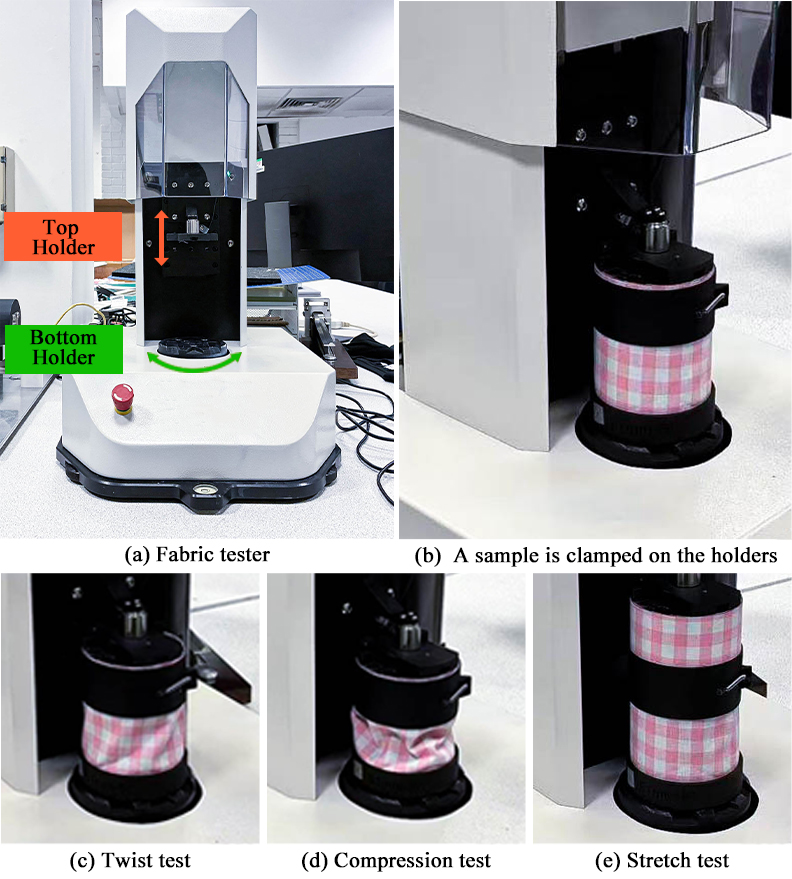}
    \caption{Fabric testing apparatus. (a) The apparatus has two cylindrical holders which are driven by two actuators. The upper holder can translate vertically (orange arrow) and the lower holder can spin (green arrow). (b) A fabric sample is rolled into a cylinder and clamped on the holders. (c) In compression, the upper holder compresses downward to exert axial compressive force on the fabric sample. (d) In twist, the lower holder rotates the fabric sample to cause shear deformation. (e) In stretch, the upper holder moves up to stretch the sample.}
    \label{fig:LUFHES}
\end{figure}

\subsection{Testing Apparatus}

We adapt the testing apparatus shown in Fig. \ref{fig:LUFHES} to measure the reaction force of cylindrical fabric samples through cyclic loading-unloading in three modes: \textbf{twist}, \textbf{compression}, and \textbf{stretch}. The apparatus consists of two cylindrical holders (radius 40 mm): the upper holder (height 40 mm) is driven by a linear actuator for compression and tension; the lower holder (height 20 mm) is driven by a rotary actuator for axial twisting. A vertical force sensor is attached to the upper holder, and a torque sensor to the lower holder. All tests were performed in an environment-controlled lab at \(20 \pm 2^\circ \text{C}\) and \(65 \pm 5\%\) relative humidity, following the textile standard ISO 139:2005. Samples were preconditioned for 24 hours under these conditions before measurement. We also measured each fabric's area density using a high-precision electronic balance (0.001 g resolution) and thickness using a fabric thickness tester. For sample preparation, each fabric sample was cut into a $260 \times 110$ mm rectangle and rolled into a cylinder (radius 40 mm, height 110 mm). Its ends were clamped onto the two holders. The upper holder was additionally wrapped with a strip of the same fabric to measure friction, while the lower was fixed with a fastener. For each fabric, separate samples were prepared for the warp and weft directions (or their equivalents for knits and non-wovens). Detailed testing documentation is provided in~\cite{mao2012wo}.

\paragraph{Testing protocol.} Three testing modes are performed sequentially: twist, compression, and stretch (Fig. \ref{fig:LUFHES}). In the \textbf{twist} mode, the lower holder rotates counter-clockwise (from top view) by \(5^\circ\) (i.e., a maximum shear strain of \(\gamma_{\text{max}} \approx 0.07\)) to incur shearing deformation, then returns to the initial orientation. This angle ensures low-stress shear deformation, where yarn jamming begins before severe wrinkling~\cite{wang2020characterization}, (as $2HG5$ in KES-F~\cite{kawabata1980standardization,kuijpers2020measurement}). The cycle is repeated 4 additional times to measure if the shearing response stabilizes, and we observed that the response stabilized after the first cycle, with subsequent cycles nearly identical. In the \textbf{compression} mode, the upper holder compresses the sample by 15 mm, inducing buckling and bending, and then returns to the initial height. Given the effective sample length of 50 mm (total height 110 mm $-$ 40 mm upper clamp $-$ 20 mm lower clamp), the maximum compressive strain is \(\varepsilon_{\text{max}} = 0.3\). This operation is also repeated 4 additional times to verify stability. Likewise, the response stabilized after the first cycle for all tested fabrics. In the \textbf{stretch} mode, the upper holder moves up until it slips out from the clamped sample. This \rev{single pull} first induces stretching deformation, followed by frictional sliding between the wrapped fabric layers around the upper holder. \rev{The transition from stretching to sliding friction occurs at the first load drop in the curve.} To minimize viscoelastic effects and maintain quasi-static conditions, the compression and stretching operations were performed at a low speed of 1 mm/s, and the rotation speed was set to \(0.1^\circ\)/s.

\begin{figure}[tb]
    \centering
    \includegraphics[width=0.98\linewidth]{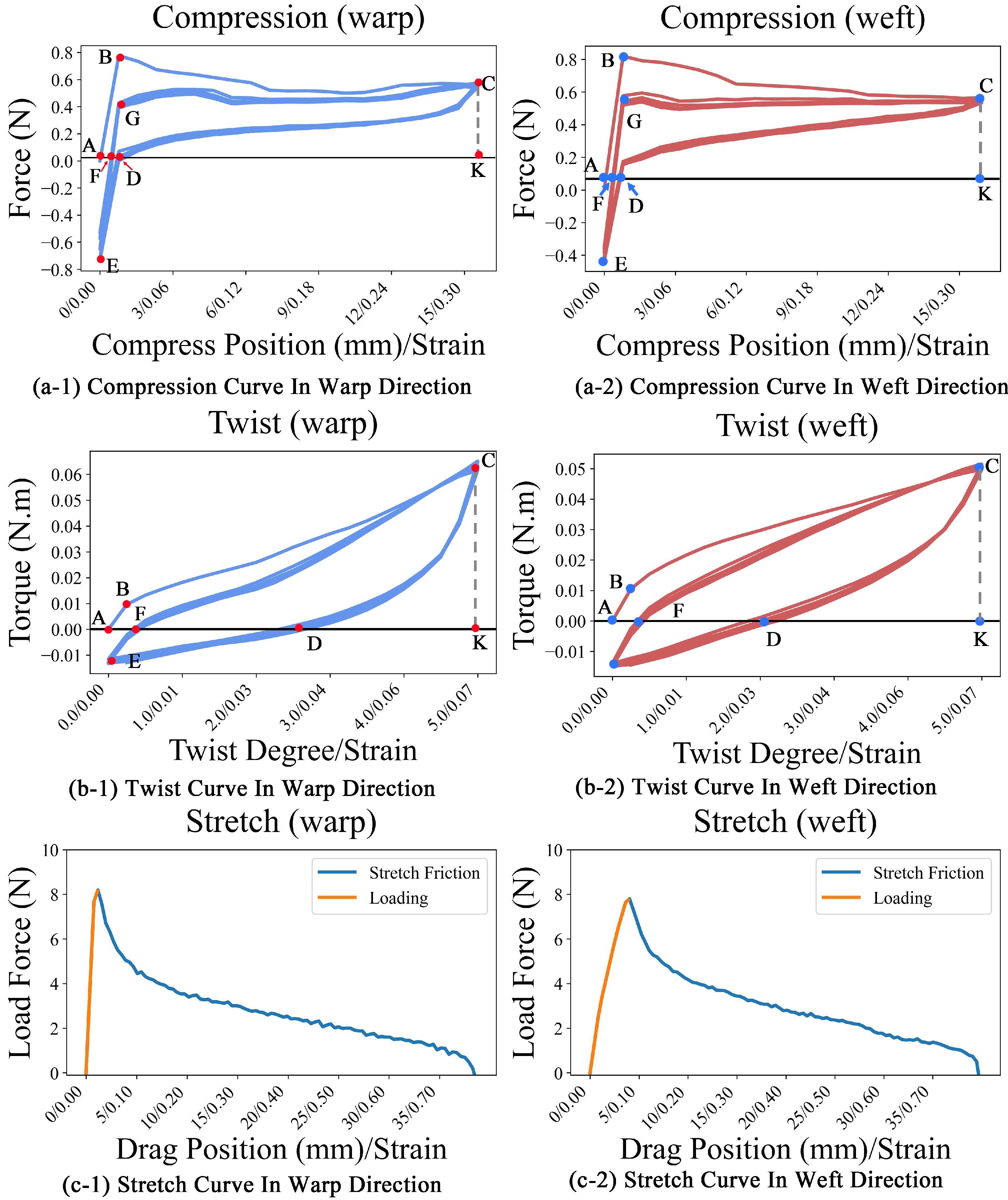}
    \caption{Measured load-deformation curves of Cotton. For each fabric, the apparatus tests a sample by recording load-deformation curves during compression (a-1,2), twisting (b-1,2), and stretching (c-1,2) along the warp and weft directions. SM includes other \rev{fabrics'} measurements.}
    \label{fig:LUFHES_test_curves}
\end{figure}
  
\paragraph{Data processing and decomposition.} The measured curves reveal distinct mechanical behaviors across the testing modes. \textbf{Loading behavior:} In compression and twist, the first loading curve goes from A to C via B (ABC for short) in Fig. \ref{fig:LUFHES_test_curves}. In compression (a-1,2), the curve exhibits a buckling instability: the load rises steeply during initial axial compression (AB), then suddenly drops at a critical displacement (B) as the cylindrical fabric transitions from compression to bending, forming wrinkles. The post-buckling (BC) is governed by bending mechanics. In twist (b-1,2), the  nonlinearity (ABC) arises from stick-slip friction (B) and yarn jamming during shear (C)~\cite{peirce19375}. In stretch (c-1,2), the curve splits into two stages: the initial stage (orange) measures tensile stiffness, followed by sliding friction around the top holder (blue). \textbf{Cyclic response:} Compression and twist conduct cyclic loading-unloading for 4 additional times (a,b-1,2). After the first load, \rev{the response forces during unloading trace the path CDE}. Then, next loading starts via EFGC (compression) or EFC (shear). The cyclic curves in (a,b-1,2) reveal consistent patterns. (1) Upon unloading, the fabric exhibits spontaneous recovery (KD), where the curve returns to a near-zero load but not to the zero-deformation point, leaving a permanent offset (AD). (2) The loading (ABC) and unloading (CDE) paths do not overlap, forming a hysteresis loop. (3) At zero displacement, a residual load remains (E). (4) After unloading, the near-zero-load point moves from D to F, reducing the permanent offset by FD. (5) The second and subsequent loading curves lie below the first and do not overlap with unloading, but all unloading curves are almost identical. (6) The first hysteresis loop differs from the later ones, which are almost identical to each other. From these cyclic curves, we decompose the energy into three components. \textbf{Energy decomposition:} The energy is quantified by areas bounded by the key points (A-K) and curves. By decomposition, the areas DCKD, EFGCDE (EFCDE for twist), and ABCGFEA (or ABCFEA for twist) correspond to the fabric's self-recoverable, recoverable, and unrecoverable deformations, respectively, as quantified by energy. The areas DCKD, DEFD, and AFEA measure the work done by the fabric's elastic force, the energy consumed to recover the recoverable deformation through unloading, and the energy consumed to undo permanent deformation~\cite{mao2014towards}, which determine persistent wrinkle formation.

\paragraph{Comparison with existing measurement systems.} Our method differs from existing approaches in key aspects. Unlike KES-F \cite{kawabata1980standardization} and FAST \cite{fast_manual}, which test flat specimens, we employ cylindrical geometry that better represents garment regions, e.g., sleeves and waists. KES-F requires four separate apparatus; our single apparatus can test compression, twist, and stretch. Unlike FAST's single-load tests, our cyclic loading-unloading captures hysteresis. For testers using cylindrical geometry, the standardized AATCC 128 \cite{aatcc128} provides only subjective rating scores or single-value crease recovery angles, which are insufficient for revealing fabric wrinkle formation. Prior cylindrical tests \cite{shaikhzadeh2009investigation} mix compression and torsion testing simultaneously without cyclic. Compared to custom and obscure setups \cite{miguel2013modeling, wang2010example}, our system uses commercially available, fully specified protocols, ensuring accuracy and reproducibility.

\section{Physical Model}

\begin{figure}[tb]
    \centering
    \includegraphics[width=\linewidth]{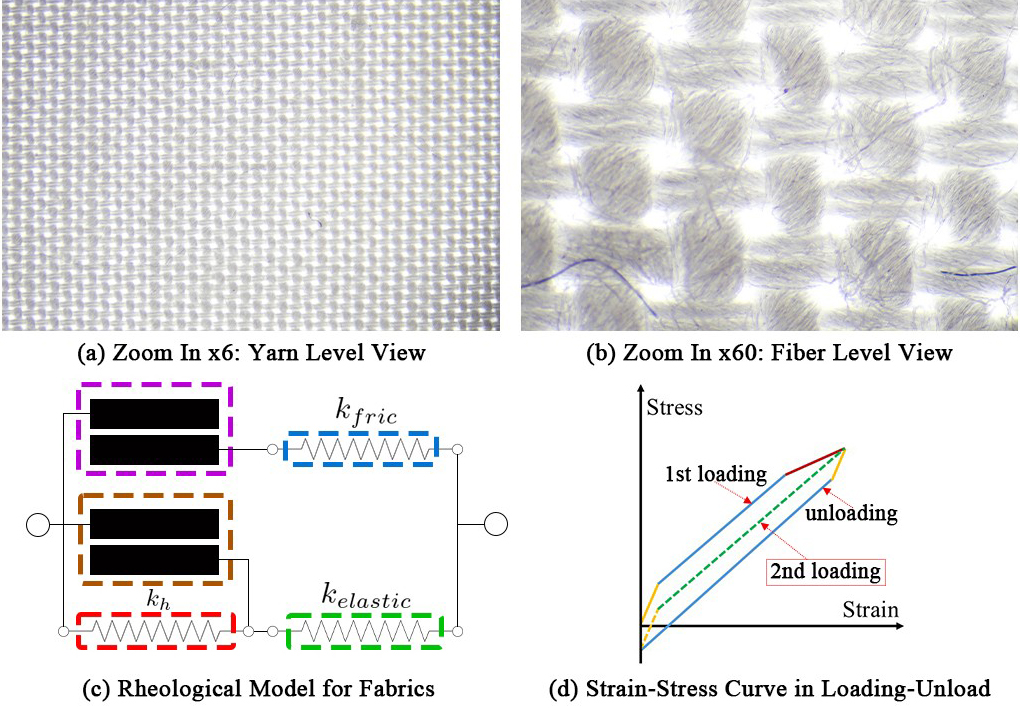}
    \caption{\rev{Rheological model (c) based on fabric micro-mechanics (a,b). The two springs (red and green) model the elasticity from geometry and material. The friction unit connected in parallel (purple) models the friction between yarns. The spring in blue models the interactions of fibers between contacting yarns. The friction unit in brown models the friction between fibers within yarns. When the model is under cyclic load-deformation, the strain-stress curve shows hysteresis (d).}}
    \label{fig:rhe_model}
\end{figure}

Based on the observation in Fig. \ref{fig:LUFHES_test_curves} and fabric micro-mechanics, our fabric constitutive law is illustrated by the rheological model in Fig. \ref{fig:rhe_model} (c). The two sequentially connected \rev{springs} (red and green) model yarn elastic deformation, decoupled into geometrical and material elasticity. Geometrical elasticity arises from yarn curvature, twisting, and repulsive contact between fibers. Material elasticity, in contrast, arises from the elasticity of the fibers themselves. They are activated simultaneously when yarns deform and are therefore connected in series. Furthermore, each yarn is composed of a bundle of fibers held together by friction, which is modeled by the friction unit (brown). When a yarn deforms and changes geometry, this friction activates to prevent fraying. Therefore, a friction unit is connected in \rev{parallel} with the spring modeling geometrical elasticity. This design enables our model to capture yarn hardening: large deformation can permanently change a yarn's fibers, making it less elastic thereafter. As yarns deform, the yarns move relative to each other. Friction between contacting yarns prevents unraveling and resists relative movement. The friction unit connected in parallel (purple) models this contact friction between adjacent yarns.  Inspired by the bristle model of contact friction, the connected elastic spring models the elastic deformation of bristles~\cite{dahl1968solid}, i.e., the fibers between two yarns deform due to their contacts (shown in Fig. \ref{fig:rhe_model} (b)). Essentially, the lower units can be considered as an elasto-plastic model with hardening plasticity\rev{~\cite{simo1998computational}}, and the upper units as an anchor-based friction model~\cite{gong2025cloth}. \rev{We acknowledge that we do not directly observe whether a specific deformation arises from yarn-level friction or plasticity. Instead, we use friction and plasticity as phenomenological modeling tools to explain the recoverable and unrecoverable components observed in the measured curves. This decomposition is based on the distinct signatures of these components in the cyclic loading-unloading data, rather than on direct physical measurement of yarn-scale mechanisms. The fabric stress is then defined as:}
\begin{equation}
    \sigma = \sigma_{elastic} + \sigma_{fric} = k_{elastic}(\varepsilon - \varepsilon_{p}) + k_{fric} \varepsilon_{fric}
    \label{eq:hysteresis_decomp}
\end{equation}
where \(k_{elastic}\) and \(k_{fric}\) denote the stiffness of elasticity and internal friction, defining the stiffness of the springs in the green and blue boxes in Fig. \ref{fig:rhe_model}, respectively. \(\varepsilon_{fric} = \varepsilon - \bar{\varepsilon}\) represents the frictional strain relative to the anchor strain \(\bar{\varepsilon}\). We adopt the additive decomposition \(\varepsilon = \varepsilon_{elastic} + \varepsilon_{p}\) to separate strain into elastic and plastic parts. \rev{Further analysis is provided in SM.}

\begin{figure}[tb]
    \centering
    \includegraphics[width=\linewidth]{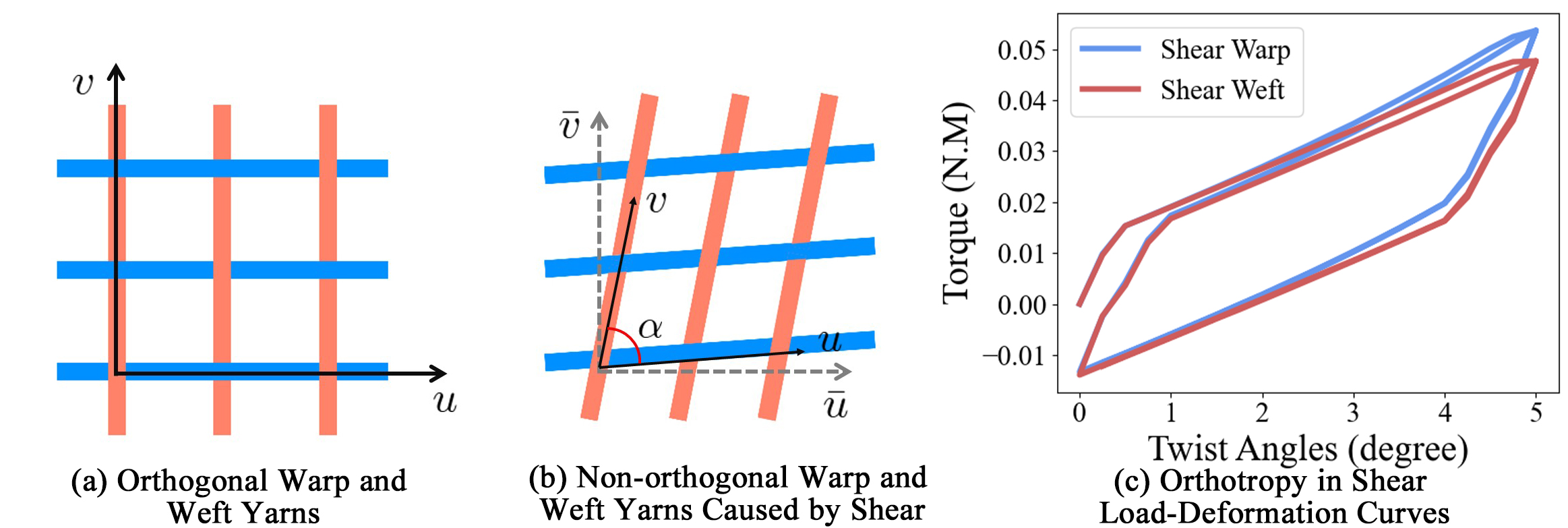}
    \caption{Modeling fabric shearing orthotropy. (a) Warp and weft yarns are perpendicular in the rest state. (b) Shearing deformation breaks the yarns' orthotropic structure. (c) This makes the measured reaction force affected by fabric stretching stiffness in warp and weft directions.}
    \label{fig:warp_weft_rotate}
\end{figure}

\paragraph{Elasticity}
We model cloth as a thin shell whose deformation is decomposed into in-plane stretching and out-of-plane bending. Our choice of constitutive laws is based on textile mechanics and observations from our measurements (Fig. \ref{fig:LUFHES_test_curves} and SM). Fabrics exhibit distinct stretching properties in the warp/wale/machine and weft/course/cross directions~\cite{quaglini2008experimental}, which we also observed in our \textbf{stretch} test. We therefore adopt the orthotropic St. Venant-Kirchhoff (StVK) model~\cite{volino2009simple,wang2010example}: $\boldsymbol{\sigma}=\tau \mathbf{K}\boldsymbol{\varepsilon} $ where the stiffness matrix \(\mathbf{K}\) distinguishes between warp, weft, and diagonal (i.e., shear) directions via diagonal components \(k_{11}\), \(k_{22}\), $k_{33}$. \(\tau\) is the fabric thickness. \cite{julio2006comparison} shows that the Yeoh model can accurately fit fabric shear nonlinearity. Inspired by this, we use a polynomial \rev{energy density} function $\psi_{shear}(\varepsilon_{uv}) = k_{33} \varepsilon_{uv}^2 = (c_{0} + c_{1} |\varepsilon_{uv}|) \varepsilon_{uv}^2$ ($\varepsilon_{uv}$ from vector $\boldsymbol{\varepsilon}$) to capture the nonlinearity observed in twist test. As shearing deformation increases, the warp and weft yarns rotate around their contact points, breaking fabric orthotropy~\cite{hu1997kes} and making shearing exhibit orthotropic behaviors. To model this phenomenon, we follow \cite{peng2005continuum} and employ a transformed stiffness matrix \(\tilde{\mathbf{K}}\):
\begin{equation}
    \mathbf{K} = \mathbf{T}^\top \tilde{\mathbf{K}} \mathbf{T}, \quad
    \mathbf{T} =
    \begin{bmatrix}
        1 & \cos^2 \alpha & 2 \cos \alpha \\
        0 & \sin^2 \alpha & 0 \\
        0 & \sin \alpha \cos \alpha & \sin \alpha
    \end{bmatrix}
\end{equation}
where \(\alpha\) is the angle between warp and weft yarns (Fig \ref{fig:warp_weft_rotate}). \rev{By doing this, the orthotropic stretching stiffness will be involved in the shearing stress when the fabric is twisted in different directions. Essentially, the difference between the twist load-deformation curves in warp and weft directions is caused by the different stretching stiffness in the warp and weft directions.}

For bending, we adopt the hinge-angle-based bending method~\cite{grinspun2003discrete} \rev{to formulate cloth bending strain}. To capture orthotropy, we decompose bending curvature $\kappa$ along the warp and weft directions. The bending stress tensor is defined as:
\begin{equation}
    \begin{bmatrix}
        \sigma_{b\_uu} \\
        \sigma_{b\_vv} \\
        \sigma_{b\_uv}
    \end{bmatrix}
    = \tau \mathbf{K}_b \boldsymbol{\kappa}
    = 
    \tau \begin{bmatrix}
        k_{b11} & k_{b12} & 0 \\
        k_{b12} & k_{b22} & 0 \\
        0 & 0 & k_{b33}
    \end{bmatrix}
    \begin{bmatrix}
        \kappa_{uu}  \\
        \kappa_{vv}  \\
        2\kappa_{uv}  
    \end{bmatrix}
    \label{eq:bending_stress}
\end{equation}
where $\kappa_{uu} = \kappa \cos^2\phi$, $\kappa_{vv} = \kappa \sin^2\phi$ and $\kappa_{uv} = \kappa \sin\phi \cos\phi$. $\phi$ is the bias angle between the bending edge and the warp direction in the fabric material coordinate. \rev{\(k_{b11}\) and \(k_{b22}\) define the bending stiffnesses in the warp and weft directions. The term \(k_{b12}\) creates spontaneous curvature in the perpendicular direction without applying an external moment there, analogous to a Poisson ratio but for curvature.The term \(k_{b33}\) is the torsional rigidity, which resists out-of-plane twisting deformation where the fabric bends in two principal directions simultaneously (i.e., bending diagonally). It controls how easily a cloth develops saddle-like curvatures.} To capture the hysteresis observed in the \textbf{twist} and \textbf{compression} test, we apply Eq. (\ref{eq:hysteresis_decomp}) to the shearing ($\varepsilon_{uv}$) and bending ($\kappa_{uu}$ and $\kappa_{vv}$).

\paragraph{Friction and Plasticity} The anchor-based friction model maintains an anchor strain \(\bar{\varepsilon}\), which updates only when slip occurs, i.e., when the strain deviates from \(\bar{\varepsilon}\) by more than a stick threshold \(\varepsilon_{thres}\):
\begin{equation}
    \bar{\varepsilon} \leftarrow \bar{\varepsilon} + \operatorname{sgn}(\varepsilon - \bar{\varepsilon}) \cdot \operatorname{ReLU}(|\varepsilon - \bar{\varepsilon}| - \varepsilon_{thres})
\end{equation}
Frictional stress then acts to restore the cloth toward the anchor strain: \(\sigma_{fric} = k_{fric} (\varepsilon - \bar{\varepsilon})\), where \(k_{fric}\) controls the friction intensity. For the hardening plastic model, plastic flow occurs only when the elastic strain exceeds a yield strain \(\varepsilon_y\), which itself hardens with accumulated plastic deformation \(\varepsilon_{hp}\):
\begin{align}
    \varepsilon_{hp} &\leftarrow \varepsilon_{hp} + \operatorname{ReLU}(|\varepsilon_e - \varepsilon_p| - \varepsilon_y) \\
    \varepsilon_y &= \varepsilon_{y0} + \frac{k_h}{k_{elastic}} \varepsilon_{hp}
\end{align}
where \(\varepsilon_{y0}\) is the initial yield strain and \(k_h\) defines the stiffness of the spring in the red box of Fig. \ref{fig:rhe_model}. The plastic strain updates as:
\begin{equation}
    \varepsilon_p \leftarrow \varepsilon_p + \operatorname{sgn}(\varepsilon_e - \varepsilon_p) \cdot \frac{k_{elastic}}{k_h + k_{elastic}} \cdot \operatorname{ReLU}(|\varepsilon_e - \varepsilon_p| - \varepsilon_y)
\end{equation}
ensuring that only a portion of the over-yield strain converts to $\varepsilon_p$.

\section{Differentiable Physical Simulation}

Our differentiable cloth simulator identifies cloth parameters $\Theta$ by minimizing the residual error between simulated and measured load-deformation curves via gradient-based optimization. It uses the Implicit Euler method~\cite{baraff2023large} to solve the motion equation \(\mathbf{f} = \mathbf{m}\ddot{\mathbf{x}}\). Given the initial state \(\mathcal{S}^{(0)} = \{\mathbf{x}^{(0)}, \dot{\mathbf{x}}^{(0)}\}\), the simulator \(f_{sim}\) advances state as \(\mathcal{S}^{(t)} = \{\mathbf{x}^{(t)}, \dot{\mathbf{x}}^{(t)}\}\), where physical force \(\mathbf{f}^{(t)} = f_{phy}(\mathcal{S}^{(t)}, \Theta)\) is computed at each step. For collision handling, we adopt the Incremental Potential Contact (IPC) method~\cite{li2020incremental}, which predicts the next state by minimizing:
\begin{equation}
    \mathbf{x}^{(t+1)} = \argmin_{\mathbf{x}^{(t+1)}} E(\mathbf{x}^{(t)})
    \label{eq:ipc}
\end{equation}
where 
\[
E(\mathbf{x}^{(t)}) = \frac{1}{2h^2} (\mathbf{x}^{(t+1)} - \hat{\mathbf{x}})^\top \mathbf{M} (\mathbf{x}^{(t+1)} - \hat{\mathbf{x}}) + \psi_e(\mathbf{x}^{(t)}) + B(\mathbf{x}^{(t)}) + D(\mathbf{x}^{(t)})
\]
\(\mathbf{M}\) is the lumped mass matrix, \(\psi_e\) the elastic potential, \(B(\mathbf{x})\) the contact barrier, \(D(\mathbf{x})\) the IPC friction energy, and \(\hat{\mathbf{x}} = \mathbf{x}^{(t)} + h \dot{\mathbf{x}}^{(t)} + h^2 \mathbf{M}^{-1} \mathbf{f}_{ext}\), with time step size \(h\) and the external force \(\mathbf{f}_{ext}\) ~\cite{kane2000variational}.

IPC was originally designed for hyperelastic materials. Incorporating plasticity and internal friction into \(\psi_e\) introduces history variables \(\boldsymbol{\varepsilon}_{p}\), \(\bar{\boldsymbol{\varepsilon}}\), and \(\boldsymbol{\varepsilon}_{hp}\) into state \(\mathcal{S}\). A fully implicit coupling would require solving the IPC minimization and the history-dependent constitutive update simultaneously, which complicates implementation and differentiation. Instead, we adopt a \textit{semi-implicit return mapping strategy}\rev{~\cite{simo1998computational,tu2009return}}, where the history variables are held fixed when solving Eq.~\eqref{eq:ipc}:
\begin{equation}
    \mathbf{x}^{(t+1)} = \argmin_{\mathbf{x}} \; E(\mathbf{x} \mid \boldsymbol{\varepsilon}_p^{(t)}, \bar{\boldsymbol{\varepsilon}}^{(t)}, \boldsymbol{\varepsilon}_{hp}^{(t)})
\end{equation}
and updated \textit{a posteriori} based on the predicted \(\mathbf{x}^{(t+1)}\) using a standard return mapping:
\begin{equation}
    \boldsymbol{\varepsilon}_p^{(t+1)}, \bar{\boldsymbol{\varepsilon}^{(t+1)}}, \boldsymbol{\varepsilon}_{hp}^{(t+1)} = \text{ReturnMapping}(\mathbf{x}^{(t+1)}, \mathcal{S}^{(t)})
\end{equation}
This semi-implicit approach suits our model well, as plastic yield and friction threshold involve non-smooth plastic flow and stick-slip transitions, which are known to be challenging for fully implicit integrators but can be naturally accommodated by freezing the history variables during the global solve~\cite{tu2009return}. The explicit a posteriori update handles non-smooth transitions correctly without compromising IPC solver convergence.

\rev{To compute gradients of the objective function $\mathcal{J}$ with respect to \(\Theta\)}, we adopt the adjoint method~\cite{givoli2021tutorial, huang2024differentiable}. The optimization problem is defined as:
\begin{equation}
    \argmin_{\Theta} \mathcal{J}(\mathbf{x}, \Theta) = \argmin_{\Theta} \sum^{T}_{t=0} \mathcal{J}(\mathbf{x}^{(t)}, \Theta)
\end{equation}
subject to \(\mathbf{M}\ddot{\mathbf{x}}^{(t)} = f_{phy}(\mathcal{S}^{(t)}, \Theta)\) and \(\mathcal{S}^{(0)}\) conforming to the given initial state \(\hat{\mathcal{S}}^{(0)}\), where \(\mathcal{J}\) is the objective function and $T$ is the number of time steps. Using the adjoint method, the gradient with respect to physical parameters, \(\frac{\partial \mathcal{J}}{\partial \Theta}\), can be computed by solving an \rev{adjoint equation} derived from the Lagrangian:
\begin{align}
    \mathcal{L}(\mathcal{S}, \Theta, \boldsymbol{\lambda}_1, \boldsymbol{\lambda}_2) &= \mathcal{J}(\mathbf{x}, \Theta) + \mathcal{L}_{cons}(\mathcal{S}, \Theta, \boldsymbol{\lambda}_1, \boldsymbol{\lambda}_2) \notag \\
    &\quad + \mathcal{L}_{init}(\mathcal{S}^{(0)}, \Theta, \boldsymbol{\lambda}_1^{(0)}, \boldsymbol{\lambda}_2^{(0)})
\end{align}
where \(\mathcal{L}_{cons}\) and \(\mathcal{L}_{init}\) enforce physical constraints and initial state constraints. We refer the reader to the SM for details \rev{about gradients}.

\section{Experiments}

Our differentiable simulator is implemented in C++/CUDA, with a PyTorch~\cite{paszke2019pytorch} interface for optimization. Physical parameter $\Theta$ includes $k_{11}, k_{22}, c_0, c_1, k_{b11}, k_{b22}, \boldsymbol{\varepsilon}_{thres}, \boldsymbol{k}_{fric}, \boldsymbol{\varepsilon}_{y0}, \boldsymbol{k}_h$ (bold symbols have three components for shearing, and bending in warp and weft direction respectively) are optimized using the Adam optimizer over 500 epochs, minimizing the mean squared error (MSE) between simulated and measured load-deformation curves. \rev{Ablation studies, training procedures, runtime profiles, and initialization strategies are also included in the SM.}

\subsection{Learn Cloth Physical Parameters}

\begin{figure}[tb]
    \centering
    \includegraphics[width=0.98\linewidth]{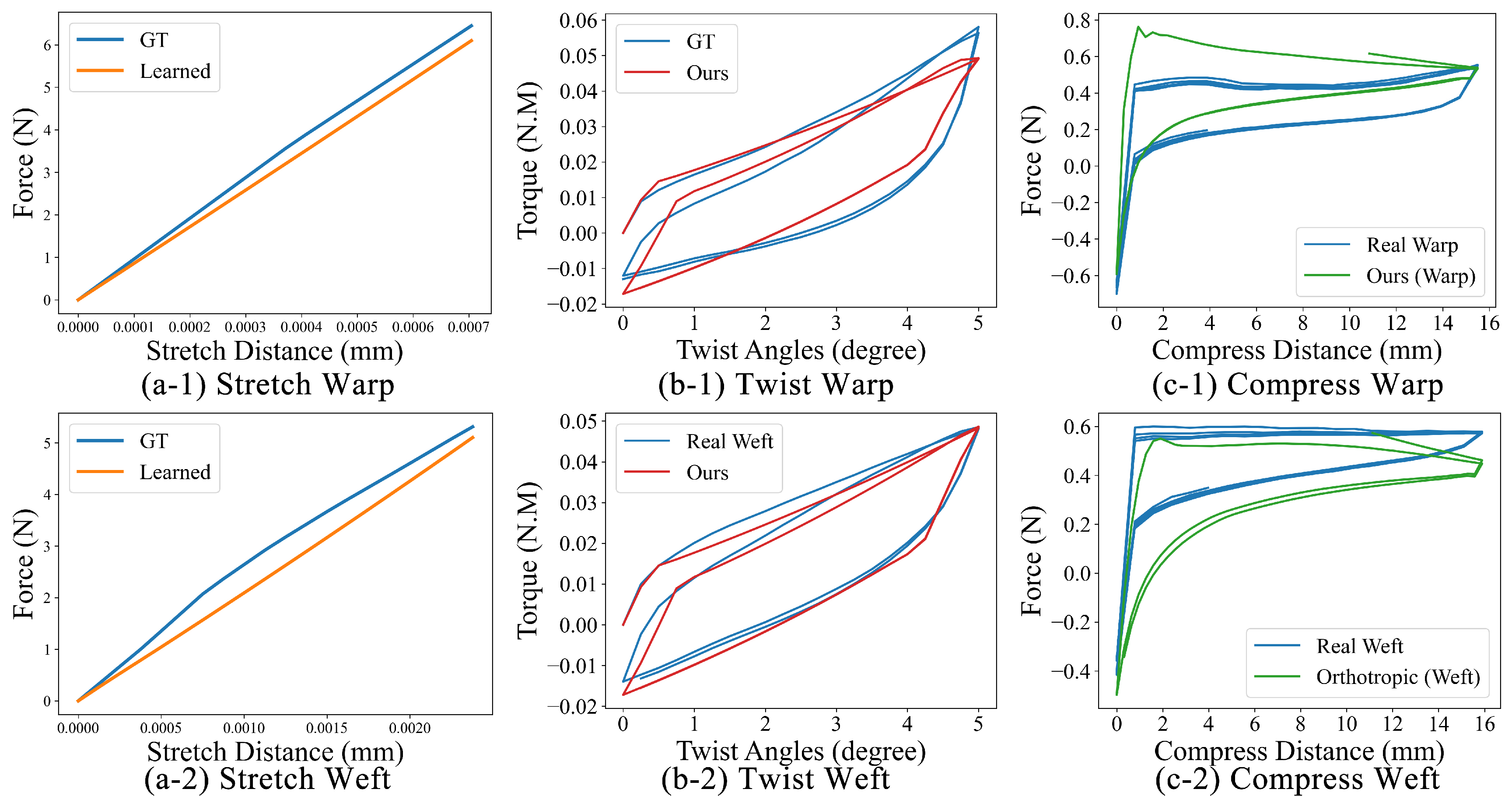}
    \caption{Our differentiable cloth simulator optimizes the physical parameters by fitting the load-deformation curves in the stretching, twisting, and compressing tests (a-1, a-2, b-1, c-1). By modeling the relative rotations of yarns, it can predict the twist load-deformation curves in the weft direction when the parameters learned from curves tested in the warp direction (b-2, c-2).}
    \label{fig:fit_curve}
\end{figure}

{\small
\begin{table}[tb]
    \centering
    \begin{tabular}{cccccc}
        \toprule
         Para/Mat  & Cotton & Linen & Wool & PET & Silk\\
         \midrule
         Str. Lr & 0.339 & 0.132 & 0.086 &0.555 &0.052 \\
         Twi. Lr & 1.124e-5 & 9.960e-7 & 1.027e-5 & 8.031e-6 & 3.980e-8 \\
         Comp. Lr & 0.086 & 0.010 & 0.005 & 0.022 & 0.006 \\
         \bottomrule
    \end{tabular}
    \caption{The MSE between the GT load-deformation curve and the curves simulated by using the learned parameters (X Lr).}
    \label{tab:fit_curve}
\end{table}
}

Our experiments demonstrate the learning capability of our differentiable simulator by optimizing physical parameters to fit measured load-deformation curves \rev{of five fabrics in our dataset, which exhibit different wrinkle resistance due to their materials}. \rev{For compression, we use only the post-buckling segment, as the initial buckling region is imperfection-sensitive and exhibits bifurcation behavior that our current model does not capture. The post-buckling response, which is more stable, still encodes the bending mechanics relevant to persistent wrinkle formation. Further details are provided in the SM.} As shown in Fig. \ref{fig:fit_curve}, optimized simulations closely match GT curves. Notably, due to our model's ability to account for yarn rotation, it accurately predicts twist deformation in the weft direction even when trained only on warp data. Quantitative results (Tab. \ref{tab:fit_curve}) confirm that parameter optimization reduces simulation error, a task prohibitively difficult via manual tuning. Using the learned parameters, we simulate persistent wrinkle formation by compressing cylindrical fabric samples and compare against real-world counterparts (Fig. \ref{fig:teaser}). The simulations reproduce material-specific behaviors: cotton and linen form pronounced wrinkles, while wool, silk, and PET show minimal permanent deformation. Our simulator further captures the difference between recoverable and unrecoverable wrinkles: dragging the sample upward flattens recoverable wrinkles, while unrecoverable wrinkles persist. This agreement between real and simulated fabrics demonstrates that our simulator, with learnable internal friction and plasticity, faithfully reproduces both recoverable and unrecoverable wrinkles across different materials. Additional results are in the SM.

\subsection{Different Wrinkles on Garments}

\begin{figure*}[h!]
    \centering
    \includegraphics[width=\textwidth]{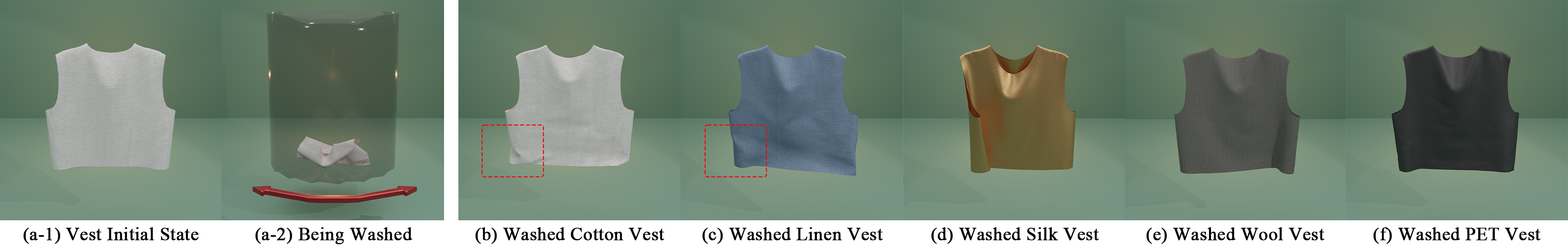}
    \caption{A comparison of the wrinkles formed on a vest after washing is shown in (a-1,2). Both the cotton (a) and linen (b) vests develop noticeable wrinkles. The wrinkles on the linen vest are larger due to its higher bending stiffness, which helps the fabric resist the flattening effect of its own weight. In contrast, silk (c) and wool (d) are wrinkle-resistant, and the vests made from these materials show no visible wrinkling. The PET vest (e) forms only very fine wrinkles, which are only noticeable when viewed closely. Garment Mesh Res: 10799 Vertices.}
    \label{fig:compare_washed_vest}
\end{figure*}

\begin{figure*}[h]
    \centering
    \includegraphics[width=\textwidth]{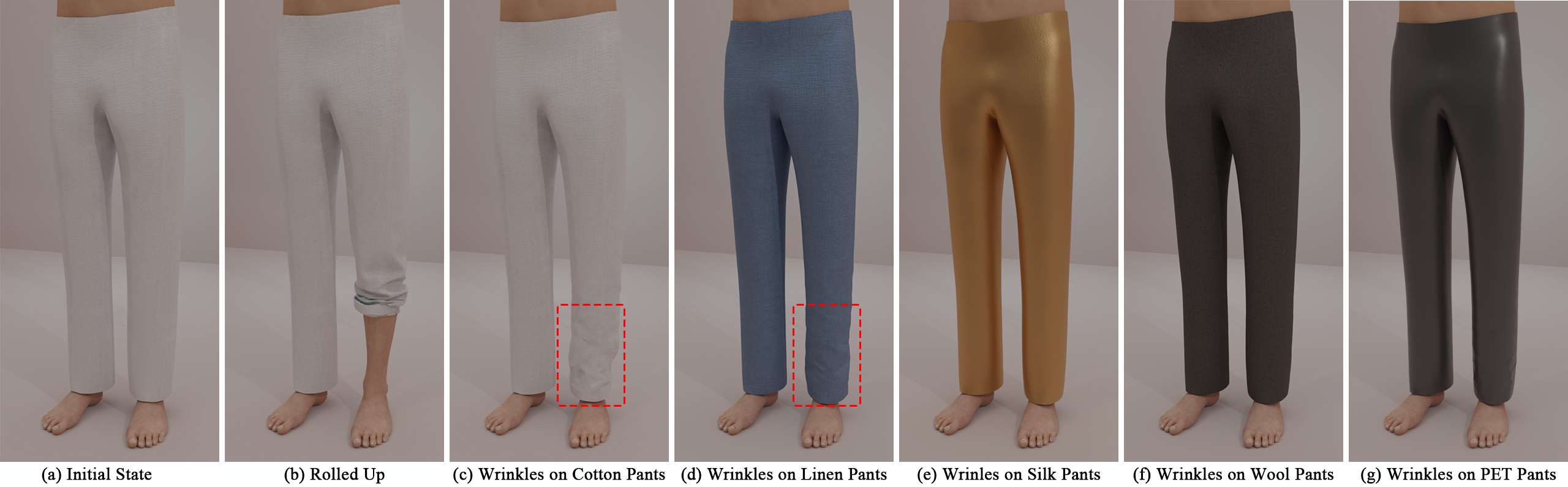}
    \caption{A comparison of the wrinkles formed on pants after one leg is rolled up. The pants initially have no wrinkles (a) and are deformed by rolling up one leg (b). Both cotton (c) and linen (d) pants develop clear persistent wrinkles after the leg is released. In contrast, the silk (e), wool (f), and PET (g) pants show greater wrinkle resistance. Due to the weight of the pant leg itself, the fine wrinkles that form on the PET pants are pulled flat and are therefore not visually noticeable. Garment Mesh Res: 73282 Vertices.}
    \label{fig:roll_up_pants}
\end{figure*}

\begin{figure*}
    \centering
    \includegraphics[width=\textwidth]{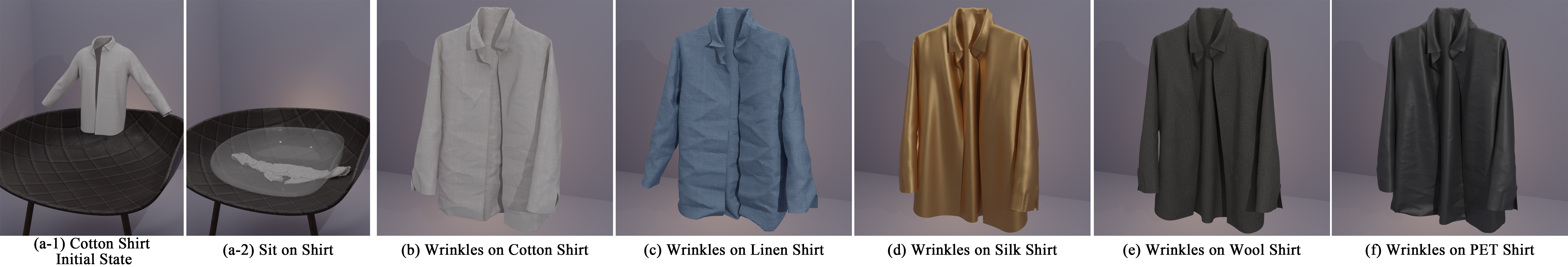}
    \caption{A comparison of the wrinkles formed on different shirts after being sat upon is shown in (a-1,2). Cotton (b) and linen (c) shirts are prone to wrinkling, with the wrinkles on the linen shirt appearing larger due to its higher bending stiffness. Silk (d) and wool (e) shirts show wrinkle resistance, consistent with real-world observations of these fabrics. The PET (f) shirt develops only fine, subtle wrinkles. Garment Mesh Res: 56250 vertices.}
    \label{fig:compare_sit_on_shirt}
\end{figure*}

\begin{figure*}
    \centering
    \includegraphics[width=\textwidth]{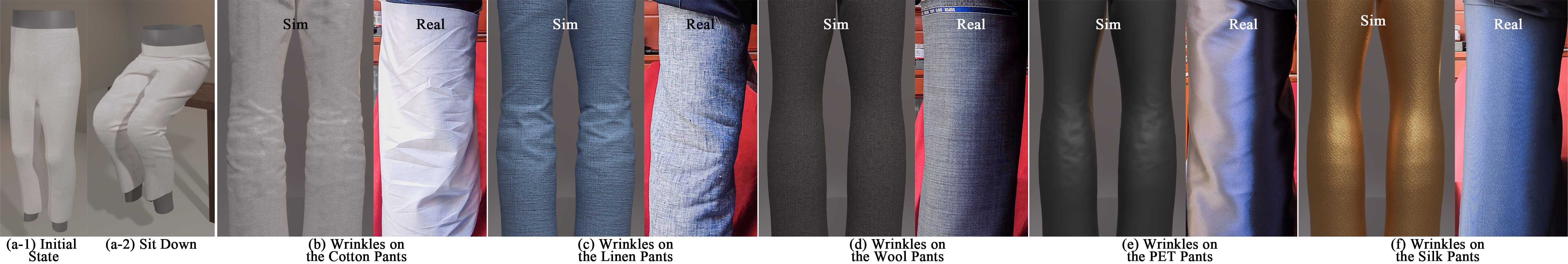}
    \caption{A comparison of persistent wrinkles formed on the back of pants after sitting down. Following the motion, the back of the cotton (b) and linen (c) pants develops distinct new wrinkles around the knee area. In contrast, the silk (d) and wool (e) pants show no visible wrinkling, as these fabrics are naturally wrinkle-resistant. The fine wrinkles that form on the PET (f) pants are not readily apparent, as the weight of the fabric itself causes them to flatten. The same wrinkles are observed on real correspondences. Garment Mesh Res: 35715 vertices.}
    \label{fig:sit_pants}
\end{figure*}

Persistent wrinkles appear as soft folds under moderate deformation or sharp creases under extreme deformation. By jointly modeling internal friction and plasticity, our simulator captures this full spectrum across different materials and deformation regimes. Although physical parameters for each fabric are learned from small samples, our simulator generalizes to larger, complex garments like vests, shirts, and pants.
\textbf{Wrinkles in Moderate Deformations.} 
Moderate wrinkles often form when garments rest under their own weight. We simulate this by spinning a vest in a virtual washing machine drum (Fig. \ref{fig:compare_washed_vest} (a)). After washing, soft wrinkles appear on cotton and linen vests (Fig. \ref{fig:compare_washed_vest} (b,c)), while PET develops finer creases (Fig. \ref{fig:compare_washed_vest} (d)). Wool and silk vests remain nearly wrinkle-free. Similarly, when rolling up one pant leg and releasing (Fig. \ref{fig:roll_up_pants} (a,b)), distinct wrinkles persist on cotton and linen (Fig. \ref{fig:roll_up_pants} (c,d)), while silk, wool, and PET show minimal wrinkling; for PET, fabric weight pulls small creases flat (Fig. \ref{fig:roll_up_pants} (e,f,g)). \textbf{Wrinkles in Extreme Deformations.} 
Under severe compression, fabrics develop sharp, lasting wrinkles. We simulate a shirt after being sat upon (Fig. \ref{fig:compare_sit_on_shirt}). Upon lifting, sharp wrinkles appear on cotton and linen (Fig. \ref{fig:compare_sit_on_shirt} (b,c)), while silk and wool remain largely smooth—though the wool collar retains slight deformation (Fig. \ref{fig:compare_sit_on_shirt} (d,e)). PET retains fine, sharp wrinkles (Fig. \ref{fig:compare_sit_on_shirt} (f)). Persistent wrinkles also form under body motion: sitting severely deforms the knee area of pants, leaving persistent creases upon standing (Fig. \ref{fig:sit_pants} (a)). Again, cotton and linen exhibit distinct wrinkles (Fig. \ref{fig:sit_pants} (b,c)), while other fabrics show subtler effects; on PET, fine wrinkles are pulled flat by the fabric's own weight (Fig. \ref{fig:sit_pants} (d,e,f)).

\begin{figure}[tb]
    \centering
    \includegraphics[width=\linewidth]{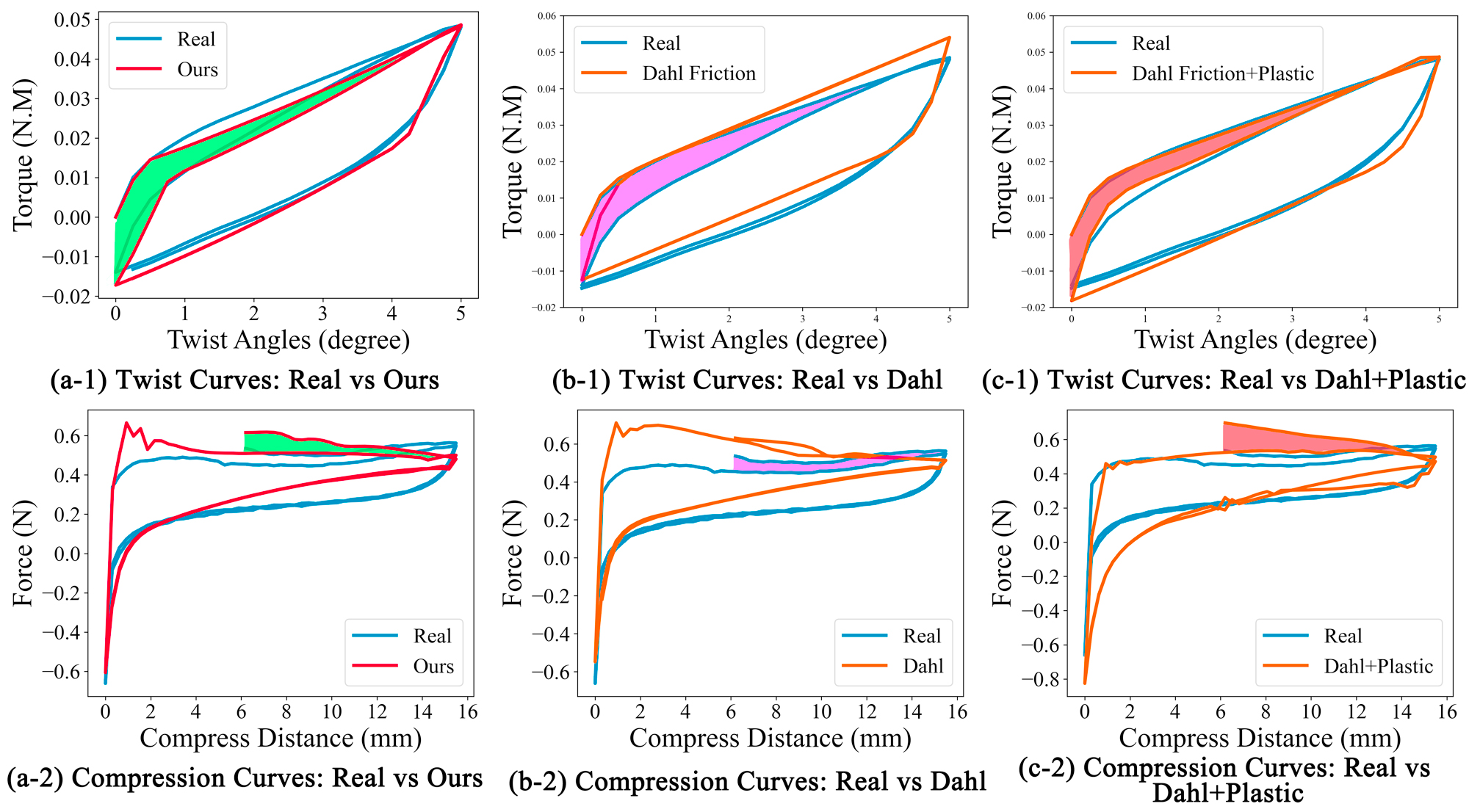}
    \caption{Dahl friction model cannot capture the non-local memory effect of hysteresis in the measurement (pink zone in (b-1,2)). By contrast, by combining friction and plasticity, our method can capture the non-local memory effect and more closely fit the measurements (green zone in (a-1,2)). Our model's friction model can be replaced by Dahl model, which can also capture the non-local memory (orange zone in (c-1,2)).}
    \label{fig:compare_dahl_curve}
\end{figure}

\subsection{Paper Wrinkles}

\begin{figure*}[htb]
    \centering
    \includegraphics[width=\linewidth]{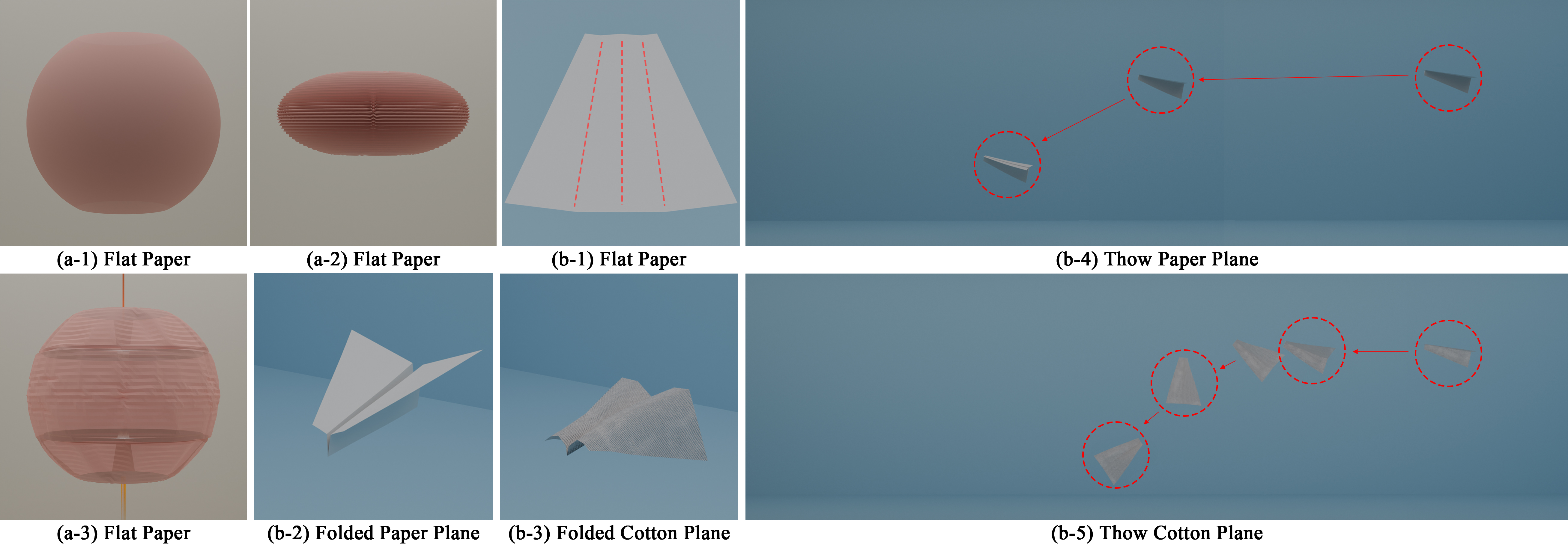}
    \caption{Crafting a paper lamp (a-1,2,3). We begin with a smooth, wrinkle-free spherical paper (a-1) and fold it to introduce wrinkles (a-2). The paper lamp is then crafted by unfolding the paper and supporting it with a frame (a-3). Furthermore, we simulate the construction of a paper plane by folding a sheet along the dashed lines (b-1). The folded paper plane retains its shape after folding (b-2). In contrast, a plane made from cotton fabric is too soft to hold its shape (b-3). Consequently, the paper plane flies forward steadily when thrown (b-4), whereas the cotton plane deforms upon release and descends more rapidly (b-5).}
    \label{fig:paper_wrinkles}
\end{figure*}

Our framework also applies beyond fabrics. We demonstrate its versatility by characterizing standard A4 printing paper, which exhibits pronounced hysteresis and forms sharp, stable wrinkles even under small deformations. We measure its load-deformation curves and estimate its physical parameters. Using these parameters, we simulate folding a spherical sheet into a paper lamp (Fig. \ref{fig:paper_wrinkles} (a)). The simulation captures paper's tendency to accumulate fine, rigid, persistent creases. For comparison, we simulate folding a paper airplane (Fig. \ref{fig:paper_wrinkles} (b)). The combined effects of internal friction and plasticity allow the paper to retain its folded shape after release, even under airflow. In contrast, a cotton fabric airplane lacks the same rigidity; it deforms more easily and descends faster. These results highlight how our simulator captures distinct wrinkle-forming behaviors, from soft textiles to stiff paper.

\subsection{Comparisons}

\begin{figure*}[h]
    \centering
    \includegraphics[width=\textwidth]{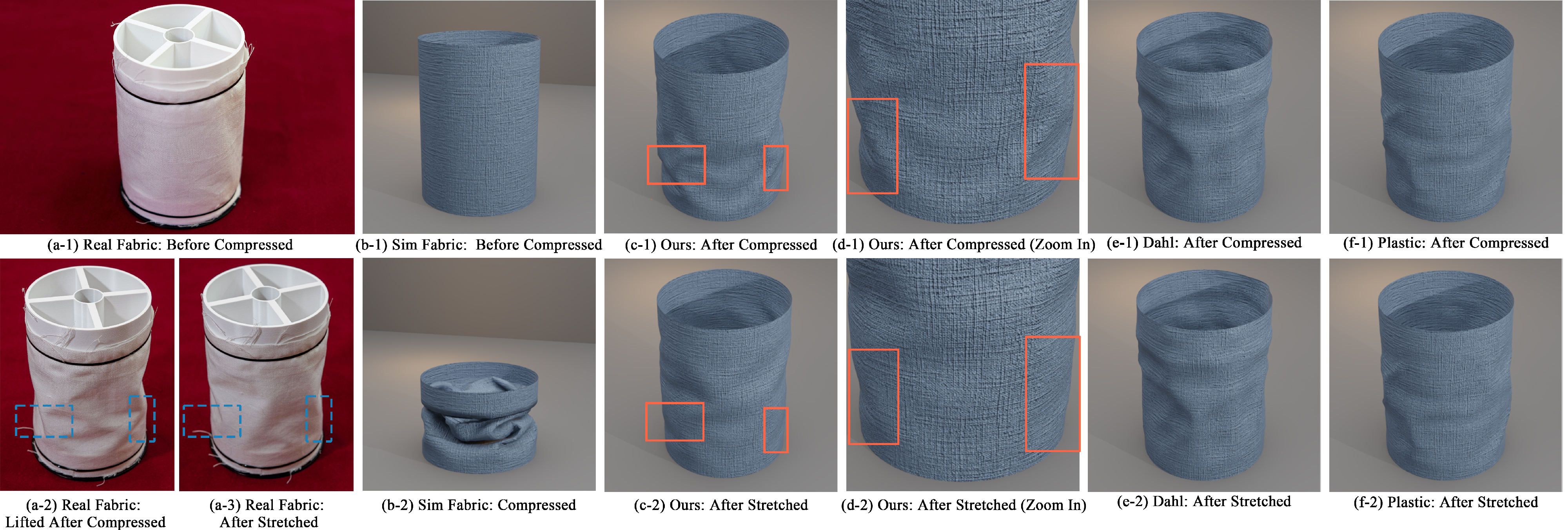}
    \caption{In real fabrics, some persistent wrinkles can be smoothed out by stretching (a-1,2,3). Our simulator reproduces this behavior through the integration of a friction model, which captures the recoverable nature of such wrinkles (b,c,d-1,2). This effect cannot be simulated using either the Dahl friction model~\cite{miguel2013modeling} or the hardening plasticity model~\cite{narain2013folding} (e,f-1,2).}
    \label{fig:comparison}
\end{figure*}

Our work advances beyond prior approaches by integrating internal friction and plasticity in a unified, differentiable framework, enabling simulation and parameter estimation of persistent wrinkles and hysteresis. We compare with \cite{miguel2013modeling} and \cite{narain2013folding}, which either use friction or plasticity to model persistent wrinkles. Compared with \cite{miguel2013modeling}, which uses a Dahl friction model for hysteresis, our method addresses three limitations. First, Dahl friction alone cannot account for plasticity. While it can retain wrinkles via large friction coefficients, it simultaneously hinders recovery, failing to match reversible wrinkle behavior (Fig. \ref{fig:comparison} (a)). Second, their framework cannot test shearing parameters due to limited experimental setup. Third, the Dahl model lacks non-local memory~\cite{mayergoyz2003mathematical}; its stress depends only on the current and immediate prior state, preventing it from reproducing measured full hysteresis loops (Fig. \ref{fig:compare_dahl_curve} (b-1,2)). Our model naturally incorporates non-local memory via plasticity, faithfully matching observed curves. Replacing our Coulomb friction (2 parameters) with Dahl (3 parameters) also retains non-local memory from plasticity (Fig. \ref{fig:compare_dahl_curve} (c-1,2)). \cite{narain2013folding} models persistent wrinkles using only hardening plasticity. Purely for simulation without parameter estimation, it requires manual tuning. Moreover, lacking internal friction, it cannot capture recoverable wrinkles (Fig. \ref{fig:comparison} (f)). To our knowledge, our simulator is the first to cohesively combine friction and plasticity in a differentiable manner, enabling accurate reproduction of measured hysteresis and faithful simulation of the full wrinkle spectrum: from recoverable folds to permanent creases. \rev{Finally, although \cite{gong2025cloth} combines internal friction and plasticity to model wrinkles, their simulator lacks learning ability and relies on tedious parameter tuning.}

\section{Discussion and Conclusion}

In this paper, we investigate learning fabric physical parameters to reproduce persistent wrinkles. We first create a dataset, \textbf{Fabric-101}, using a high-precision fabric testing apparatus. With this data, we present the first differentiable cloth simulator that integrates internal friction and plasticity to model persistent wrinkles. Our experiments show that the simulator accurately fits measured hysteresis and, using learned parameters, realistically simulates persistent wrinkles across diverse fabrics, without relying on manual parameter tuning. By unifying friction and plasticity, our model captures both recoverable and permanent deformation, advancing beyond prior works that rely solely on friction or plasticity. We further demonstrate generality on paper, where our simulator estimates physical parameters and simulates stable, sharp folds. 

\rev{One limitation is that our dataset and model do not yet include time, temperature, viscosity, or moisture. Additionally, we estimate bending parameters only from the post-buckling curve, as thin-wall buckling is imperfection-sensitive and bifurcates. Since replicating real imperfections is difficult, we use a perfect cylindrical mesh and cannot model buckling bifurcation, which leads to larger compression fitting errors. Our work only phenomenologically decouples friction and plasticity by modeling recoverable and unrecoverable deformations. Future work will explore decoupling contact yarn/fiber friction from permanent yarn/fiber deformation. Our shearing model attributes orthotropy to yarn rotation and stretching, which is insufficient for knitted fabrics. We discuss these in SM. Finally, predictive validation on unseen deformation amplitudes is left for future work.}

\section{Acknowledgments}
This work was partially funded by the HKSAR ITC under ITSP-Platform grants (Ref: ITS/335/23FP, ITS/469/24FP). We also thank Roaches (\url{https://www.roaches.co.uk}) and their team for their technical support.

\bibliographystyle{ACM-Reference-Format}
\bibliography{sample-bibliography}


%% file: siga_app_arxiv.tex
\pagebreak
\clearpage

\appendix

\section{Fabric Dataset}

The \textbf{Fabric-101} dataset comprises 104 fabrics in total, classified by fiber type: Cotton (38), Linen (6), Wool (10), Silk (16), Synthetic (24), and others such as bamboo, modal, and Lyocell (10). For each fabric, we test both warp/wale and weft/course directions to capture material orthotropy. In addition, we include 4 types of paper (2 for printing, 2 for folding) with different thickness and density. For every sample, we provide measured load-deformation curves, thickness, area density, as well as macro and micro photographs. The full dataset, including all raw measurements and metadata, will be released upon acceptance. A complete listing is provided in \Cref{tab:fabric1,tab:fabric2,tab:fabric3}. \rev{Our dataset is
\begin{itemize}[leftmargin=*]
    \item \textbf{Inclusive:} Including common commercial fabrics across diverse materials, woven patterns, thickness, and density;
    \item \textbf{Extensible:} The dataset can be extended in future work for learning the physical parameters of new fabrics as we adapt the commercially available testing apparatus and follow the documented testing standards;
    \item \textbf{Accurate:} The professional apparatus, textile testing standards, and controlled environment ensure the accuracy of the measurement.
\end{itemize}}

The five fabrics and paper used in our experiments are highlighted by bold font in the tables, where we refer to the Acetate Lining (R04) as PET for short. Compared with fabrics, paper is thinner and lighter. This is also why the folded paper plane can fly: the lift force from the air flow can counteract the paper plane's self-weight. In addition to the cotton fabric's load-deformation curves and photos provided in the main paper, \Cref{fig:linen_curve,fig:silk_curve,fig:wool_curve,fig:PET_curve} provide all the measured load-deformation curves and yarn/fiber-level view photos of the linen, silk, wool, and PET fabrics used in our paper. 

\begin{revsection}

\section{Distinguish Elasticity, Friction, and Plasticity}

\begin{figure*}[htb]
    \centering
    \includegraphics[width=\linewidth]{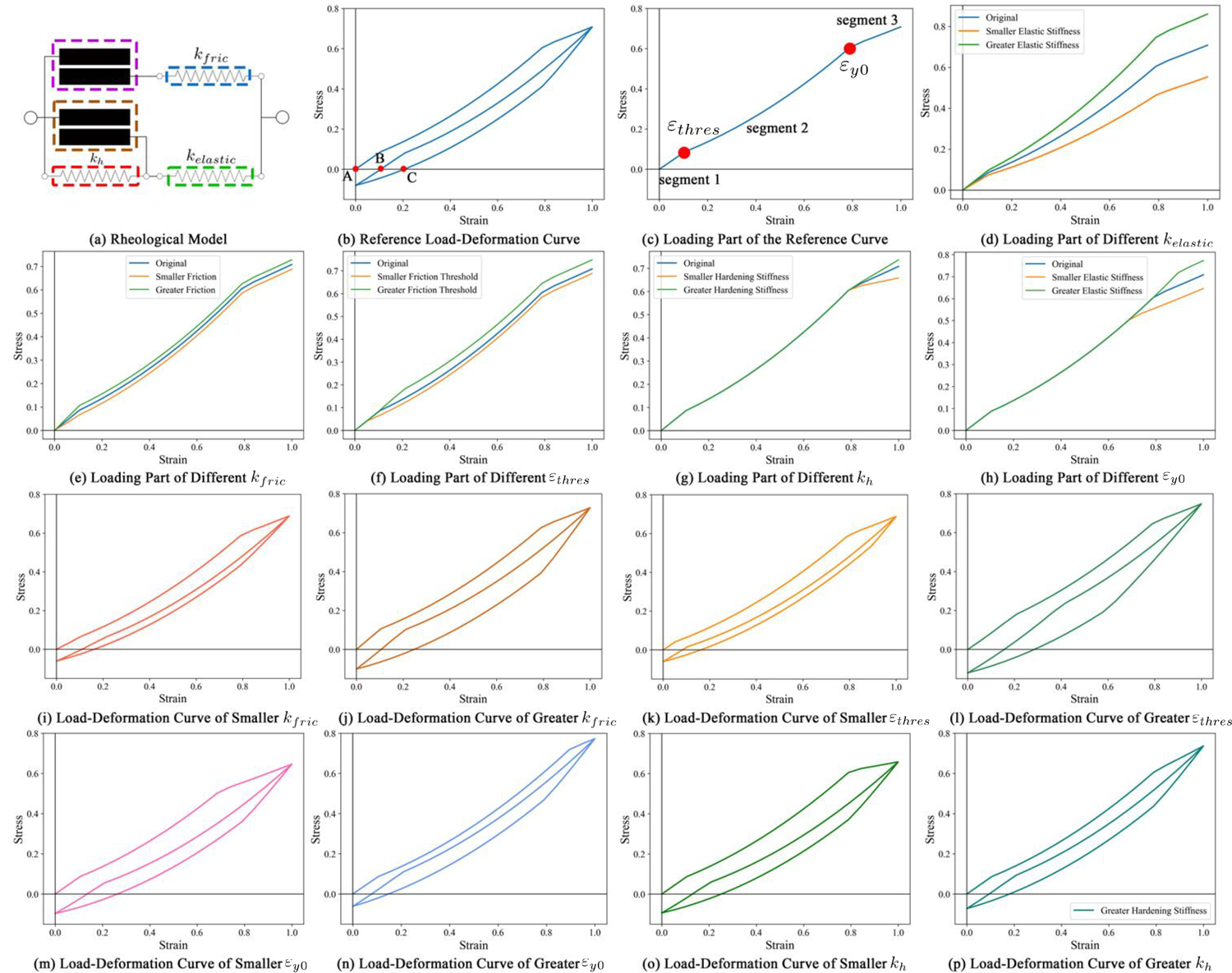}
    \caption{\rev{Distinguish the elastic, frictional, and plastic components from the cycle load-deformation curve (b) in our 1-dimensional rheological model (a). The stiffness parameters $k_{elastic}$, $k_{fric}$, and $k_h$ determine the slope of the three segments delimited by the friction threshold $\varepsilon_{thres}$ and yield strain $\varepsilon_{y0}$ (c-h). The frictional parameters $k_{fric}$ and $\varepsilon_{thres}$ affect the area between the second loading curve and the unloading curve. The plastic parameters $k_h$ and $\varepsilon_{y0}$ affect the area between the first and second loading curve.}}
    \label{fig:tell_friction_plastic}
\end{figure*}

The three deformation components—elastic, frictional, and plastic—can be distinguished from cyclic loading-unloading curves through their distinct effects on different segments of the curve, as illustrated in \Cref{fig:tell_friction_plastic}.

The first loading curve can be divided into three segments, as shown in \Cref{fig:tell_friction_plastic} (c). The slope of the first segment is governed by both friction and elastic stiffness (\Cref{fig:tell_friction_plastic} (d, e)), the second segment by elastic stiffness alone (\Cref{fig:tell_friction_plastic} (d)), and the third segment by hardening parameters (\Cref{fig:tell_friction_plastic} (g)). Varying the friction stiffness or hardening parameters affects only the first (\Cref{fig:tell_friction_plastic} (d)) or last segment (\Cref{fig:tell_friction_plastic} (g)), respectively, while changing the elastic stiffness alters the slopes of both the first and second segments simultaneously (\Cref{fig:tell_friction_plastic} (e)). The boundaries between these segments are determined by the friction threshold and plastic yield strain (\Cref{fig:tell_friction_plastic} (f, h)).

The intersections between the abscissa axis and the curves indicate the strain where the reaction force is zero. The gap between the first intersection point and the third intersection point means that loading changes the rest strain. The gap between the second and third intersection points means unloading to the initial strain returns the rest strain to the original rest strain, i.e., the rest strain is recovered. The gap between the first and second intersection points means that the deformation caused by the loading is not fully recovered by the unloading. The friction and plastic components in the rheological model account for the recoverable and unrecoverable deformations in the cycle loading-unloading. Only the first loading incurs unrecoverable deformation governed by the plastic model. The unloading and subsequent loading curves are governed purely by elastic and friction parameters, each exhibiting a single turning point determined by the friction threshold. Consequently, as shown in \Cref{fig:tell_friction_plastic} (i-l), the hysteresis between unloading and the second loading is purely frictional, while the hysteresis between the first loading and unloading reflects both friction and plasticity (\Cref{fig:tell_friction_plastic} (m-p)). This distinction allows the three components to be identified from the different roles they play across the cycles.

\begin{figure}[tb]
    \centering
    \includegraphics[width=\linewidth]{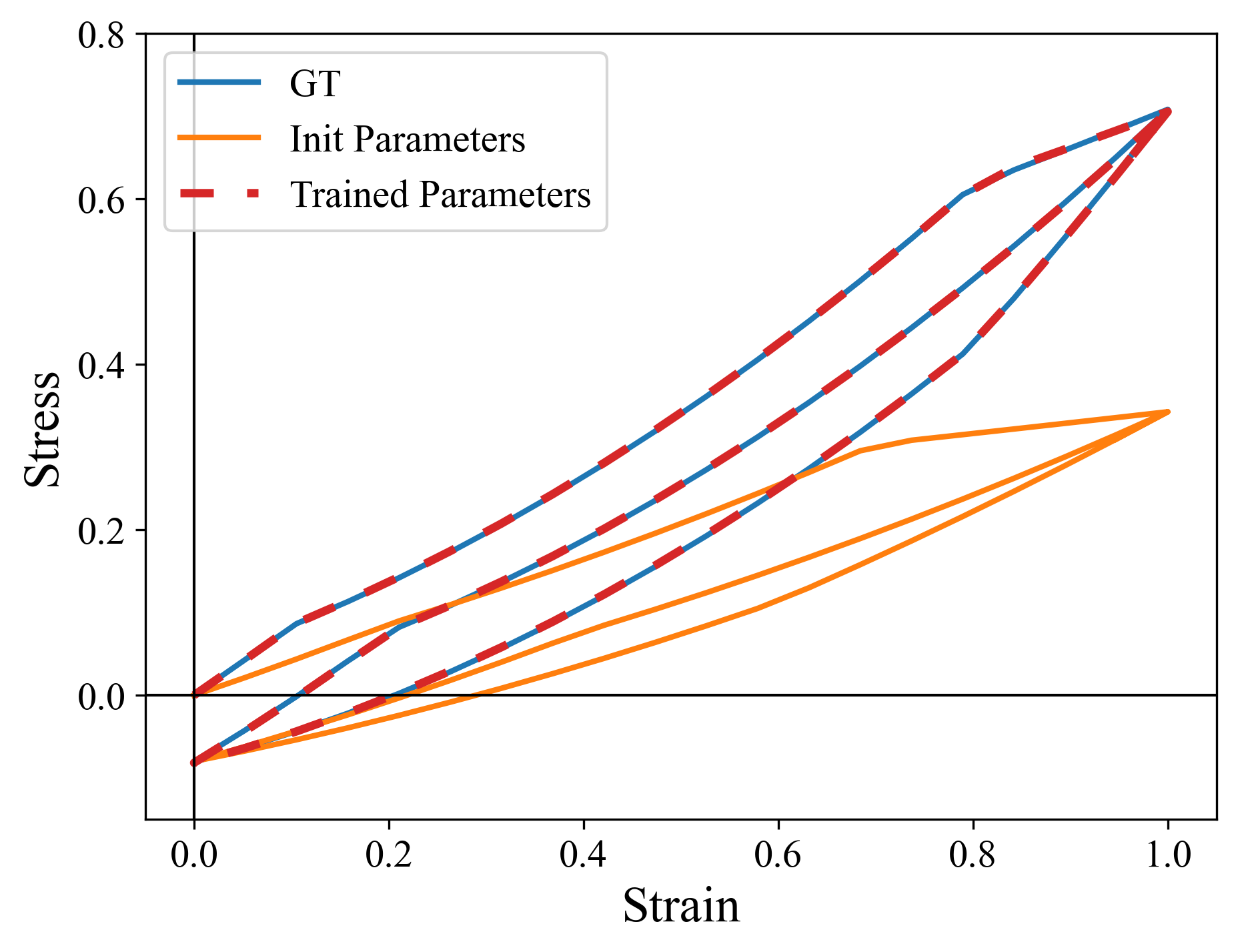}
    \caption{\rev{Fit the ground truth load-deformation curve simulated in the 1-dimensional rheological model by estimating the physical parameters with gradient-based optimization.}}
    \label{fig:train_1d}
\end{figure}

\begin{table}[bt]
    \centering
    \caption{\rev{Gradient-based optimization can accurately estimate the elastic, frictional, and plastic physical parameters by fitting the ground truth load-deformation curve. The elastic spring is nonlinear whose stiffness is controlled by two parameters $c_0$ and $c_1$: $\sigma = c_0 \varepsilon + c_1 \varepsilon^2$.}}
    \begin{tabular}{ccccccc}
        \toprule
         Params &  $c_0$ & $c_1$ &$k_{fric}$ & $\varepsilon_{fric}$ & $\varepsilon_{y0}$ & $k_h$\\
         \midrule
         GT & 0.4 & 1.0 & 0.4 & 0.1 & 0.8 & 0.3 \\
         Init & 0.3 & 0.5 & 0.1 & 0.2 & 0.7 & 0.1 \\
         Trained & 0.409 & 0.947 & 0.391 & 0.101 & 0.808 & 0.272 \\
         \bottomrule
    \end{tabular}
    \label{tab:train_1d}
\end{table}

Given the simulated load-deformation curve in this one dimensional rheological model, we use gradient descent to optimize the elastic, frictional, and plastic parameters $k_{elastic}$, $k_{fric}$, $\bar{\varepsilon}_{fric}$, $\varepsilon_{y0}$, and $k_h$. $k_{elastic}$ consists of two parameters ($c_0$ and $c_1$) to model nonlinear elasticity: i.e., $\sigma= (c_0 + c_1|\varepsilon|)\varepsilon$, as our shearing model. \Cref{fig:train_1d} and \Cref{tab:train_1d} demonstrate that gradient-based optimization can estimate these physical parameters and closely fit the ground truth load-deformation curve. These observations confirm that the elastic, frictional, and plastic components can be distinguished from the three curves. 

\end{revsection}

\begin{revsection}

\section{Compared with Bingham-Norton Model}

\begin{figure*}[htb]
    \centering
    \includegraphics[width=\textwidth]{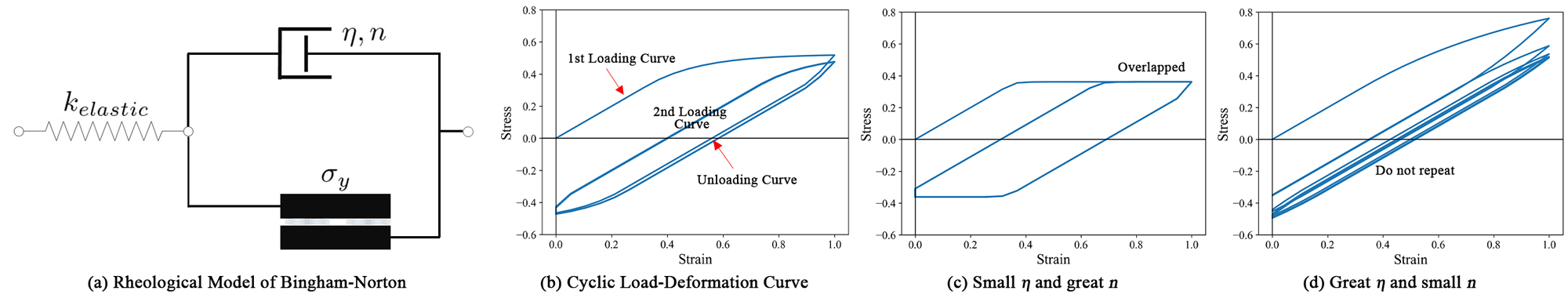}
    \caption{\rev{(a) The Bingham-Norton model is an elastic-viscoplastic model where the dashpot can model time-dependent mechanics. (b) The load-deformation curve of the Bingham-Norton model under uniaxial cyclic load-unloading. (c) When $\eta$ is small and $n$ is great, plastic flows fast. The first and second loading curves overlap after plastic deformation occurs. (d) When $\eta$ is great and $n$ is small, plastic flows slow. The cyclic load-unloading curves are not repeated.}}
    \label{fig:bn_model}
\end{figure*}

The Bingham model is the most basic elastic-viscoplastic model; the Bingham-Norton model extends it by incorporating Norton's creep law, which introduces nonlinearity in the plastic flow rate (shown in \Cref{fig:bn_model} (a)). The plastic strain rate is governed by:
\begin{equation}
    \dot{\varepsilon}_p = \frac{1}{\eta} \left( \frac{|\sigma|}{\eta} \right)^{(n-1)} \mbox{ReLU}(|\sigma| - \sigma_y) \operatorname{sgn}(\sigma),
\end{equation}
where \(\eta\) is the viscosity, \(n\) controls the strain-rate sensitivity, and \(\sigma_y\) is the yield stress. In the Bingham-Norton model, hysteresis arises purely from viscoplasticity: the viscosity \(\eta\) and exponent \(n\) determine how much strain exceeding \(\sigma_y\) is converted into plastic strain over time.

As illustrated in \Cref{fig:bn_model} (b), the gap between the first and second loading curves originates from the time-dependent nature of plastic flow. If \(\eta\) is reduced and \(n\) increased, the strain exceeding \(\sigma_y\) is converted almost immediately, and the first and second loading curves overlap after plastic deformation occurs (shown in \Cref{fig:bn_model} (c)). Conversely, increasing \(\eta\) and decreasing \(n\) slows the plastic flow, breaking the repeatability of subsequent cycles. The strain exceeding $\sigma_y$ does not have time to fully convert into plastic strain in the first loading. The unconverted part will be postponed to the second loading procedure, which, however, does not have time to fully convert either. The plastic strain continuously evolves in cyclic loading-unloading (shown in \Cref{fig:bn_model} (d)). Repeatability only emerges when the strain increments per cycle become sufficiently small, allowing the plastic flow to balance within each cycle. These two cases are rarely observed in our dataset.

Moreover, a key distinction between the Bingham-Norton model and ours is rate-dependence: Bingham-Norton is inherently time-dependent, while our model is rate-independent and quasi-static. We acknowledge that fabrics exhibit visco-plasticity, which can influence hysteresis and creeping behavior, and visco-plastic models have been used to simulate permanent deformations in graphics \cite{bargteil2007finite,li2022energetically,schreck2020practical}. However, our apparatus is designed to operate at a deliberately slow speed to minimize rate effects and evaluate quasi-static fabric mechanics. Moreover, Bingham-Norton overlooks two important aspects of fabric behavior: plastic hardening and the coexistence of internal friction. 


\end{revsection}

\section{Additional Experiments}

\subsection{Wrinkles Formed in Different Deformation}

\begin{figure*}[htb]
    \centering
    \includegraphics[width=\textwidth]{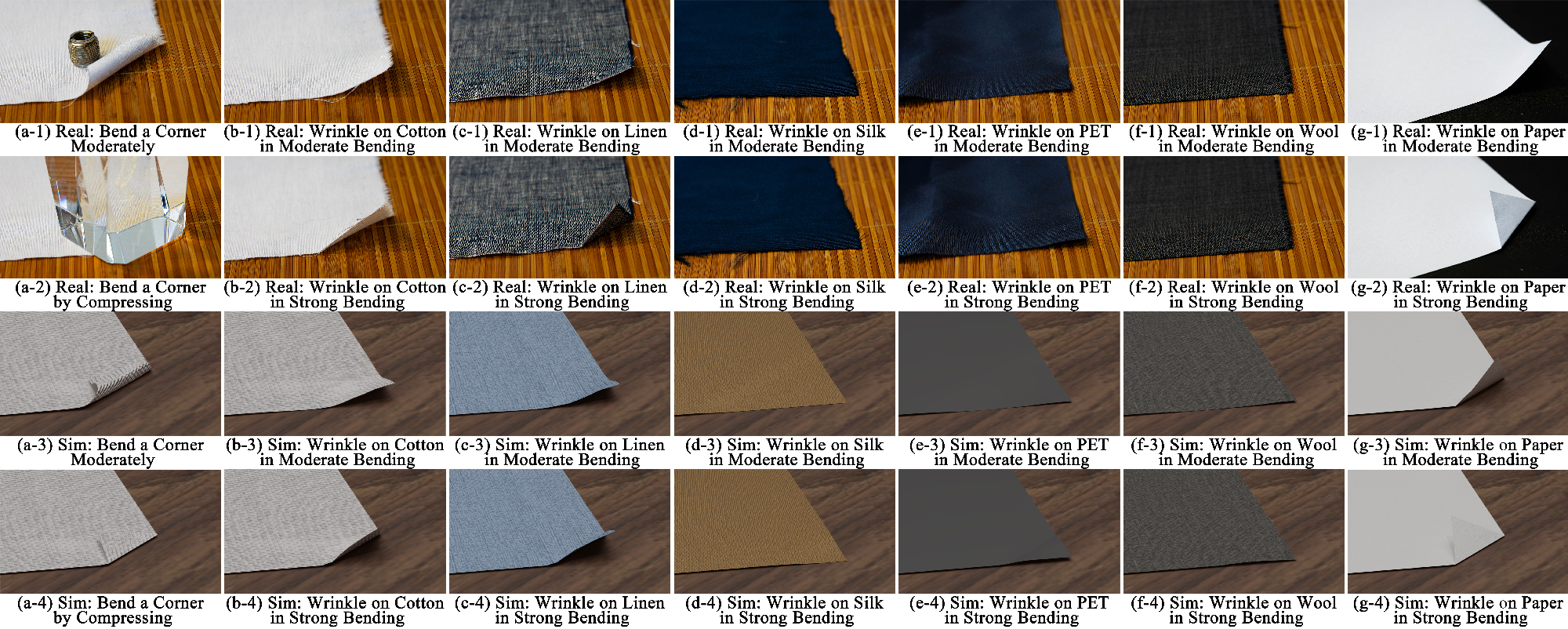}
    \caption{Persistent wrinkles vary under moderate (a-1) or strong deformations (a-2). Furthermore, due to differences in material properties, persistent wrinkles differ across fabric types (b,c,d,e,f-1,2). Our differentiable cloth simulator integrates both internal friction and plasticity to model how such wrinkles form under varying deformation conditions. Using gradient-based optimization, the simulator estimates the physical parameters of each fabric and closely reproduces the wrinkles observed in real materials (b,c,d,e,f-3,4). Additionally, our method is also capable of simulating persistent folds in paper (g-1,2,3,4).}
    \label{fig:app_teaser}
\end{figure*}

Persistent wrinkles formed under different deformation levels are visually distinct: \textit{mild} deformation typically produces soft, recoverable folds, while \textit{severe} deformation results in sharp, permanent creases ( \Cref{fig:app_teaser}). This difference arises because internal friction and plasticity jointly govern persistent wrinkle formation. As deformation increases, stick-slip transitions between contacting yarns occur first; plastic flow activates only under extreme deformation. Our differentiable simulator learns both internal friction and plasticity from the recoverable and unrecoverable components of the measured hysteresis, and then faithfully reproduces persistent wrinkles under mild and severe deformations across six different materials (five fabrics plus paper).

\begin{figure}[tb]
    \centering
    \includegraphics[width=\linewidth]{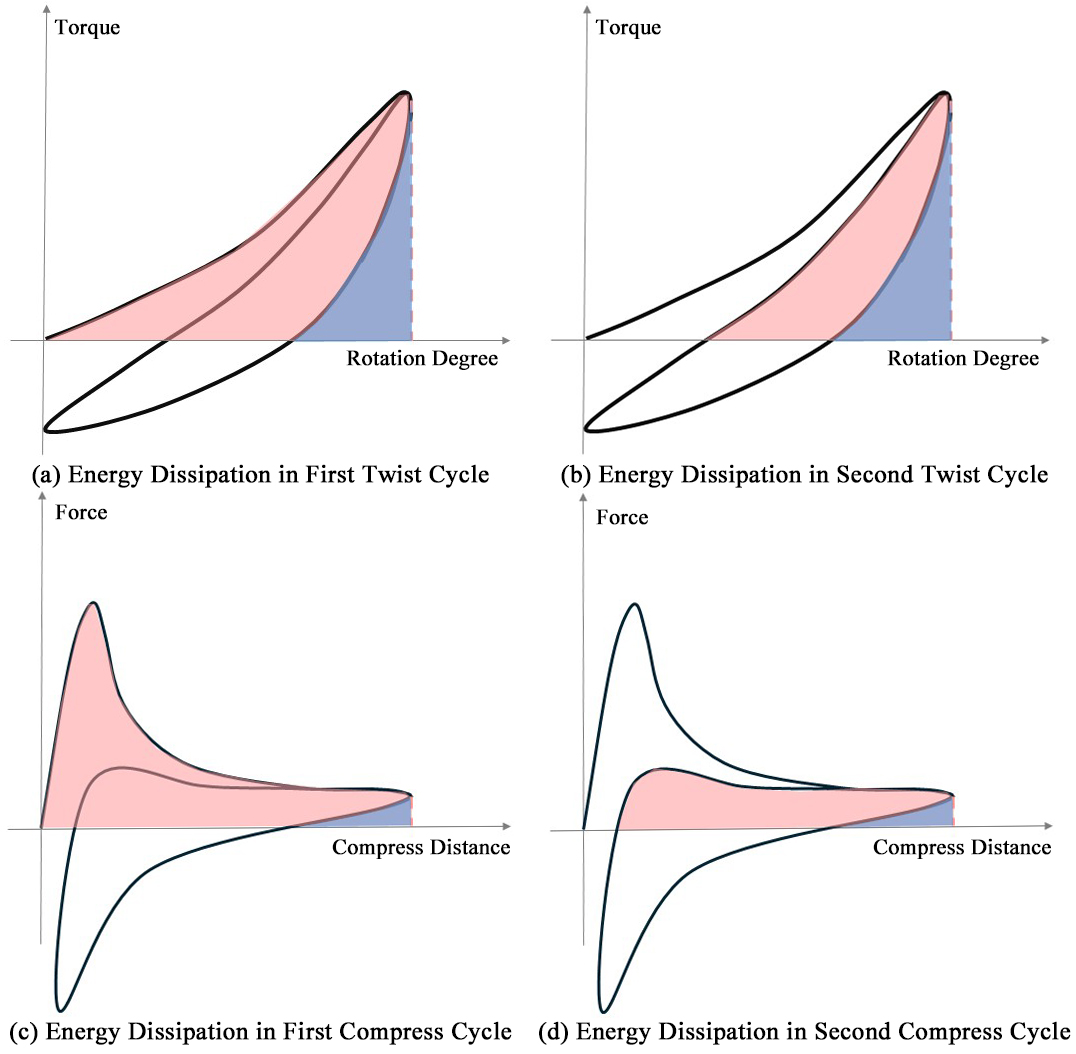}
    \caption{The energy dissipation in the twisting and compressing load-deformation curves is the integration of the area highlighted by the pink color.}
    \label{fig:dissipation}
\end{figure}

\begin{table}[tb]
    \centering
    \begin{tabular}{cccccc}
    \toprule
         Test/Fab & Cotton & Linen & PET & Wool & Silk \\
         \midrule
         Sim 1st-cycle& 0.0879 & 0.0075 & 0.0232 & 0.0276 & 0.0061 \\
         Sim 2nd-cycle& 0.0750 & 0.0059 & 0.0200 & 0.0188 & 0.0044 \\
         Real 1st-cycle& 0.0986 & 0.0080 & 0.0270 & 0.0312 & 0.0064 \\
         Real 2nd-cycle& 0.0747 & 0.0052 & 0.0096 & 0.0229 & 0.0048 \\
         \bottomrule
    \end{tabular}
    \caption{Compare the energy dissipation of different fabrics in both simulation and real measurements occurring in the twisting test.}
    \label{tab:twist_dissipation}
\end{table}

\begin{table}[tb]
    \centering
    \begin{tabular}{cccccc}
    \toprule
         Test/Fab & Cotton & Linen & PET & Wool & Silk \\
         \midrule
         Sim 1st-cycle & 8.1336 & 3.4651 & 1.9475 & 1.6484 & 0.5276 \\
         Sim 2nd-cycle & 7.9675 & 2.1885 & 1.2700 & 1.3101 & 0.4782 \\
         Real 1st-cycle & 4.1513 & 3.3336 & 0.5850 & 2.2878 & 0.7450 \\
         Real 2nd-cycle & 3.4966 & 2.0929 & 1.0605 & 2.4559 & 0.7293 \\
         \bottomrule
    \end{tabular}
    \caption{Compare the energy dissipation of different fabrics in both simulation and real measurements occurring in the compressing test.}
    \label{tab:comp_dissipation}
\end{table}

\subsection{Learned Energy Dissipation}

\begin{figure}[tb]
    \centering
    \includegraphics[width=\linewidth]{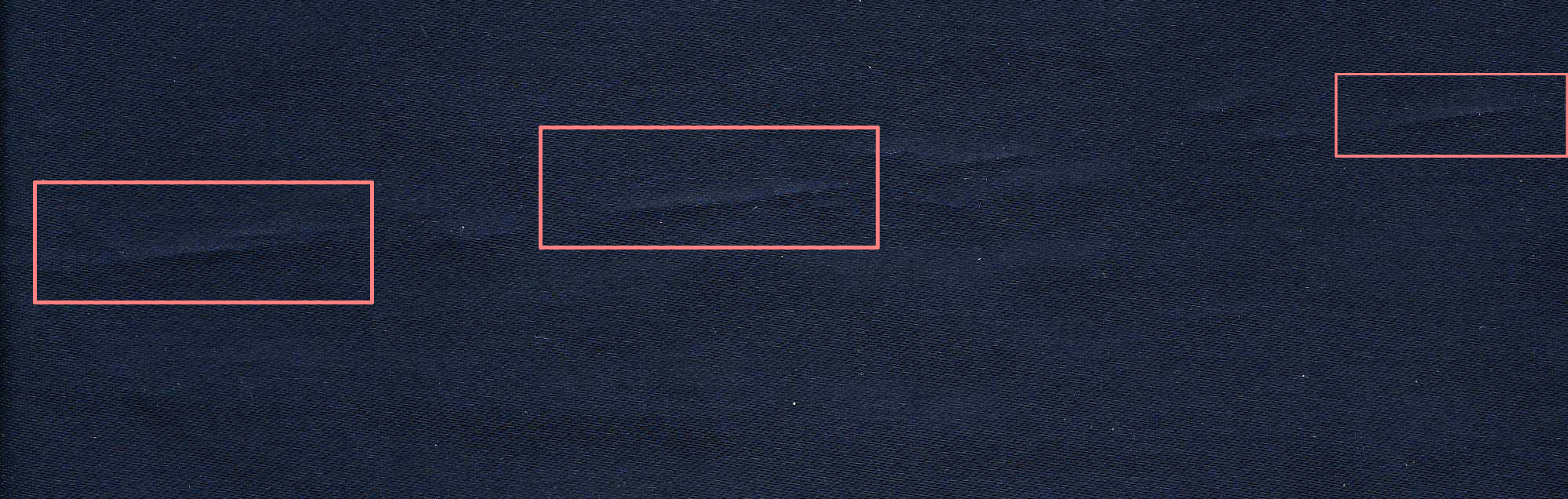}
    \caption{Real PET fabric can form tiny and sharp persistent wrinkles (marked by the pink boxes) in strong deformations.}
    \label{fig:PET_wrinkles}
\end{figure}

Fabric hysteresis is reflected by the energy dissipation in the load-deformation curves. As shown in  \Cref{fig:dissipation}, the energy dissipation in the first and second load-unload cycles in the twisting and compressing tests is reflected by areas highlighted in pink. Quantitatively,  \Cref{tab:twist_dissipation} and \Cref{tab:comp_dissipation} compare the measured and simulated dissipation across fabrics, showing close agreement. Critically, these results reveal a direct relationship: higher energy dissipation correlates with a greater propensity for persistent wrinkles. Cotton and linen, which dissipate the most energy during compression, form the most pronounced wrinkles. PET exhibits moderate dissipation and consequently forms only fine, sharp wrinkles (shown in  \Cref{fig:PET_wrinkles}), aligning with real-world observation.

\subsection{Plasticity from Yarns}

\begin{figure}[tb]
    \centering
    \includegraphics[width=\linewidth]{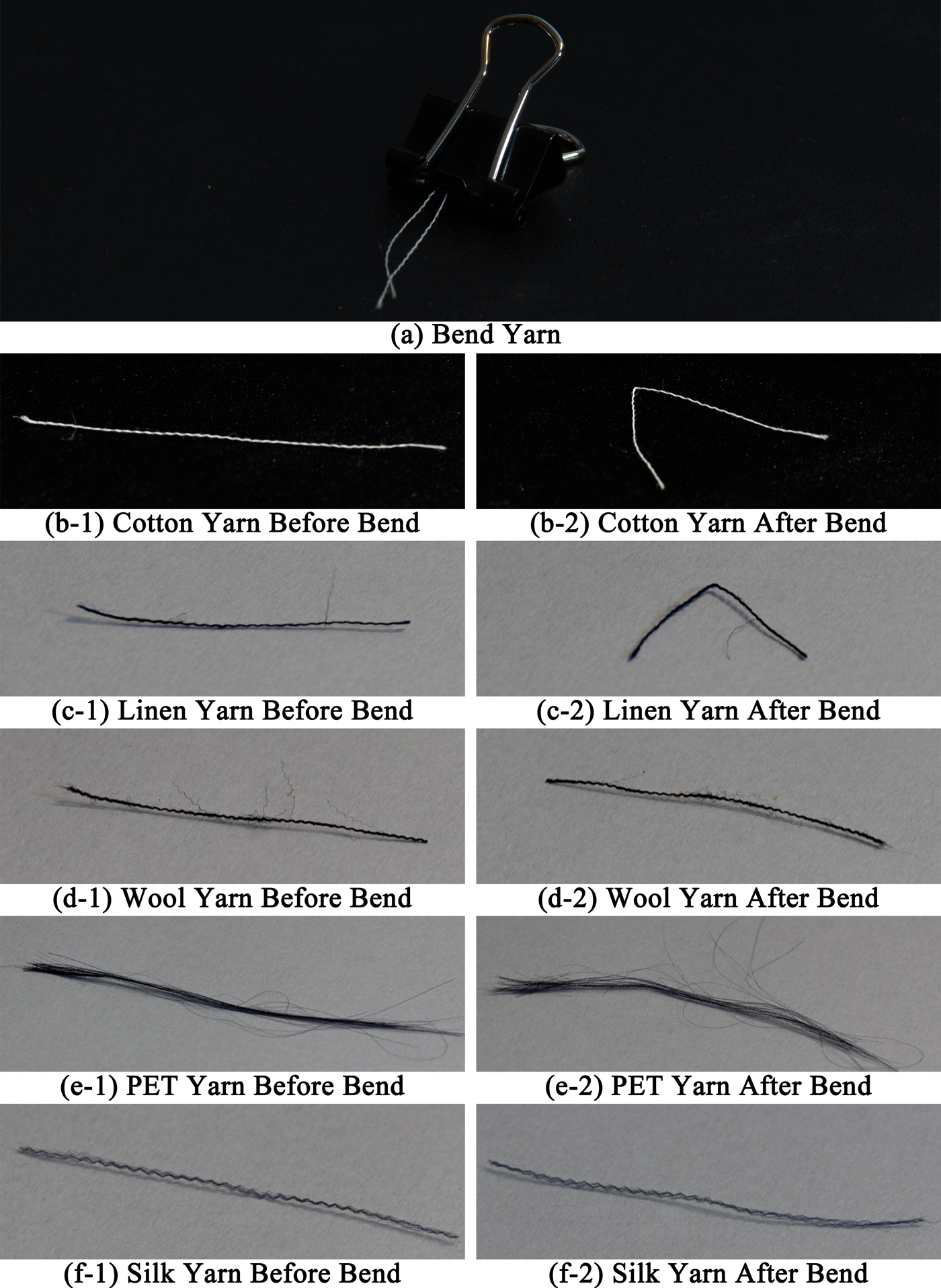}
    \caption{The persistent deformation after bending a yarn of different fibers. The cotton and linen yarns have obvious permanent deformation (b,c-2) after the bending deformation (a). This explains why the persistent wrinkles derive from plasticity of the cotton and linen fabrics. The PET yarn forms smaller permanent bending deformation which conforms to the tiny wrinkles formed on the PET garments after they are deformed extremely. By contrast, the wool (d-2) and silk (f-2) yarns do not have obvious permanent bending deformation, so the wool and silk fabrics are very wrinkle resistant.}
    \label{fig:bend_yarn}
\end{figure}

The plastic model in our differentiable simulator accounts for the persistent wrinkles caused by the permanent deformations of yarns. \Cref{fig:bend_yarn} illustrates the permanent bending deformation exhibited by different yarns after a bending manipulation (a). Cotton and linen yarns (b,c-2) retain significant, unrecoverable deformation, providing a microscopic basis for the sharp persistent wrinkles observed in these fabrics after extreme compression (shown in Figure 1 of the main paper). PET yarn (e-2) exhibits only mild permanent bending, correlating with the fine, sharp creases characteristic of PET fabric. In contrast, wool and silk yarns demonstrate nearly full recovery to their initial shape (d,f-2), explaining their high wrinkle resistance observed in their fabrics. These results validate the necessity of a plasticity component in our differentiable cloth simulator. 

\subsection{Intendedly Made Wrinkles}

\begin{figure}[htb]
    \centering
    \includegraphics[width=\linewidth]{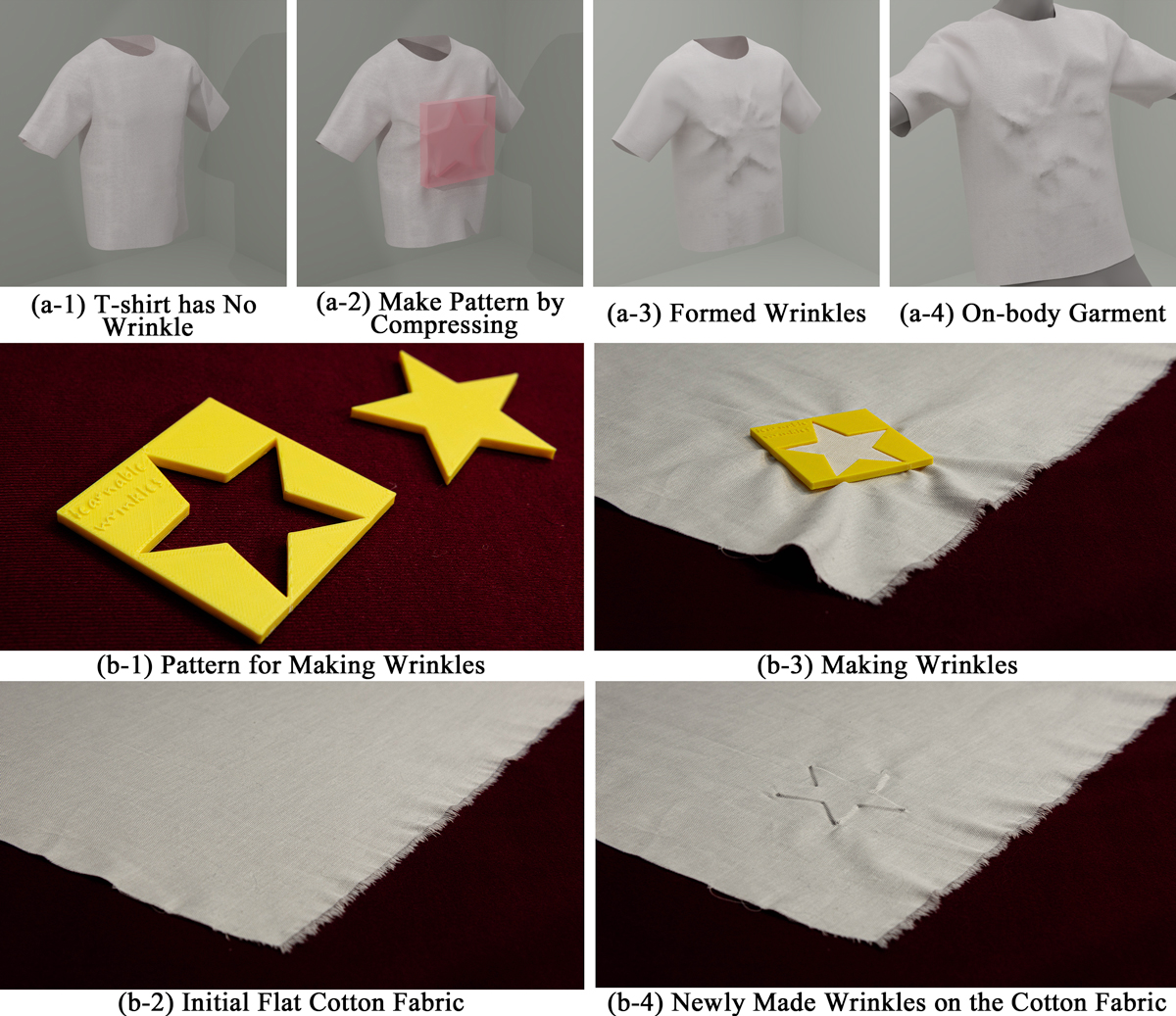}
    \caption{As cotton is likely to accumulate persistent wrinkles, we can use this property to deliberately add patterns on cotton fabrics: as a start pattern (b-4) through compressing an initially flat cotton fabric (b-2) by the mold (b-1,3). Our simulator can learn these properties and simulate the deliberately made pattern on a cotton t-shirt (a-1,2,3,4).}
    \label{fig:make_pattern}
\end{figure}

Beyond unintentional wrinkling, persistent wrinkles can be intentionally designed to shape garments or create stylistic effect. Since cotton is a wrinkle-prone fabric, we can impress a permanent design onto real cotton using a custom mold: a star-shaped logo (shown in  \Cref{fig:make_pattern} (b)). Given the estimated cotton's physical parameters, our simulator can faithfully reproduce this design process in simulation. Given a virtual star mold and cotton t-shirt, the simulator simulates the embossed firm wrinkle patterns by compressing the mold on the shirt (as illustrated  \Cref{fig:make_pattern} (a)).

\subsection{Crumpling  Paper}

\begin{figure}[tb]
    \centering
    \includegraphics[width=\linewidth]{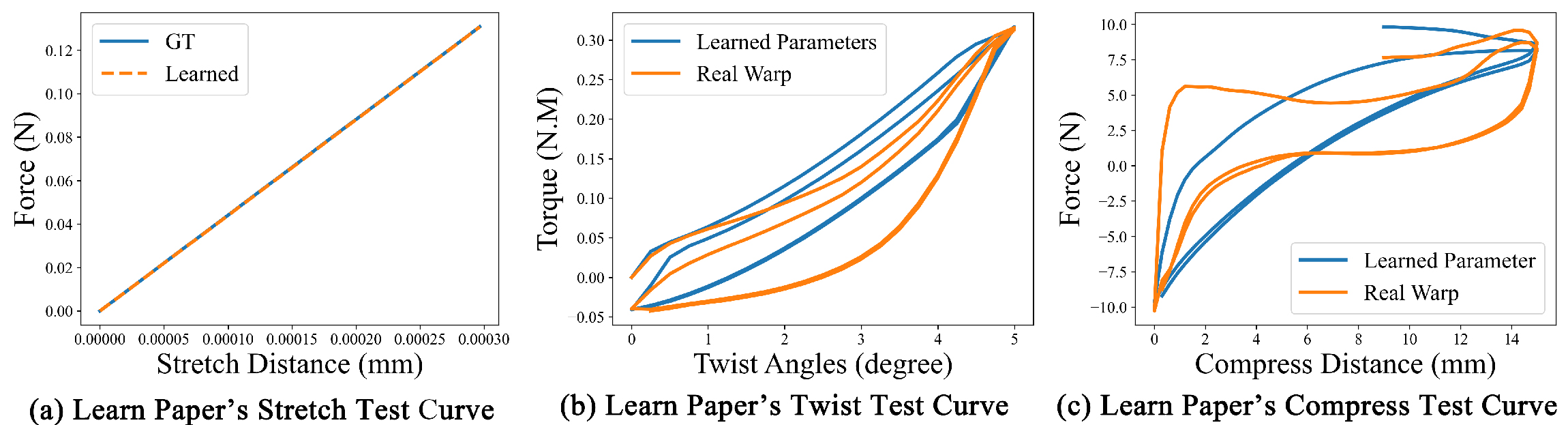}
    \caption{Our differentiable cloth simulator can fit the stretching, twisting, and compressing load-deformation curves of paper by optimizing the physical parameters.}
    \label{fig:learn_paper}
\end{figure}

\begin{figure}
    \centering
    \includegraphics[width=\linewidth]{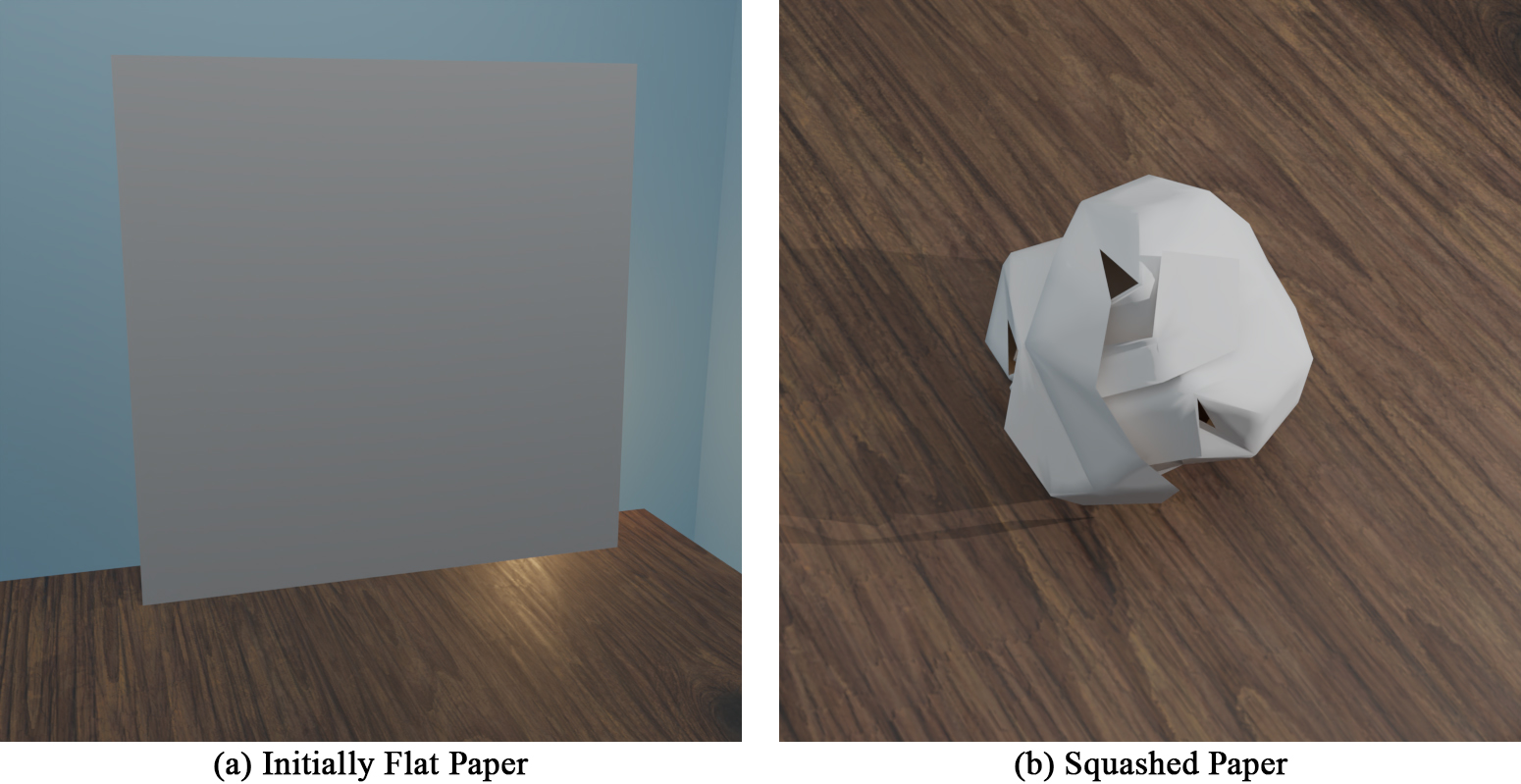}
    \caption{An initially flat paper (a) is crumpled into a ball. (b) Due to the formed persistent folds, the paper cannot unfold by itself.}
    \label{fig:squash_paper}
\end{figure}

Our adapted fabric tester can also measure paper. We only measure and learn paper in a single direction because paper is an isotropic material and does not have warp and weft directions. Given the measured load-deformation curves, our differentiable cloth simulator learns paper physical parameters by fitting the measurements (shown in  \Cref{fig:learn_paper}). Given the estimated parameters, we simulate crumpling a paper into a ball (shown in  \Cref{fig:squash_paper}). The paper forms firm persistent wrinkles and cannot return to its initial shape.

\begin{revsection}

\subsection{Learning Knitted Fabrics}

\begin{figure}[htb]
    \centering
    \includegraphics[width=\linewidth]{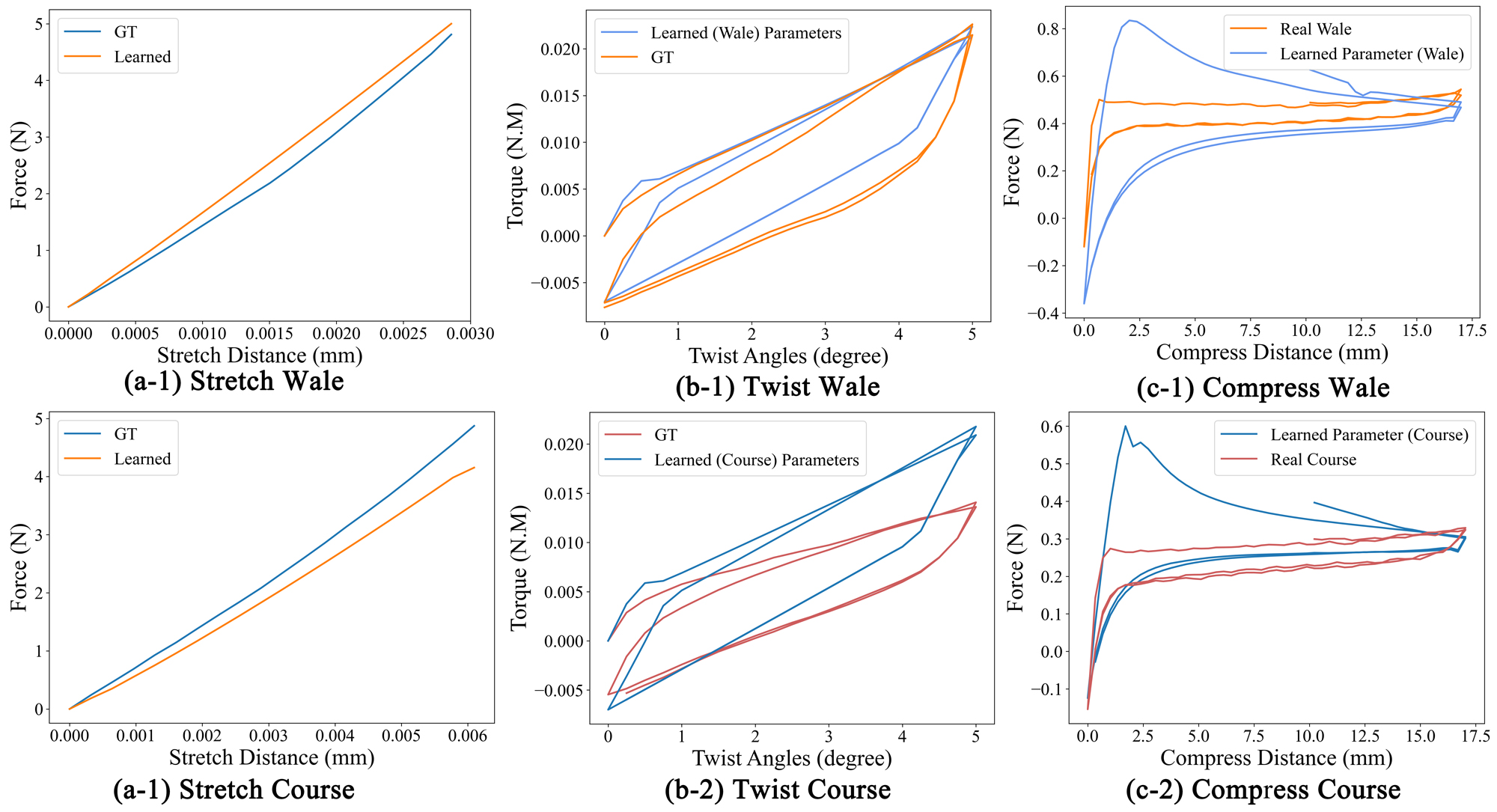}
    \caption{\rev{Fit the load-deformation curves of the knitted fabrics (KN-01 in our dataset) by optimizing the physical parameters. The predicted twist curve in course direction is different from the ground truth because the shearing orthotropy is not only caused by the different stretching stiffness in wale and course direction.}}
    \label{fig:kn01}
\end{figure}

\begin{table}[htb]
    \centering
    \caption{\rev{Mean squared error (MSE) between the simulated and measured curve of the fabric KN-01 with initial parameters and learned parameters.}}
    \begin{tabular}{cccc}
        \toprule
         MSE/Curves & Str. Warp & Str. Weft & Twt. Warp \\
         \midrule
         Initial & 1.783 & 10.045 & 9.639e-6 \\
         Trained & 0.059 & 0.380 & 4.005e-6 \\
         \midrule
         MSE/Curves & Twt. Weft & Comp. Warp & Comp. Weft \\
         \midrule
         Initial & 1.003e-5 & 0.051 & 0.053 \\
         Trained& 1.347e-5 & 0.006 & 0.010 \\
         \bottomrule
    \end{tabular}
    \label{tab:kn01_mse}
\end{table}

We also estimate the parameters of knitted fabrics (KN-01 in our data). \Cref{fig:kn01} compares the learned curve to the ground truth. The difference between the twist curve in course direction is large because our shearing model is not sufficient to account for the shearing mechanics in knitted fabrics, where the shearing orthotropy not only derives from the different stretching stiffness in wale and course directions. 

\subsection{Fit Synthesized Data}

\begin{table}[htb]
    \centering
    \caption{\rev{The estimated parameters from the synthesized load-deformation curves.}}
    \begin{tabular}{cccc}
    \toprule
         Parameters &  $k_{11}$ & $k_{22}$ & $c_0$\\
     \midrule
         GT &  300000.0 & 200000.0 & 100000.0 \\
         Init & 400000.0 & 400000.0 & 50000.0 \\
         Lr & 300008.1 & 199999.9 & 123299.9 \\
     \midrule
          Parameters &  $c_1$ & $k_{fric\_shear}$ & $\bar{\varepsilon}_{thres\_shear}$ \\
    \midrule
         GT &  100000.0 & 100000.0 & 0.0018 \\
         Init & 50000.0 & 30000.0 & 0.001 \\
         Lr & 101577.2 & 96458.9 & 0.0015 \\
     \midrule
          Parameters &  $k_{h\_shear}$ & $\varepsilon_{y0\_shear}$ & $k_{b11}$ \\
    \midrule
         GT &  70000.0 & 0.5 & 200.0 \\
         Init & 20000.0 & 0.8 & 300.0 \\
         Lr & 67617.7 & 0.4406 & 199.9 \\
     \midrule
          Parameters &  $k_{b11\_fric}$ & $\varepsilon_{b11\_thres}$ & $k_{h\_bend\_warp}$ \\
    \midrule
         GT &  300.0 & 0.08 & 100.0 \\
         Init & 400.0 & 0.05 & 200.0 \\
         Lr & 299.2743 & 0.0798 & 99.599 \\
    \midrule
          Parameters &  $\varepsilon_{y0\_bend\_warp}$ & $k_{b22}$ & $k_{b22\_fric}$ \\
    \midrule
         GT &  0.5 & 200.0 & 240.0 \\
         Init & 0.8 & 120.0 & 300.0 \\
         Lr & 0.4406 & 120.0003 & 240.0106 \\
    \midrule
    Parameters &  $\varepsilon_{b22\_thres}$ & $k_{h\_bend\_weft}$ & $\varepsilon_{y0\_bend\_weft}$ \\
    \midrule
         GT &  0.06 & 50.0 & 0.7 \\
         Init & 0.08 & 100.0 & 0.5 \\
         Lr & 0.06 & 49.873 & 0.7 \\
    \bottomrule
    \end{tabular}
    \label{tab:fit_param}
\end{table}

We first simulate the stretch, twist, and compression curves by using manually defined parameters and then recover these parameters by fitting the simulated curves. As shown in the  \Cref{tab:fit_param}, our differentiable cloth simulator can optimize the parameters from the initial values to the ground truth. 

\end{revsection}

\section{Validation in Cusick Drape}

\begin{figure}[htb]
    \centering
    \includegraphics[width=\linewidth]{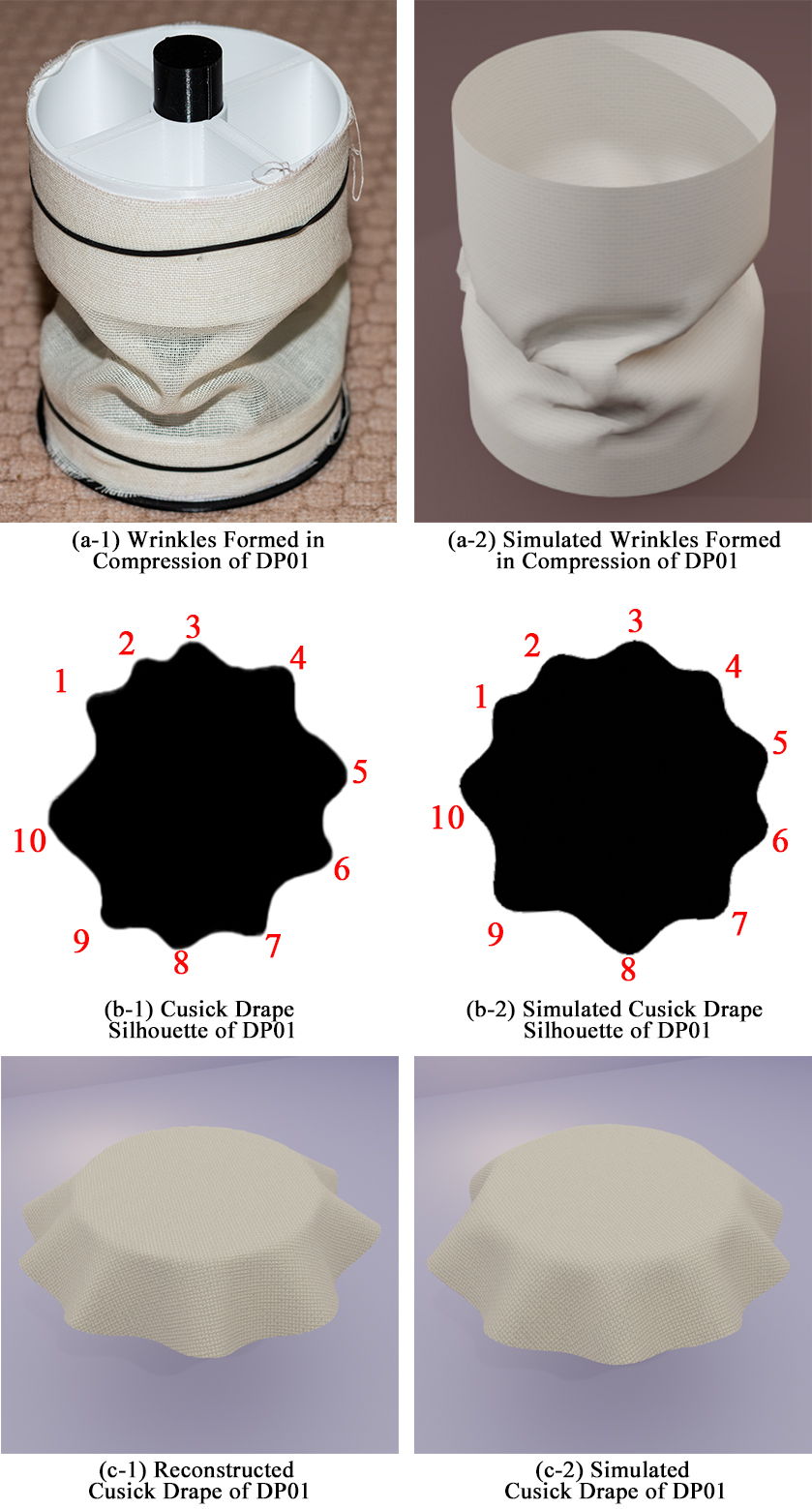}
    \caption{\rev{We further validate our estimated parameters in simulating Cusick drape. Given the parameters estimated from our load-deformation measurements, we can plausibly simulate the compression folds (a-2) which are similar to the real counterpart (a-1) formed in the compression. Furthermore, the parameters can also be applied in simulating Cusick drape. The drape shape and silhouette of the simulated cloth is similar to the real one, i.e., have an identical number of corners.}}
    \label{fig:drape}
\end{figure}

To validate that the estimated parameters generalize beyond the testing modes (stretch, twist, and compression), we compare the Cusick drape of real fabrics with simulations using parameters learned from the load-deformation curves. The Cusick drape test is a standard method for evaluating fabric drapeability in textile engineering. It captures the silhouette of a circular fabric sample under its own weight using a drapemeter~\cite{chu1950mechanics}. Previous studies have established that drape shape and physical parameters are closely related~\cite{collier1991measurement, collier1989effects}. In our dataset, we tested fabrics DP 1 to DP 12 following the BS EN ISO 9073-9:2008 standard. As shown in \Cref{fig:drape}, using parameters estimated from the load-deformation curves of fabric DP 1, our simulator not only reproduces the wrinkles in the axial compression test but also generates Cusick drape shapes that plausibly match the real fabric counterparts. The number of nodes (or folds) in a draped fabric is governed by its physical properties~\cite{lo2002modeling}; softer fabrics tend to produce more corners, while stiffer fabrics produce fewer. In \Cref{fig:drape} (b-1,2), both the real and simulated drapes exhibit 10 nodes, as highlighted by red dots. These results demonstrate that the estimated parameters capture the underlying physical characteristics of the fabric and generalize to different simulation scenarios.

\section{Ablation Studies}

\subsection{Constitutive Model}

\begin{figure}[tb]
    \centering
    \includegraphics[width=\linewidth]{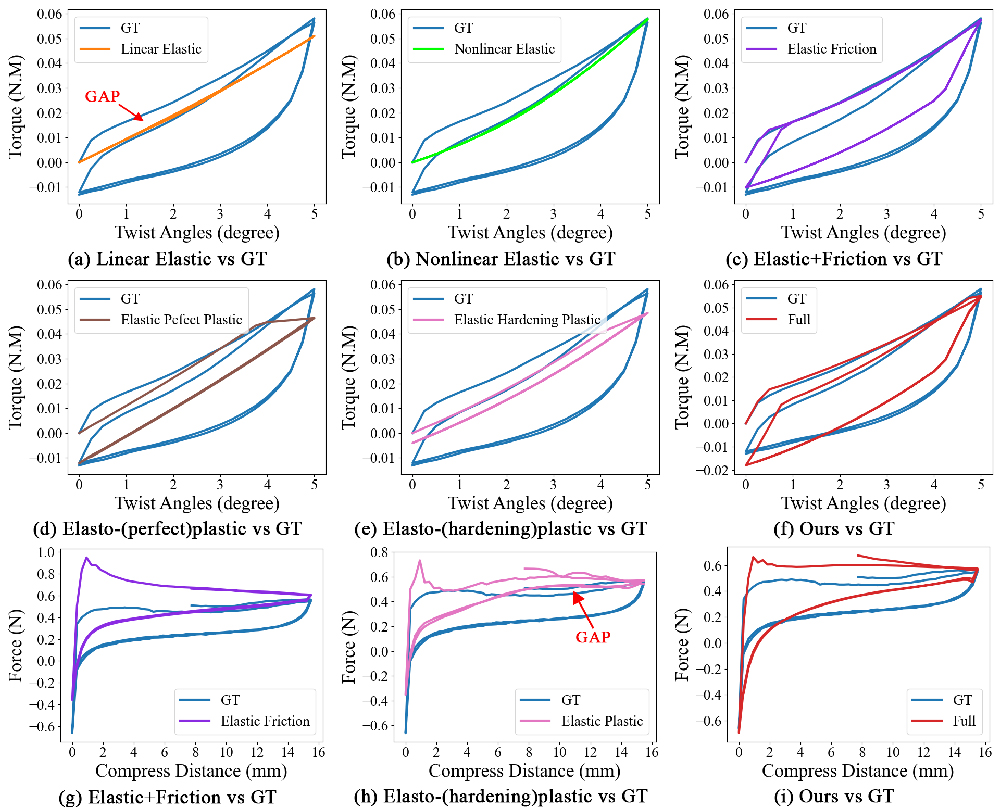}
    \caption{Our constitutive model is carefully designed to capture the features in real fabric's twist and compression load-deformation curves.}
    \label{fig:ablation_shear}
\end{figure}

\Cref{fig:ablation_shear} (a-f) demonstrate the necessity of our model's nonlinearity, friction, and hardening plasticity for capturing GT twist curves, \rev{where all models are trained from scratch independently}. The Yeoh model captures increasing stiffness (b), unlike linear elasticity (a). Elastic+Friction captures hysteresis but cannot reproduce the gap between first and subsequent cycles (c). Elastic+Perfect Plasticity captures residual deformation but not post-first-cycle hysteresis (d). Perfect plasticity stops torque increase at yield (d), while hardening plasticity continuously increases torque, closely fitting GT (e). Only the full shear model simultaneously captures: (1) the gap between first and second loading curves; (2) constant hysteresis from the second cycle onward; and (3) increasing stiffness with twist angle. The same applies to compression (\Cref{fig:ablation_shear} (f-i)): friction alone cannot capture the first-to-second cycle difference (f), and ignoring friction yields smaller hysteresis and residual force than GT (h). Our model closely fits GT and captures all features (i). 

\subsection{Orthotropic Bending}

\begin{figure}[htb]
    \centering
    \includegraphics[width=\linewidth]{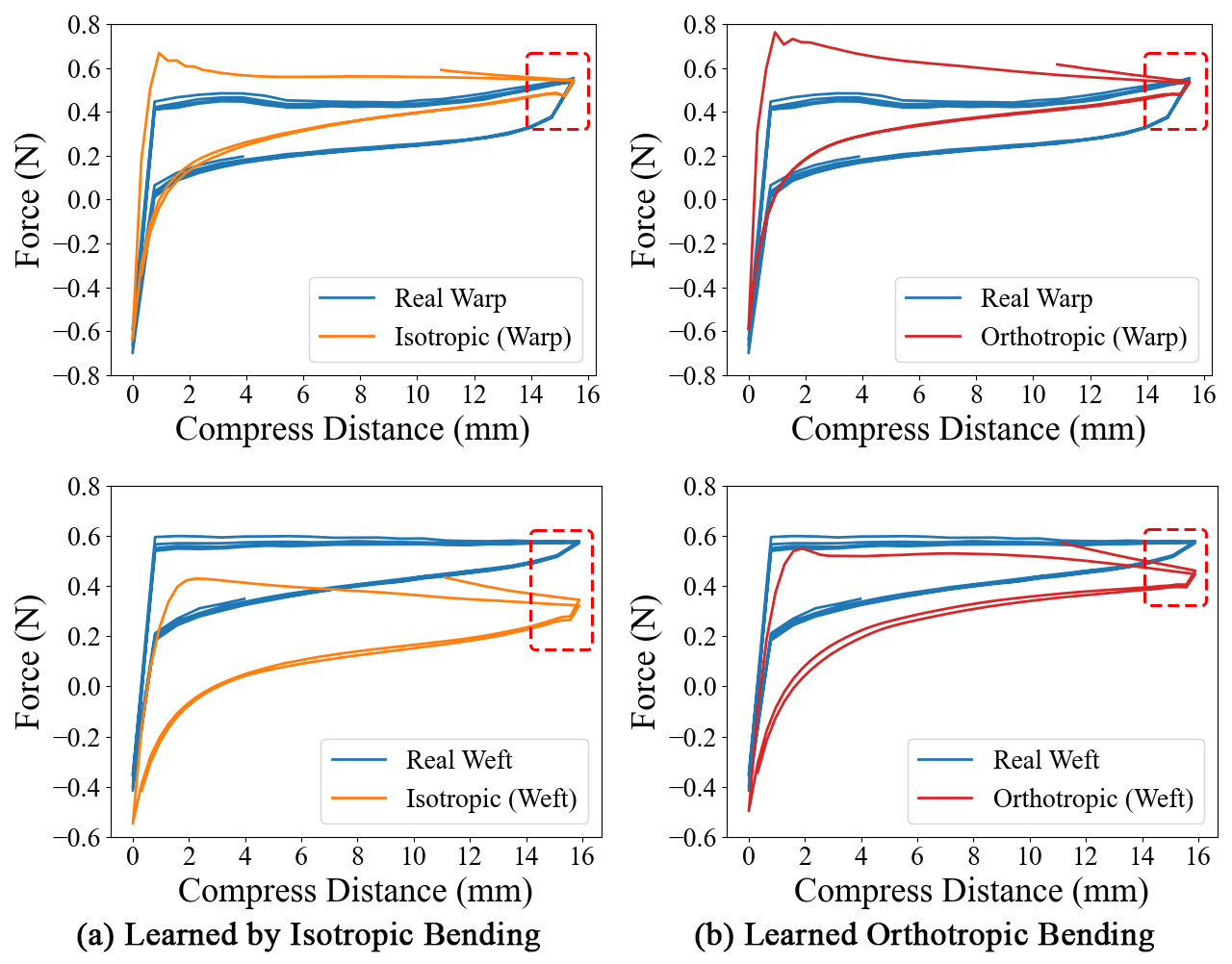}
    \caption{Compared with isotropic bending model, our orthotropic bending model can closely fit the measured compression curves in both warp and weft direction. }
    \label{fig:comp_bend}
\end{figure}

Our bending model decomposes curvature along fabric warp and weft direction so that it can capture bending material orthotropy. Compared with isotropic bending, it can more closely fit the measured compression curves. As shown in  \Cref{fig:comp_bend}, although the isotropic bending model can closely fit the curve in warp direction, it cannot generalize well to the curve in the weft direction. The mean squared error between the real and simulated curve in the warp direction is 0.092, but increases to 0.135 in the weft direction. By contrast, the orthotropic bending model can closely fit to both warp and weft curve: average mean squared error in warp and weft directions is 0.0895. Nevertheless, orthotropic bending model uses more physical parameters and needs to fit both warp and weft compression curves.

\subsection{Time Step Size}

\begin{figure}
    \centering
    \includegraphics[width=\linewidth]{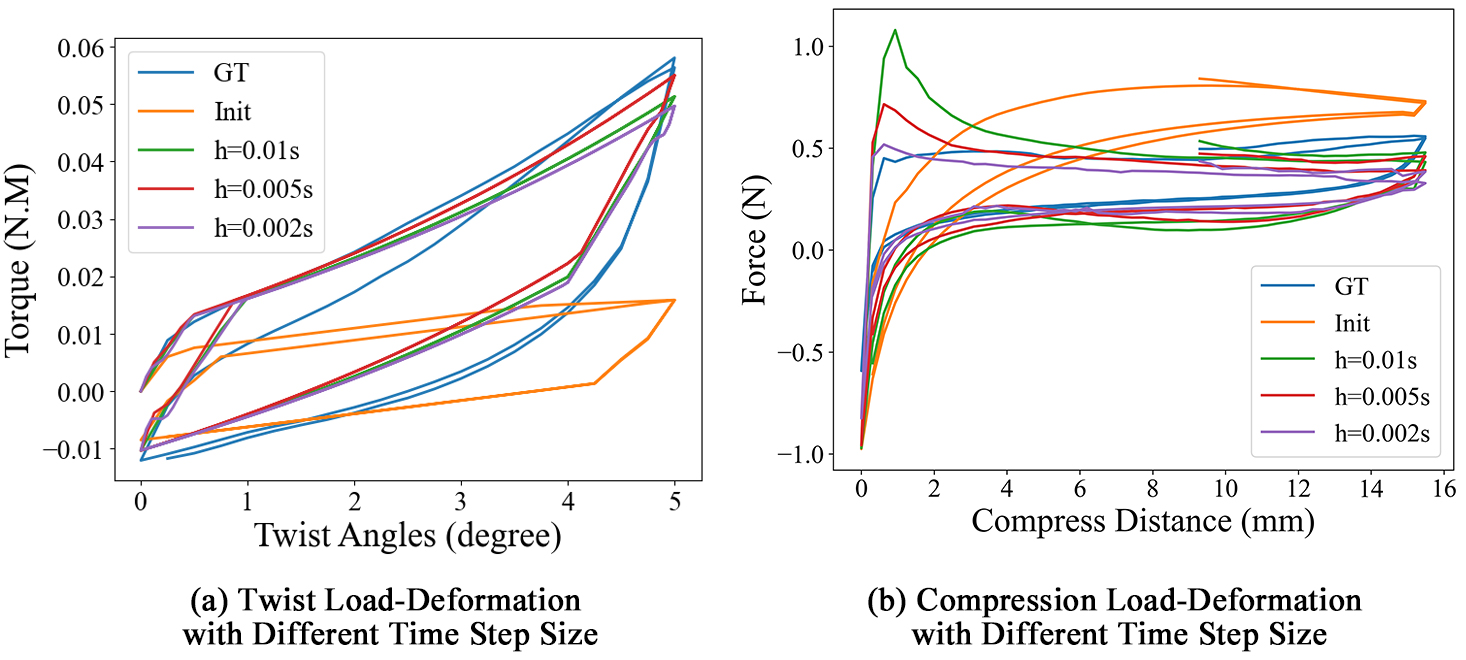}
    \caption{The influence from the time step size is far smaller (the difference between blue, green, red, and purple curves) than that from the physical parameters (the difference between the blue and orange curves).}
    \label{fig:time_step}
\end{figure}

We use the \textit{semi-implicit return mapping strategy}~\cite{tu2009return} to update the stick-slip friction and plastic flow, which can make the simulation affected by the time step size. To evaluate this, we simulate the compression and twist testing modes with different time step sizes: $0.01$, $0.005$, and $0.002$.  \Cref{fig:time_step} shows that, compared to the difference caused by the physical parameters, the influence from the time step size is far smaller. In our experiment, we use $0.01$ s as the time step in learning for training efficiency, i.e., using fewer simulation steps to simulate and fit the measurements.

\begin{revsection}

\subsection{Mesh Resolution}

\begin{figure}
    \centering
    \includegraphics[width=\linewidth]{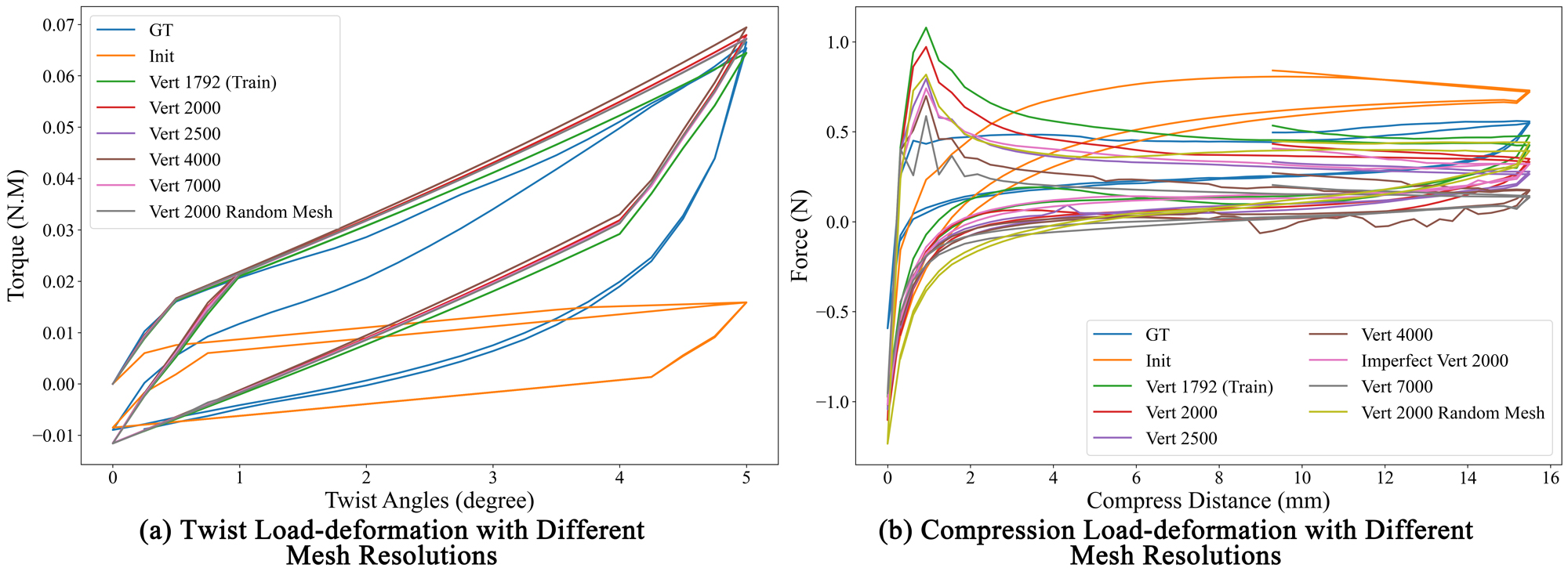}
    \caption{\rev{For the twist, the influence from the mesh resolution is far smaller than that from the physical parameters. For the compression, the reaction force reduces as the mesh resolution increases because the mesh is more flexible and can encode finer wrinkles.}}
    \label{fig:mesh_resolution}
\end{figure}

\begin{figure*}[htb]
    \centering
    \includegraphics[width=\textwidth]{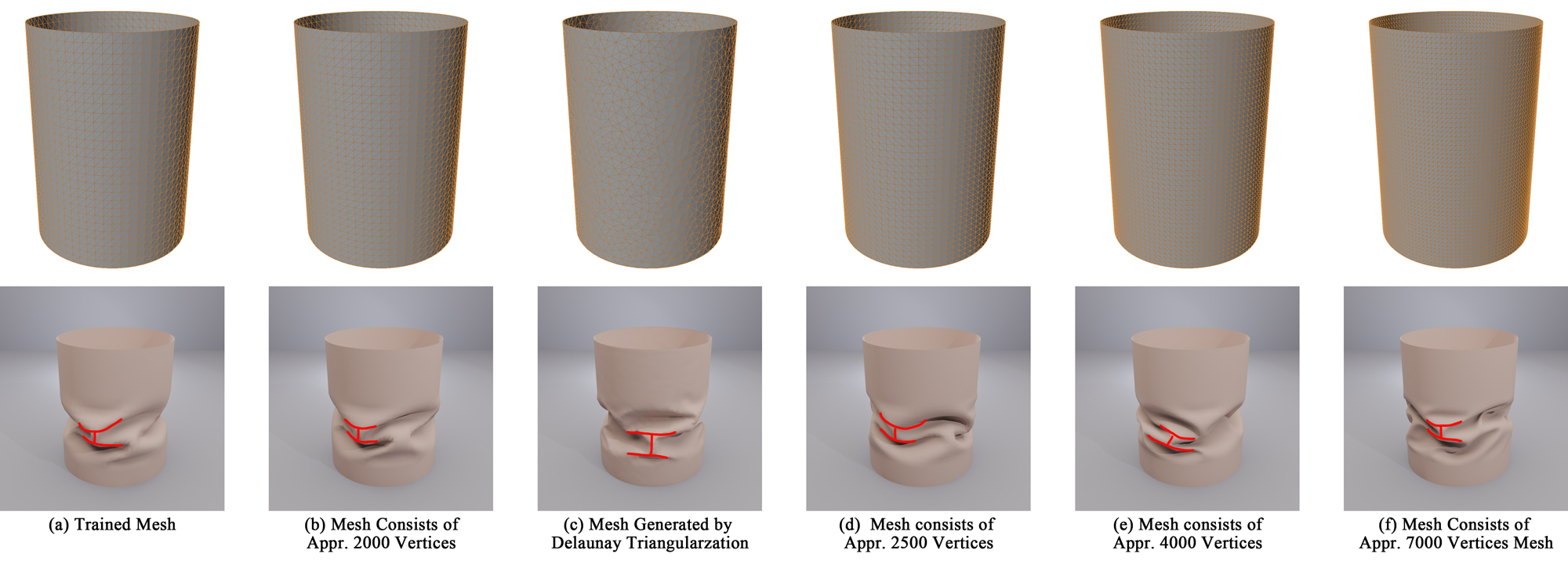}
    \caption{\rev{Compare the wrinkles formed in axial compression using meshes with different resolution and discretization (b-f) from the mesh used in the training (a). Moderately increasing the number of mesh vertices (c, d) or varying mesh discretization (c) has an influence on the distribution of wrinkles. Increasing the number of mesh vertices allows the mesh to accommodate finer wrinkles (e,f). Due to the material properties, the formed wrinkle patterns in different meshes consistently show a similar shape (highlighted in red).}}
    \label{fig:compress_wrinkles_diff_meshes}
\end{figure*}

\begin{table*}[htb]
    \centering
    \caption{\rev{The relative mean curvature (MC) of the simulated compressed cloth with respect to the perfect cylinder mesh. Changing mesh discretization and increasing mesh resolution introduce a subtle influence on the relative MC of the compressed state. By contrast, using an imperfect initial mesh, whose initial relative MC is not 1.0, results in a greater variance in the relative mean curvature of the compressed mesh.}}
    \begin{tabular}{ccccccccc}
    \toprule
         Meshes & Train & Vert. 2000 & Vert. 2500 & Vert. 4000 & Vert. 7000 & Delaunay Vert. 2000 & Imperfect & Silk  \\
         \midrule
         Init MC & 1.0 & 1.0 & 1.0 & 1.0 & 1.0 & 1.0 &  1.004 & 1.0 \\
         Compressed MC & 0.802 & 0.789 & 0.795 & 0.798 & 0.792 & 0.791 & \textbf{0.777} & \textbf{0.855} \\
         \bottomrule
    \end{tabular}
    \label{tab:compression_mc}
\end{table*}

\begin{table*}[htb]
    \centering
    \caption{\rev{The mean squared error (MSE) reduces after optimizing the initial parameters, i.e., columns Init Param Ori Mesh vs. Trained Param Ori Mesh. The MSE error increases as we change the mesh resolution, topology, and perturb the initial mesh from the perfect cylinder. However, the increase of MSE caused by mesh is smaller than that caused by the physical parameters. As buckling is imperfection sensitive, the MSE around the buckling stage between the simulated and measure curve is greater. Also, the buckling is more sensitive to the mesh resolution, where the critical buckling force decreases as the mesh resolution increases because the mesh has more degrees of freedom and is more flexible.}}
    \begin{tabular}{ccccccccc}
    \toprule
        Curve/Mesh &  \makecell{Init Param \\ Ori Mesh} & \makecell{Trained Param \\Ori Mesh} & Vert. 2000 & Vert. 2500 & Vert. 4000 & Vert. 7000 & Delaunay Vert. 2000 & Imperfect \\
        \midrule
        Post-Buckling & 0.257 & 0.019 &	0.035 &	0.038 &	0.067&	0.079 &	0.037 & 0.023\\
        Full & 0.873 & 0.757 & 0.461 & 0.826 &	0.502 &	0.131 &	0.091 &	0.029 \\
        \bottomrule
    \end{tabular}
    \label{tab:mse_mesh}
\end{table*}

\begin{table}[htb]
    \centering
    \caption{\rev{Varying time step size has subtle influence on the relative mean curvature of the simulated compressed cloth.}}
    \begin{tabular}{ccccc}
    \toprule
         Time Step Size & $0.01s$ & $0.005s$ & $0.002s$ & Silk\\
         \midrule
         Init MC & 1.0 & 1.0 & 1.0 & 1.0 \\
         Compressed MC & 0.802 & 0.798 & 0.794 & \textbf{0.855} \\
         \bottomrule
    \end{tabular}
    \label{tab:comress_mc_time_step}
\end{table}

\begin{figure}[htb]
    \centering
    \includegraphics[width=\linewidth]{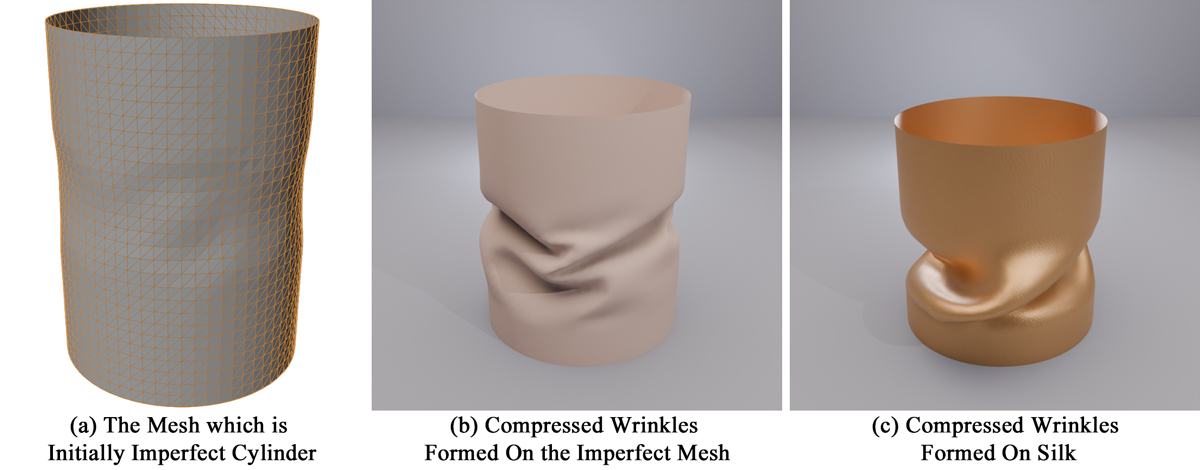}
    \caption{\rev{(a) is an initially imperfect cylinder which has some mild deformations before compression. (b) The folds formed by compressing the imperfect mesh. (c) The folds formed on the silk cylinder are dramatically different.}}
    \label{fig:imperfect_silk}
\end{figure}

We simulate load-deformation curves at different mesh resolutions and topologies using parameters estimated from the 1,792-vertex mesh. As shown in \Cref{fig:mesh_resolution} (a), mesh resolution has only a subtle influence on the twist curves. For compression, however, the reaction force decreases as mesh resolution increases (\Cref{fig:mesh_resolution} (b)), because finer meshes have more degrees of freedom and are more flexible. Nevertheless, the influence of mesh resolution is smaller than that of physical parameters (shown in \Cref{fig:compress_wrinkles_diff_meshes}). \Cref{tab:compression_mc} and \Cref{tab:comress_mc_time_step} report the relative mean curvature of compressed fabric under varying mesh and time step sizes. The relative mean curvature remains similar across these variations, indicating that the resulting wrinkle patterns are consistent. In contrast, using an initially imperfect cylinder mesh leads to a larger change in relative mean curvature (shown in \Cref{fig:imperfect_silk} (b)). Moreover, altering physical parameters produces far more pronounced changes in both relative mean curvature (\Cref{tab:compression_mc}, \Cref{tab:comress_mc_time_step}, silk entry) and wrinkle morphology (\Cref{fig:imperfect_silk} (c)).

\subsection{Axial Compression Buckling}

\begin{figure}[htb]
    \centering
    \includegraphics[width=\linewidth]{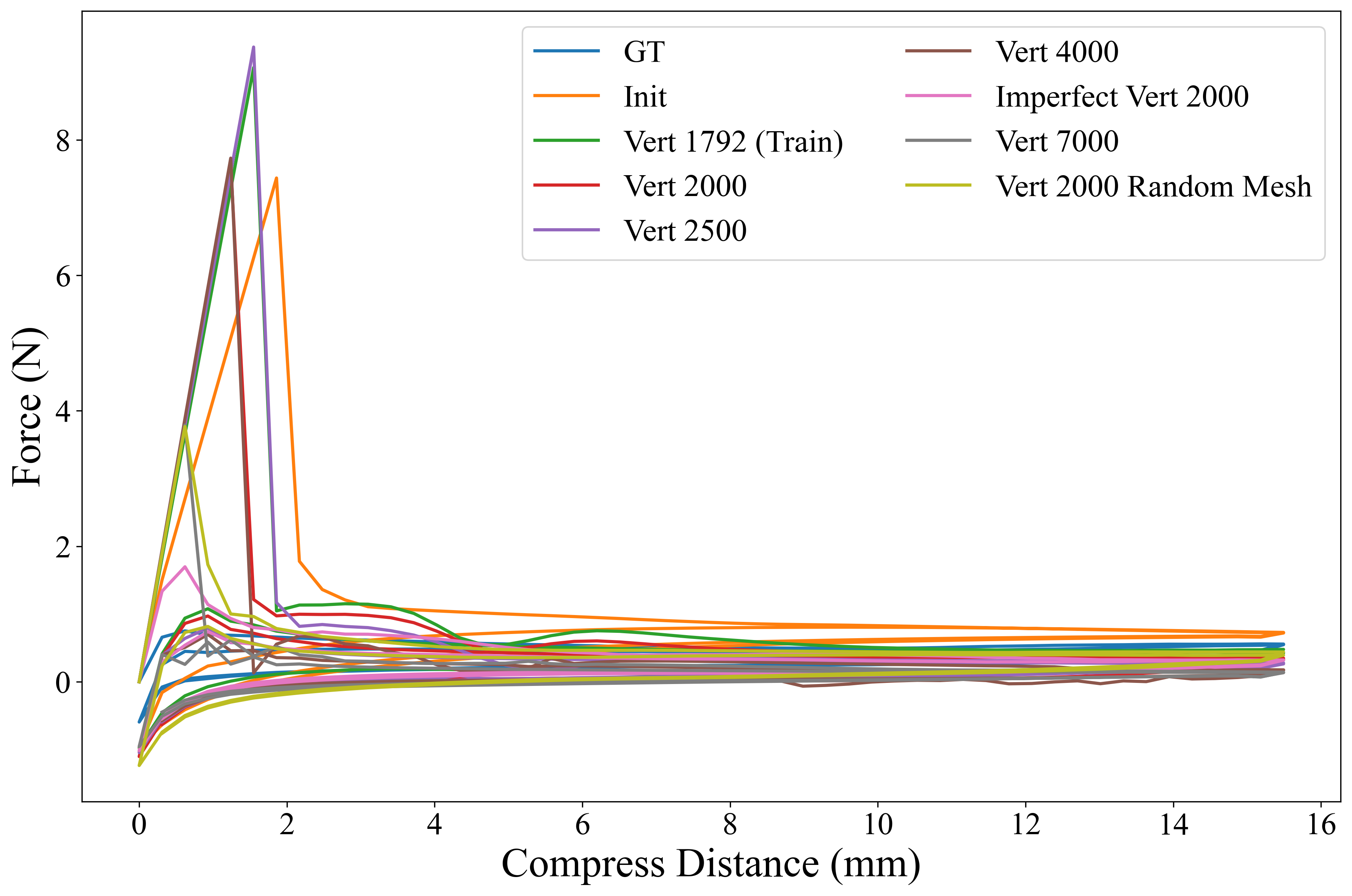}
    \caption{\rev{The simulated critical buckling force is sensitive to mesh resolution, mesh discretization, and imperfection.}}
    \label{fig:full_buckle_curve}
\end{figure}

Thin-wall buckling is a transient transition from in-plane compression to out-of-plane bending, while a cylinder shell is under axial compression. Since bending stiffness is smaller than in-plane compression stiffness, the measured load drops immediately after buckling occurs. The force at which buckling initiates is the critical buckling force, which is related to material stiffness. However, thin-wall buckling is highly sensitive to geometric imperfections, as real cylindrical samples are rarely perfect cylinders. Consequently, the measured critical buckling force is typically lower than theoretical predictions, such as those from Euler theory~\cite{harutyunyan2023buckling}. Although our tester applies pre-tension to minimize initial imperfections, the measured critical force remains lower than simulations using perfect cylindrical meshes.

As shown in \Cref{fig:full_buckle_curve}, the measured critical buckling force is significantly smaller than simulated values. The simulated critical force is also sensitive to mesh resolution, topology, and initial mesh imperfections. For instance, increasing mesh resolution reduces the critical buckling force, and using an imperfect mesh lowers it further. In contrast, the post-buckling response is considerably less sensitive to these factors. Mesh resolution, discretization, and imperfections have a much stronger influence on the buckling stage than on the post-buckling stage, where the critical force varies noticeably across different meshes. For this reason, we use only the post-buckling stage for parameter estimation. We acknowledge that the buckling stage contains important material information, but our current differentiable cloth simulator cannot utilize it, i.e., cannot simulate buckling bifurcation and capture geometric imperfection. We plan to address this in future research.

\end{revsection}

\begin{revsection}

\subsection{Run Time Profile}

We compare run time and Newton iterations (Newton Iter.) in the axial compression simulation. As shown in  \Cref{tab:run_time}, compared with elastic models, our more complex constitutive models incur small extra cost.

\begin{table}[htb]
    \centering
    \caption{\rev{Run time profile of elastic model and our model when simulating the LUFHES compression. We use two meshes with different resolutions: approximately 7000 and 15000 vertices respectively.}} 
    \begin{tabular}{ccc}
    \toprule
        Models \& Mesh & Avg. ms/step & Avg. Newton Iter. \\
        \midrule
         Elastic Mesh Vert 7000 &  155.3 & 3.7 \\
         Elastic Mesh Vert 15000 &  461.9 & 9.0 \\
         Ours Mesh Vert 7000 & 168.4 & 3.0 \\
         Ours Mesh Vert 15000 & 781.0 & 8.7 \\
         \bottomrule
    \end{tabular}
    \label{tab:run_time}
\end{table} 
\end{revsection}

\section{Full Physical Model}

\begin{figure}[tb]
    \centering
    \includegraphics[width=\linewidth]{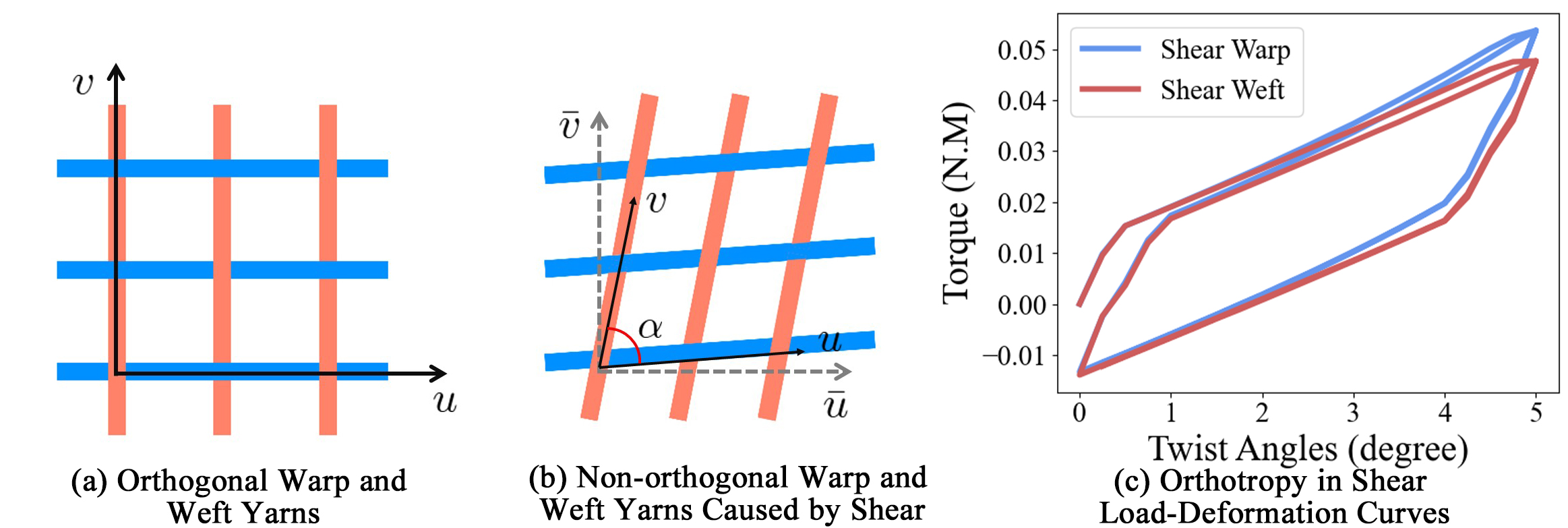}
    \caption{Modeling fabric shearing orthotropy. (a) Warp and weft yarns are perpendicular to each other in the initial rest state. (b) Shearing deformation breaks the orthotropic structures of the yarns. (c) This makes the measured reaction force affected by the fabric stretching stiffness in warp and weft direction.}
    \label{fig:warp_weft_rotate_app}
\end{figure}

Our simulator models clothes as a thin shell whose deformation can be decoupled into in-plane tensile and out-of-plane bending. To capture fabric orthotropic tensile properties observed in the tests, we use the cloth warp ($u$) and weft direction ($v$) as the principal directions of the deformation gradient (illustrated in~\Cref{fig:warp_weft_rotate_app} (a)):
\begin{equation}
    \mathbf{F} = 
    \begin{bmatrix}
        \frac{\partial x}{\partial u} & \frac{\partial x}{\partial v} \\
        \frac{\partial y}{\partial u} & \frac{\partial y}{\partial v} \\
        \frac{\partial z}{\partial u} & \frac{\partial z}{\partial v} \\
    \end{bmatrix}
\end{equation}
whose two columns encode fabric tensile deformation along warp and weft direction respectively. We employ the Lagrange-Green strain to exclude the rotational component of $\mathbf{F}$, i.e., $\mathbf{F}=\mathbf{RQ}$:
\begin{equation}
    \boldsymbol{\varepsilon} = \frac{1}{2}(\mathbf{FF}^\top - \mathbf{I}) = 
    \begin{bmatrix}
        \varepsilon_{uu} & \varepsilon_{uv} \\
        \varepsilon_{uv} & \varepsilon_{vv} 
    \end{bmatrix}
\end{equation}
where $\varepsilon_{uu}$, $\varepsilon_{vv}$, and $\varepsilon_{uv}$ measure fabric's tensile deformation along cloth warp, weft, and diagonal directions respectively. We use the St. Venant-Kirchhoff (StVK) hyper-elastic model to simulate cloth tensile physical behaviors:
\begin{equation}
    \psi_{tensile}(\RNum{1}, \RNum{2}) = \frac{1}{2}(\lambda + 2 \mu) \RNum{1}^2 - 2 \mu \RNum{2}
\end{equation}
where $\RNum{1} = Tr(\boldsymbol{\varepsilon})$ and $\RNum{2}= \frac{1}{2}(Tr(\boldsymbol{\varepsilon}) - Tr(\boldsymbol{\varepsilon}^2))$ are the first and second principal invariants. $\lambda$ and $\mu$ are the Lam\'e parameters which determine fabric stretching physical properties. The tensile stress is the derivative of the energy w.r.t. the strain: $\boldsymbol{\sigma}_{tensile} = \frac{d \int \psi_{tensile} d \Omega}{d \boldsymbol{\varepsilon}}$ where $\Omega$ denotes cloth's geometrical region. This can be more concisely denoted by Viogt form
\begin{equation}
    \begin{bmatrix}
        \sigma_{uu} \\
        \sigma_{vv} \\
        \sigma_{uv}
    \end{bmatrix}
    =
    \tau \mathbf{K} \boldsymbol{\varepsilon}
    =
    \begin{bmatrix}
        k_{11} & k_{12} & 0 \\
        k_{12} & k_{22} & 0 \\
        0 & 0 & k_{33}
    \end{bmatrix}
    \begin{bmatrix}
        \varepsilon_{uu}  \\
        \varepsilon_{vv}  \\
        \varepsilon_{uv}  
    \end{bmatrix}
    \label{eq:tensile_stress}
\end{equation}
where the entries $k_{11}$, $k_{22}$, $k_{12}$, $k_{33}$ composing the stiffness matrix, $\mathbf{K}$, denote cloth stretching stiffness in warp and weft direction, Poisson ratio, and shearing stiffness~\cite{wang2011data}. $\tau$ is fabric's thickness. Cloth orthotropic material properties observed in the stretching test are modeled by the difference between the $k_{11}$ and $k_{22}$. The twisting tests also show different load-deformation curves when measuring the cloth sample in warp and weft direction. The reason is that, as cloth shearing deformation increases, the warp and weft yarns rotate at the cross overs and the fabric is no longer an orthotropic material~\cite{hu1997kes}. We adopt the transformation method proposed by~\cite{peng2005continuum} to transform the stiffness matrix $\mathbf{K}$ by
\begin{equation}
    \mathbf{K}= \mathbf{T}^\top \tilde{\mathbf{K}} \mathbf{T} \text{, }
    \mathbf{T}=
    \begin{bmatrix}
        1 & \cos^2 \alpha & 2 \cos \alpha \\
        0 & \sin^2 \alpha & 0 \\
        0 & \sin \alpha  \cos \alpha & \sin \alpha
    \end{bmatrix}
    \label{eq:trans}
\end{equation}
where $\alpha$ in the transform matrix $\mathbf{T}$ is the angle between the warp and weft yarns (\Cref{fig:warp_weft_rotate_app}). The orthotropic stretching stiffness will be involved in the shearing stress when the fabric is twisted in different directions. Essentially, the difference between the twist load-deformation curves in warp and weft directions is caused by the different stretching stiffness in the warp and weft directions. 

To model the material nonlinearity absent in the StVK model~\cite{volino2009simple} but observed in the twisting test due to shear lock~\cite{peirce19375}, we adopt the Yeoh model to encode fabric material nonlinearity based on the discovery of~\cite{julio2006comparison}. It captures material nonlinearity by polynomial function: $\psi_{shear} = \sum_{i=1}^I c_{i} \RNum{1}^i$. We use the second-order polynomial function to fit the observed nonlinear shearing deformation and define the shear stress by 
\begin{equation}
    \sigma_{uv} = 2 c_0 \varepsilon_{uv} + 3 c_1 \varepsilon_{uv}^2
    \label{eq:shear_stress}
\end{equation}
where $c_0$ and $c_1$ decide cloth shearing stiffness.

Cloth bending energy derives from the departure of the cloth's current curvature from its rest curvature. We adopt the hinge-based model for describing cloth bending deformation across the triangle mesh due to its simplicity~\cite{grinspun2003discrete}. It approximates cloth curvature by the local mean curvatures based on the dihedral angle between two adjacent triangle faces in mesh: $\kappa = (\theta - \bar{\theta}) \frac{l}{h}$ where $l$ is the rest length of the common edge between the two triangles and $h$ is the one third of the sum of the two triangles original height incident to the common edge, i.e., $\frac{h_1 + h_2}{6}$. To capture material orthotropy, our simulator defines the bending energy density as
\begin{equation}
    \psi_{bending}(\boldsymbol{\kappa}) = \frac{1}{2} k_{b11} \kappa_{uu}^2 + \frac{1}{2} k_{b22} \kappa_{vv}^2 + 2 k_{b33} \kappa_{uv}^2 +  k_{b12} \kappa_{uu} \kappa_{vv}
\end{equation}
The curvature of a bending edge is decomposed along the principal directions which are aligned to cloth warp and weft direction: $\kappa_{uu} = \kappa \cos^2\phi$, $\kappa_{vv} = \kappa \sin^2\phi$ and $\kappa_{uv} = \kappa \sin\phi \cos\phi$. $\phi$ is the bias angle between the bending edge and the warp direction in the fabric material coordinate. Therefore, \(k_{b11}\) and \(k_{b22}\) define the bending stiffnesses in the warp and weft directions. The term \(k_{b12}\) creates spontaneous curvature in the perpendicular direction without applying an external moment there, analogous to a Poisson ratio but for curvature.The term \(k_{b33}\) is the torsional (warping) rigidity, which resists out-of-plane twisting deformation where the fabric bends in two principal directions simultaneously (i.e., bending diagonally).

\paragraph{Internal friction} Cloth internal friction derives from the frictional contacts of fibers that compose the cloth. We employ Coulomb friction model because, on the one hand, it can capture hysteresis observed in the measurement; on the other hand, it clearly delimits stick and slip friction so that it is more controllable in simulating wrinkles than the friction models without stick friction, e.g., Dahl model. Our simulator defines an anchor strain, $\bar{\varepsilon}$, to simulate stick and slip friction: $\bar{\varepsilon}$ only varies when slip friction occurs. To simulate the persistent wrinkles caused by internal friction, $\bar{\varepsilon}$ varies when the cloth strain departs from the current anchor strain farther than the maximum stick friction threshold, $\varepsilon_{thres}$:
\begin{equation}
    \bar{\varepsilon}\leftarrow
    \begin{cases}
        \bar{\varepsilon},  &|\varepsilon - \bar{\varepsilon}| < \varepsilon_{thres}\\
        \bar{\varepsilon} + sgn(\varepsilon - \bar{\varepsilon})(|\varepsilon - \bar{\varepsilon}| - \varepsilon_{thres}),  &\mbox{otherwise}
    \end{cases}
\end{equation}
The friction stress tends to keep the cloth in the anchor strain:
\begin{equation}
    \sigma_{fric} = 
    \begin{cases}
        \tau k_{fric} (\varepsilon - \bar{\varepsilon}), \quad &|\varepsilon - \bar{\varepsilon}| < \varepsilon_{thres}\\
        \tau k_{fric} sgn(\varepsilon - \bar{\varepsilon}) \varepsilon_{thres}, \quad &\mbox{otherwise}
    \end{cases}
\end{equation}
where $k_{fric}$ determines the strength of friction force. Therefore, as the anchor strain $\bar{\varepsilon}$ departs from a flat state, the friction stress will tend to retain the cloth in a non-flat state which forms persistent wrinkles. We use the Rectified Linear (ReLU) function to tell stick-slip friction and then the friction stress is defined by
\begin{gather}
    \bar{\varepsilon} \leftarrow \bar{\varepsilon} + sgn(\varepsilon - \bar{\varepsilon})\mbox{ReLU}(|\varepsilon - \bar{\varepsilon}| - \varepsilon_{thres}) \label{eq:anchor_update}\\
    \sigma_{fric} = \tau k_{fric} (\varepsilon - \bar{\varepsilon}) \label{eq:fric_stress}
\end{gather}

\paragraph{Plasticity} Cloth plasticity models the irreversible deformation resulting from severe cloth deformations. The cloth strain is decoupled into an elastic and a plastic part by linear addition, i.e., $\varepsilon = \varepsilon_e + \varepsilon_p$. Cloth tensile and bending stresses tend to restore the cloth to a state where $\varepsilon_e = 0$. Consequently, introducing plastic strain causes the cloth to tend toward a state with non-zero total strain. Plastic strain evolves only when the elastic strain exceeds the current yield strain $\varepsilon_y$. We adopt a hardening plasticity model in which the yield strain increases with the accumulated plastic deformation $\varepsilon_{hp}$:
\begin{equation}
    \varepsilon_{hp} \leftarrow
    \begin{cases}
        \varepsilon_{hp}, &\mbox{if } |\varepsilon_e - \varepsilon_p | < \varepsilon_y \\
        \varepsilon_{hp} + (|\varepsilon_e - \varepsilon_p| - \varepsilon_y), &\mbox{otherwise}
    \end{cases}
\end{equation}
\begin{equation}
    \varepsilon_y = \varepsilon_{y0} + \frac{k_h}{k} \varepsilon_{hp}
    \label{eq:yeild_update}
\end{equation}
where $\varepsilon_{y0}$ defines the cloth initial yield strain. $k_h$ is the plastic hardening stiffness and records the accumulated plastic deformations. Also, the strain exceeding the yield strain only partially flows to the plastic strain:
\begin{equation}
    \varepsilon_p \leftarrow
    \begin{cases}
        \varepsilon_p &\mbox{if } |\varepsilon_e - \varepsilon_p | < \varepsilon_y \\
        \varepsilon_p + sgn(\varepsilon_e - \varepsilon_p) \frac{k}{k_h + k} (|\varepsilon_e - \varepsilon_p| - \varepsilon_y ) &\mbox{otherwise} \notag
    \end{cases}
\end{equation}
We use ReLU function to model the step function:
\begin{gather}
    \varepsilon_{hp} \leftarrow \varepsilon_{hp} + \mbox{ReLU} (|\varepsilon_e - \varepsilon_p| - \varepsilon_y) \label{eq:hardening_update} \\
    \varepsilon_p \leftarrow \varepsilon_p + sgn(\varepsilon_e - \varepsilon_p) \frac{k}{k_h + k} \mbox{ReLU} (|\varepsilon_e - \varepsilon_p| - \varepsilon_y) \label{eq:plastic_update}
\end{gather}

\section{Optimization Strategy}

\begin{figure*}[htb]
    \centering
    \includegraphics[width=\textwidth]{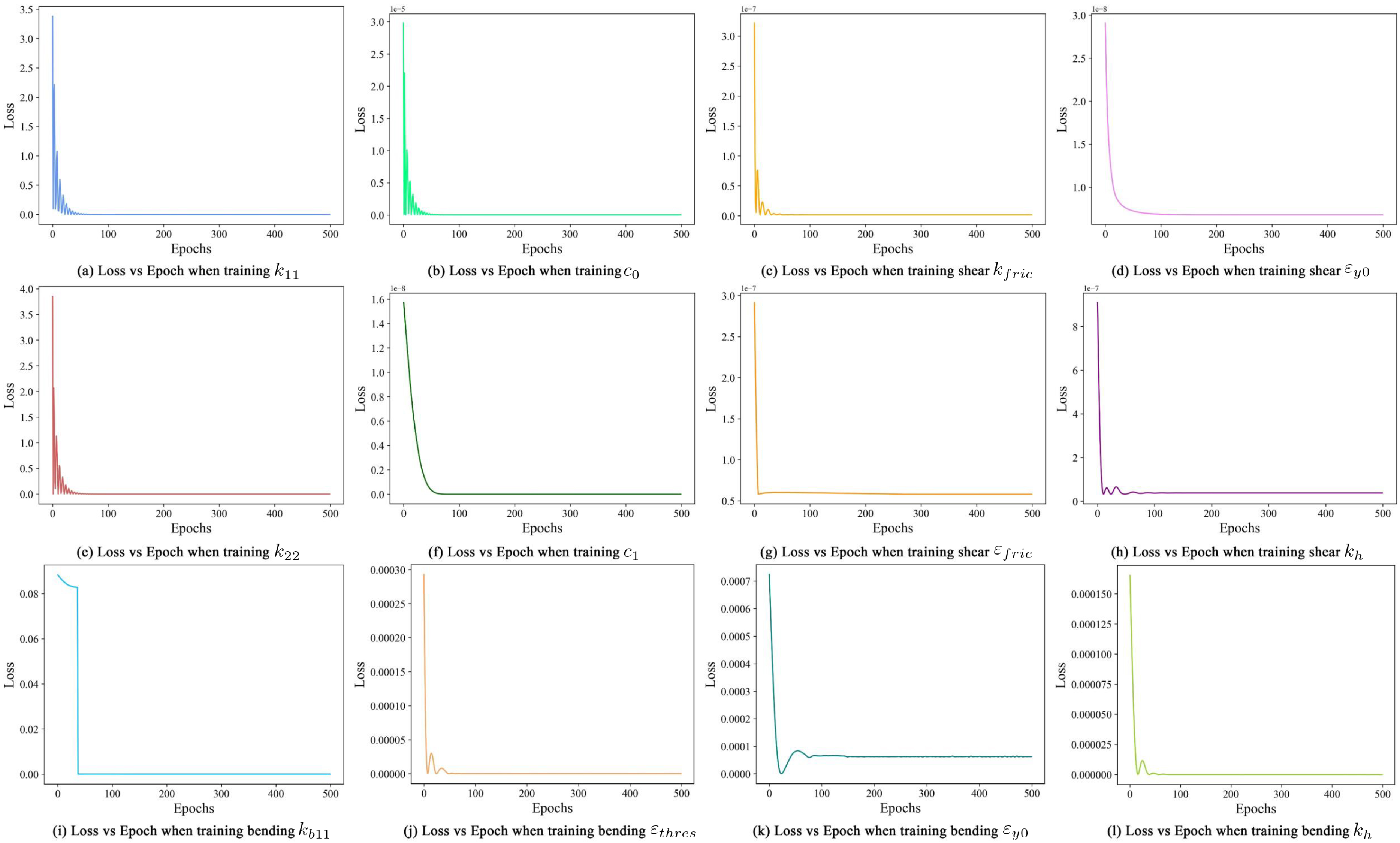}
    \caption{\rev{Training loss vs. epoch when optimizing different parameters. The losses are all plateaus within 500 epochs.}}
    \label{fig:train_loss}
\end{figure*}

The learned parameters are $k_{11}, k_{22}, c_0, c_1, k_{b11}, k_{b22}, \boldsymbol{\varepsilon}_{thres}, \boldsymbol{k}_{fric}, \boldsymbol{\varepsilon}_{y0}$, $\boldsymbol{k}_h$, where bold symbols denote two-element vectors for friction and plasticity in bending and shearing deformations, respectively. \rev{We use a cylindrical mesh consisting of 1,792 vertices in training.} Our differentiable cloth simulator replicates the compression, twisting, and stretching operations performed by the fabric tester and computes the reaction force from the simulated cloth sample, thereby generating simulated load–deformation curves. The loss function is defined as the mean squared error (MSE) between the simulated and measured force or torque:

\begin{equation}
\mathcal{L} = \frac{1}{T}\sum_{i=1}^T (\mbox{f}_i - \hat{\mbox{f}}_i)^2,
\label{eq:loss}
\end{equation}
where $\mbox{f}$ and $\hat{\mbox{f}}$ represent the simulated and measured force or torque, respectively, and $T$ is the number of simulation steps. Parameter estimation is performed in three sequential stages:

First, we estimate the cloth stretching stiffness by optimizing $k_{11}$ and $k_{22}$ to fit the stretching curves in the warp and weft directions, respectively. These parameters are determined first because they affect the fabric’s shearing stress through yarn rotation.

Second, we estimate parameters related to cloth shearing deformation. We assume that the tester's twisting operation incurs negligible lateral contraction and set the Poisson ratio \rev{$\nu$} to zero, \rev{so $k_{12} = 0$}. Also, the manipulator's twisting angle is used to approximate the rotation angle between warp and weft, i.e., $\alpha$ in~\Cref{eq:trans} is approximated by $\pi - \mbox{twisting angle}$. We begin by considering fabric sample as a purely elastic material and optimize $c_0$ to fit the first loading curve. If nonlinearity is observed in the measured data, $c_1$ is also optimized. Next, internal friction parameters $k_{fric}$ and $\varepsilon_{thres}$ are included to fit the first loading–unloading cycle, as the hysteresis in the cycle reveals cloth internal friction properties. Finally, plasticity parameters $k_h$ and $\varepsilon_{y0}$ are introduced by fitting the first loading–unloading cycle and the second loading curve, where the unrecoverable deformations are shown after including the second loading curves. We approximate the Poisson ratio by \rev{$\nu = \frac{0.5(k_{11}+k_{22})}{2c_0} - 1$}.

Third, we optimize bending parameters using compression curves as GT. Compression testing effectively performs an axial compression test on a thin-walled cylindrical shell. The critical buckling force depends not only on material stiffness but also on the shell’s geometric structure~\cite{weaver2003interactive}. Due to imperfect geometry, measured critical forces are typically lower than theoretical predictions based on Euler shell model~\cite{harutyunyan2023buckling}. In our setup, the soft fabric sample can hardly be clamped into a perfect cylindrical shape, so simulations using an ideal cylinder yield higher critical buckling force predictions than real-world measurements. If we assumed the tested sample is perfectly cylindrical, the learned bending stiffness would become unrealistically small to match the measured buckling force, which would then fail to capture post-buckling behavior and violate physical consistency. Since the exact initial shape of the sample is unavailable (due to occlusion by the testing apparatus), our simulator cannot accurately predict the critical buckling force. Therefore, we exclude the buckling region in the first compression cycle and use the post-buckling part to estimate cloth bending parameters. Optimization proceeds as follows:

\begin{enumerate}
    \item Fit $k_{b11}$ (or $k_{b22}$) to the first loading curve in compression.
    \item Include friction parameters $k_{fric}$ and $\varepsilon_{thres}$ by fitting the first loading-unloading cycle.
    \item Include plastic parameters $k_h$ and $\varepsilon_{y0}$ by fitting the first loading–unloading cycle and the second loading curve.
\end{enumerate}
Note that $k_{fric}$, $\varepsilon_{thres}$, $k_h$ and $\varepsilon_{y0}$ are independent for warp and weft directions, and should be learned by fitting the compression curves in warp and weft directions. For simplicity, we assume the bending in fabric principal directions (i.e., warp and weft) is independent and set $k_{b12}=0$. The rationale is that bending a cloth in its warp direction usually only bends the warp yarns and does not bend its weft yarns, and vice versa. To the torsional stiffness $k_{b33}$, we approximate it by the linear components of shearing stiffness, i.e., $k_{b33}=\frac{c_0 \tau^3}{12}$ \cite{jones1998mechanics}. For both compression and twisting, the loading–unloading cycles beyond the second are repetitive and are therefore excluded from optimization. \rev{In summary, the optimized parameters include 
\begin{itemize}[leftmargin=*]
    \item Stretching stiffness in warp and weft directions $k_{11}$ and $k_{22}$;
    \item Shearing stiffness $c_0$ and $c_1$
    \item Shearing friction stiffness $k_{fric\_shear}$ and threshold $\varepsilon_{thres\_shear}$
    \item Shearing yield strain $\varepsilon_{y0\_shear}$ and hardening parameter $k_{h\_shear}$
    \item Bending stiffness in warp and weft $k_{b11}$ and $k_{b22}$
    \item Bending friction stiffness $k_{b11\_fric}$ and $k_{b22\_fric}$, and thresholds $\varepsilon_{b11\_thres}$ and $\varepsilon_{b22\_thres}$
    \item Bending yield strains $\varepsilon_{y0\_bend\_warp}$ and $\varepsilon_{y0\_bend\_weft}$, and hardening parameters $k_{h\_bend\_warp}$ and $k_{h\_bend\_weft}$
\end{itemize}
and the assumed parameters include
\begin{itemize}[leftmargin=*]
    \item Stretching stiffness $k_{12}$
    \item Bending parameters $k_{b12}$ and $k_{b33}$
\end{itemize}} 
All parameters are optimized using the Adam optimizer over 500 epochs. \rev{\Cref{fig:train_loss} shows the losses when training the parameters by the synthesized data, where the results show that the training losses have become stable before 500 epochs. The learning rate for the stiffness parameters is 100.0, and learning rate for the friction threshold and yield strain is 0.001. During training, the time step size $h=0.01s$, the $\hat{d}$ is set to the measured fabric thickness. For frictional contact, the parameters $\kappa=1e8$ and $\mu=0.5$ control the barrier potential and defines friction coefficient. We use elastic soft constraint to simulate the apparatus's manipulations, where the constraint energy 
of the manipulated vertices is $E = \frac{1}{2}k_{cons} m \|\mathbf{x} - \mathbf{x}_{target}\|^2$, where $k_{cons}$ defines the stiffness of the soft constraint, $m$ is the mass of the vertex, $\mathbf{x}$ is the vertex's current position, and $\mathbf{x}_{target}$ defines where the vertex should be. $k_{cons}$ is $200$Pa for compression and twist, and $2000$Pa for stretching.}

\section{Adjoint Method}

Our differentiable cloth simulator employs the adjoint method~\cite{givoli2021tutorial, huang2024differentiable} to do the gradient calculation for its efficiency. Our differentiable cloth simulator solves a constrained optimization problem
\begin{align}
    \min_{\Theta} \quad &\mathcal{J}(\mathbf{x}, \Theta) = \int_{t=0}^T \mathcal{J} (\mathbf{x}, \Theta, t) dt \\
    s.t. \quad &g(\mathbf{x}, \Theta, t) = 0, \\
    &\mathbf{x}(0) = \mathbf{x}^{(0)} \mbox{ and } \dot{\mathbf{x}}(0) = \dot{\mathbf{x}}^{(0)}
\end{align}
where $\mathbf{x}^{(0)}$ and $\dot{\mathbf{x}}^{(0)}$ define the cloth initial state $\mathcal{S}_0$. Thus, they can be excluded from the constraints. The objective/loss function $\mathcal{J}$ measures the difference between the simulation and measurements by
\begin{equation}
    \mathcal{J} (\mathbf{x}, \Theta, t) = (f_{phy}(\mathbf{x}, \Theta, t) - \hat{\mathbf{f}})^2
    \label{eq:adjoint_loss}
\end{equation}
, which is the continuous form of  \Cref{eq:loss}. $g$ essentially encodes the Newton's Second law, $ \rho \ddot{\mathbf{x}} = \sigma$ (or $\mathbf{M}\ddot{\mathbf{x}} = \mathbf{f}$ for geometrically discretized system), to model the simulator's ordinary differential equation (ODE)
\begin{equation}
    g (\mathbf{x}, \Theta) =  \int_{t=0}^T g (\mathbf{x}, \Theta, t) dt = \mathbf{M} \int_{t=0}^T \ddot{\mathbf{x}}(t) - f_{phy} (\mathbf{x}, \Theta, t) dt
    \label{eq:continous_constrain}
\end{equation}
where $f_{phy}$ calculates the forces imposed on the cloth, e.g., stretching, bending, gravity, collision, and external friction, etc. Our physics-based simulator employs the implicit Euler method \cite{baraff2023large} to approximate the ODE numerically by
\begin{gather}
    \mathbf{x}^{(t)} - \mathbf{x}^{(t-1)} = h \dot{\mathbf{x}}^{(t)} \\
    \mathbf{M} ({\mathbf{x}}^{(t)} - {\mathbf{x}}^{(t-1)}) = h f_{phy}(\mathbf{x}^{(t)}, \Theta)
\end{gather}
where $h$ is the time interval for approximate temporal integration. Therefore, the constraints imposed by $g$ ( \Cref{eq:continous_constrain}) are replaced by
\begin{gather}
    \sum_{t=1}^T \mathbf{M} ({\mathbf{x}}^{(t)} - {\mathbf{x}}^{(t-1)}) - h f_{phy}(\mathbf{x}^{(t)}, \Theta) = 0 \\
    \sum_{t=1}^T \mathbf{x}^{(t)} - \mathbf{x}^{(t-1)} - h \dot{\mathbf{x}}^{(t)} = 0
\end{gather}
and so does the loss function:
\begin{equation}
    \int_{t=0}^T \mathcal{J} (\mathbf{x}, \Theta, t) dt = \sum_{t=0}^{T} \mathcal{J} (\mathbf{x}^{(t)}, \Theta).
\end{equation}
In every testing, the tested sample is manipulated from the same initial state. Therefore, $\mathbf{x}^{(0)}$ is known and $\dot{\mathbf{x}}^{(0)}$ is always $\mathbf{0}$, and they are not determined by the cloth physical parameters $\Theta$. The optimization problem can be converted to the \textit{Lagrangian}:
\begin{align}
    &\mathcal{L}(\mathcal{S}, \Theta, \boldsymbol{\lambda}_1, \boldsymbol{\lambda}_2)  \notag\\
    =&\mathcal{J}(\mathbf{x}, \Theta) + \mathcal{L}_{cons}(\mathcal{S}, \Theta, \boldsymbol{\lambda}_1, \boldsymbol{\lambda}_2)
    \quad + \mathcal{L}_{init}(\mathcal{S}^{(0)}, \Theta, \boldsymbol{\lambda}_1^{(0)}, \boldsymbol{\lambda}_2^{(0)}) \notag \\
    =& \sum_{t=1}^{T} \big( \mathcal{J} (\mathbf{x}^{(t)}, \Theta) + 
    \boldsymbol{\lambda}_1^{(t)} (\mathbf{M} (\dot{\mathbf{x}}^{(t)} - \dot{\mathbf{x}}^{(t-1)}) - h f_{phy}(\mathbf{x}^{(t)}, \Theta)) + \notag \\
    &\boldsymbol{\lambda}_2^{(t)} (\mathbf{x}^{(t)} - \mathbf{x}^{(t-1)} - h \dot{\mathbf{x}}^{(t)}) \big) \notag \\
    =& \sum_{t=1}^{T} \big( \mathcal{J} (\mathbf{x}^{(t)}, \Theta) + 
    \boldsymbol{\lambda}_1^{(t)} (\mathbf{M} \dot{\mathbf{x}}^{(t)} - h f_{phy}(\mathbf{x}^{(t)}, \Theta)) +
    \boldsymbol{\lambda}_2^{(t)} (\mathbf{x}^{(t)}  \notag \\
    &- h \dot{\mathbf{x}}^{(t)}) \big) - \sum_{t=0}^{T-1} \big( \boldsymbol{\lambda}_1^{(t+1)} \mathbf{M} \dot{\mathbf{x}}^{(t)} + \boldsymbol{\lambda}_{2,t+1} \mathbf{x}_{t} \big) \notag \\
   =&\sum_{t=0}^{T-1} \big( \mathcal{J} (\mathbf{x}^{(t)}, \Theta) + \boldsymbol{\lambda}_1^{(t)} (\mathbf{M} \dot{\mathbf{x}}^{(t)} - h f_{phy}(\mathbf{x}^{(t)}, \Theta)) + \notag \\
   & \boldsymbol{\lambda}_2^{(t)} (\mathbf{x}^{(t)} - h \dot{\mathbf{x}}^{(t)}) - \boldsymbol{\lambda}_1^{(t+1)} \mathbf{M} \dot{\mathbf{x}}^{(t)} - \boldsymbol{\lambda}_2^{(t+1)} \mathbf{x}^{(t)} \big) + \mathcal{J} (\mathbf{x}^{(T)}, \Theta) + \notag \\
   & \boldsymbol{\lambda}_1^{(T)} (\mathbf{M} \dot{\mathbf{x}}^{(T)} - h f_{phy}(\mathbf{x}^{(T)}, \Theta)) + \boldsymbol{\lambda}_2^{(T)} (\mathbf{x}^{(T)} - h \dot{\mathbf{x}}^{(T)}) \notag \\
   =& \sum_{t=0}^{T-1} \big( \mathcal{J} (\mathbf{x}, \Theta, t) - \lambda_{1,t} h f_{phy}(\mathbf{x}_{t}, \Theta, t) + (\boldsymbol{\lambda}_2^{(t)} - \boldsymbol{\lambda}_2^{(t+1)}) \mathbf{x}^{(t)} \notag \\
   & + ((\boldsymbol{\lambda}_1^{(t)} - \boldsymbol{\lambda}_1^{(t+1)})\mathbf{M} - \boldsymbol{\lambda}_2^{(t)} h) \dot{\mathbf{x}}^{(t)} \big) + \mathcal{J} (\mathbf{x}^{(T)}, \Theta) \notag \\
   &- \boldsymbol{\lambda}_1^{(T)} h f_{phy}(\mathbf{x}^{(T)}, \Theta) + \boldsymbol{\lambda}_2^{(T)} \mathbf{x}^{(T)} +
   (\boldsymbol{\lambda}_1^{(T)} \mathbf{M}  - \boldsymbol{\lambda}_2^{(T)} h ) \dot{\mathbf{x}}^{(T)}
\end{align}
where the two Lagrangian parameters $\boldsymbol{\lambda}_1$ and $\boldsymbol{\lambda}_2$ are the adjoint variables. The derivative of the \textit{Lagrangian} w.r.t. the cloth physical parameter $\Theta$ is
\begin{align}
    \frac{\partial \mathcal{L}}{\partial \Theta} =& \sum_{t=1}^{T-1} 
    \Big( 
        \frac{\partial \mathcal{J}(\mathbf{x}^{(t)}, \Theta)}{\partial \Theta} +
        \boldsymbol{\lambda}_1^{(t)} h \frac{\partial f_{phy}(\mathbf{x}^{(t)}, \Theta)}{\partial \Theta}\notag \\
        & (\boldsymbol{\lambda}_1^{(t)} - \boldsymbol{\lambda}_1^{(t+1)}) \frac{\partial \mathbf{M}}{\partial \Theta} \dot{\mathbf{x}}^{(t)}
    \Big) +
    \sum_{t=1}^{T-1} 
    \Big( 
        \big( 
            \boldsymbol{\lambda}_2^{(t)} - \boldsymbol{\lambda}_2^{(t+1)} + \notag \\ 
            & \frac{\partial \mathcal{J}(\mathbf{x}, \Theta, t)}{\partial \mathbf{x}_t} -
            \boldsymbol{\lambda}_1^{(t)} h \frac{\partial f_{phy}(\mathbf{x}^{(t)}, \Theta)}{\partial \mathbf{x}^{(t)}}
        \big)
        \frac{\partial \mathbf{x}^{(t)}}{\partial \Theta} - \notag \\
        &\boldsymbol{\lambda}_1^{(t)} h \frac{\partial f_{phy}(\mathbf{x}^{(t)}, \Theta)}{\partial \mathbf{x}^{(t-1)}}\frac{\partial \mathbf{x}^{(t-1)}}{\partial \Theta} + \notag \\
        &\big(
            (\boldsymbol{\lambda}_1^{(t)} - \boldsymbol{\lambda}_1^{(t+1)}) \mathbf{M} - \boldsymbol{\lambda}_2^{(t)} h 
        \big)
        \frac{\partial \dot{\mathbf{x}}^{(t)}}{\partial \Theta}
    \Big) +
    \frac{\partial \mathcal{J}(\mathbf{x}^{(T)}, \Theta)}{\partial \Theta} + \notag \\
    & \boldsymbol{\lambda}_1^{(T)} h \frac{\partial f_{phy}(\mathbf{x}^{(T)}, \Theta)}{\partial \Theta} + 
    \big(\frac{\partial \mathcal{J}(\mathbf{x}^{(T)}, \Theta)}{\partial \mathbf{x}^{(T)}} - \notag \\
    & \boldsymbol{\lambda}_1^{(t)} h \frac{\partial f_{phy}(\mathbf{x}^{(T)}, \Theta)}{\partial \mathbf{x}^{(T)}} + \boldsymbol{\lambda}_2^{(t)} h \big) \frac{\partial \mathbf{x}^{(T)}}{\partial \Theta} - \notag \\
    & \boldsymbol{\lambda}_1^{(T)} h \frac{\partial f_{phy}(\mathbf{x}^{(T)}, \Theta)}{\partial \mathbf{x}^{(T-1)}} \frac{\partial \mathbf{x}^{(T-1)}}{\partial \Theta} + 
    (\boldsymbol{\lambda}_1^{(T)} \mathbf{M} - \boldsymbol{\lambda}_{2,T} h) \frac{\partial \dot{\mathbf{x}}^{(T)}}{\partial \Theta} + \notag \\
    & \boldsymbol{\lambda}_1^{(T)} \frac{\partial \mathbf{M}}{\partial \Theta} \dot{\mathbf{x}}^{(T)}
\end{align}
Rearrange the terms to merge the $\frac{\partial \mathbf{x}}{\partial \Theta}$ and $\frac{\partial \dot{\mathbf{x}}}{\partial \Theta}$ together:
\begin{align}
    \frac{\partial \mathcal{L}}{\partial \Theta} 
    =& \sum_{t=1}^{T-1} 
    \Big( 
        \frac{\partial \mathcal{J}(\mathbf{x}^{(t)}, \Theta)}{\partial \Theta} +
        \boldsymbol{\lambda}_1^{(t)} h \frac{\partial f_{phy}(\mathbf{x}^{(t)}, \Theta)}{\partial \Theta} \notag \\
        & (\boldsymbol{\lambda}_1^{(t)} - \boldsymbol{\lambda}_1^{(t+1)}) \frac{\partial \mathbf{M}}{\partial \Theta} \dot{\mathbf{x}}_t
    \Big) +
    \sum_{t=1}^{T-1} 
    \Big( 
        \big( 
            \boldsymbol{\lambda}_2^{(t)} - \boldsymbol{\lambda}_2^{(t+1)} + \notag \\ 
            & \frac{\partial \mathcal{J}(\mathbf{x}^{(t)}, \Theta)}{\partial \mathbf{x}^{(t)}} -
            \boldsymbol{\lambda}_1^{(t)} h \frac{\partial f_{phy}(\mathbf{x}^{(t)}, \Theta)}{\partial \mathbf{x}^{(t)}} - \notag \\
            &\boldsymbol{\lambda}_1^{(t+1)} h \frac{\partial f_{phy}(\mathbf{x}^{(t+1)}, \Theta)}{\partial \mathbf{x}^{(t)}}
        \big)
        \frac{\partial \mathbf{x}^{(t)}}{\partial \Theta} - \notag \\
        &\big(
            (\boldsymbol{\lambda}_1^{(t)} - \boldsymbol{\lambda}_1^{(t+1)})\mathbf{M} - \boldsymbol{\lambda}_2^{(t)} h 
        \big)
        \frac{\partial \dot{\mathbf{x}}^{(t)}}{\partial \Theta}
    \Big) +
    \frac{\partial \mathcal{J}(\mathbf{x}^{(T)}, \Theta)}{\partial \Theta} - \notag \\
    & \boldsymbol{\lambda}_1^{(T)} h \frac{\partial f_{phy}(\mathbf{x}^{(T)}, \Theta)}{\partial \Theta} + 
    \big(\frac{\partial \mathcal{J}(\mathbf{x}^{(T)}, \Theta)}{\partial \mathbf{x}^{(T)}} - \notag \\
    & \boldsymbol{\lambda}_1^{(t)} h \frac{\partial f_{phy}(\mathbf{x}^{(T)}, \Theta)}{\partial \mathbf{x}^{(T)}} + \boldsymbol{\lambda}_2^{(T)} h \big) \frac{\partial \mathbf{x}^{(T)}}{\partial \Theta} - \notag \\
    & (\boldsymbol{\lambda}_1^{(T)} \mathbf{M} - \boldsymbol{\lambda}_2^{(T)} h) \frac{\partial \dot{\mathbf{x}}^{(T)}}{\partial \Theta} +
    \boldsymbol{\lambda}_1^{(T)} \frac{\partial \mathbf{M}}{\partial \Theta} \dot{\mathbf{x}}^{(T)}
\end{align}
Adjoint method cancels the high-dimensional derivatives $\frac{\partial \mathbf{x}}{\partial \Theta}$ and $\frac{\partial \dot{\mathbf{x}}}{\partial \Theta}$ by solving the adjoint equations
\begin{align}
    (\boldsymbol{\lambda}_1^{(t)} - \boldsymbol{\lambda}_1^{(t+1)})\mathbf{M} =& \boldsymbol{\lambda}_2^{(t)} h\\
    \boldsymbol{\lambda}_2^{(t)} - \boldsymbol{\lambda}_2^{(t+1)} =&
    \boldsymbol{\lambda}_1^{(t)} h \frac{\partial f_{phy}(\mathbf{x}^{(t)}, \Theta)}{\partial \mathbf{x}^{(t)}} + \notag \\
    &\boldsymbol{\lambda}_1^{(t+1)} h \frac{\partial f_{phy}(\mathbf{x}^{(t+1)}, \Theta)}{\partial \mathbf{x}^{(t)}} -
    \frac{\partial \mathcal{J}(\mathbf{x}^{(t)}, \Theta)}{\partial \mathbf{x}^{(t)}} 
\end{align}
to make the coefficients of the derivatives equal to zero. This can be simplified by introducing $\lambda_2 = \mathbf{M} v$:
\begin{align}
    \boldsymbol{\lambda}_1^{(t)} - \boldsymbol{\lambda}_1^{(t+1)} =& \dot{\mathbf{x}}^{(t)} h\\
    \mathbf{M}(v_t - v_{t+1}) =&
    \boldsymbol{\lambda}_1^{(t)} h \frac{\partial f_{phy} (\mathbf{x}^{(t)}, \Theta)}{\partial \mathbf{x}^{(t)}} + \notag \\
    &\boldsymbol{\lambda}_1^{(t+1)} h \frac{\partial f_{phy}(\mathbf{x}^{(t)}, \Theta)}{\partial \mathbf{x}^{(t)}} -
    \frac{\partial \mathcal{J}(\mathbf{x}^{(t)}, \Theta)}{\partial \mathbf{x}^{(t)}} 
\end{align}
As the constraint requires $g(\mathbf{x}, \Theta) = 0$, we can arbitrarily choose $\boldsymbol{\lambda}_1$ and $\boldsymbol{\lambda}_2$ to let $\mathcal{L}(\mathbf{x}, \Theta, \boldsymbol{\lambda}) = \mathcal{J}$, and then the derivative of the loss function  \Cref{eq:loss} is
\begin{equation}
    \frac{\partial \mathcal{J}}{\partial \Theta} = \frac{\partial \mathcal{L}}{\partial \Theta}
\end{equation}
Then, the derivative $\frac{\partial \mathcal{J}}{\partial \Theta}$ can be obtained by dropping all the terms that contain $\frac{\partial \mathbf{x}}{\partial \Theta}$ and $\frac{\partial \dot{\mathbf{x}}}{\partial \Theta}$. $\frac{\partial \mathbf{M}}{\partial \Theta}$ can also be removed as its derivative is $\mathbf{0}$. Then, the derivative
\begin{equation}
    \frac{\partial \mathcal{J}}{\partial \Theta} =    
    \sum_{t=1}^{T} 
    \big( 
        \frac{\partial \mathcal{J}(\mathbf{x}^{(t)}, \Theta)}{\partial \Theta} +
        \mathbf{\lambda}_{1,t} h \frac{\partial f_{phy}(\mathbf{x}^{(t)}, \Theta)}{\partial \Theta}
    \big)
\end{equation}
can be used to update the cloth physical parameters $\Theta$. To our differentiable cloth simulator, $\mathcal{J}$ and $f_{phy}$ also take as input the anchor strain of the friction model, the plastic strain, and hardening plastic strain, i.e., $f_{phy}(\mathbf{x}^{(t)}, \Theta, \bar{\varepsilon}^{(t)}, \varepsilon_{p}^{(t)}, \varepsilon_{hp}^{(t)})$ and $\mathcal{J}(\mathbf{x}^{(t)}, \Theta, \bar{\varepsilon}^{(t)}, \varepsilon_{p}^{(t)}, \varepsilon_{hp}^{(t)})$.

In forward simulation, the semi-implicit scheme freezes the variables $\bar{\varepsilon}^{(t)}$, $\varepsilon_{p}^{(t)}$ and $\varepsilon_{hp}^{(t)}$ when predicting $\mathbf{x}^{(t+1)}$: 
\begin{equation}
    \mathbf{x}^{(t+1)} = \argmin_{\mathbf{x}} \; E(\mathbf{x} \mid \boldsymbol{\varepsilon}_p^{(t)}, \bar{\boldsymbol{\varepsilon}}^{(t)}, \boldsymbol{\varepsilon}_{hp}^{(t)})
\end{equation}
and updated \textit{a posteriori} based on the predicted \(\mathbf{x}^{(t+1)}\) using a standard return mapping:
\begin{equation}
    \boldsymbol{\varepsilon}_p^{(t+1)}, \rev{\bar{\boldsymbol{\varepsilon}}^{(t+1)}}, \boldsymbol{\varepsilon}_{hp}^{(t+1)} = \text{ReturnMapping}(\mathbf{x}^{(t+1)}, \mathcal{S}^{(t)})
\end{equation}
The adjoint method needs to save these frozen variables to calculate $\frac{\partial \mathcal{J}}{\partial \Theta}$ and $\frac{\partial f_{phy}}{\partial \Theta}$. Instead of directly saving $\boldsymbol{\varepsilon}_p^{(t)}, \bar{\boldsymbol{\varepsilon}}^{(t)}, \boldsymbol{\varepsilon}_{hp}^{(t)}$, we save the right hand side terms with superscript $(t-1)$ in the equations below:
\begin{align}
    \bar{\varepsilon}^{(t)} =& \bar{\varepsilon}^{(t-1)} + sgn(\varepsilon^{(t-1)}-\bar{\varepsilon}^{(t-1)}) \notag \\ &\mbox{ReLU} (|\varepsilon^{(t-1)}-\bar{\varepsilon}^{(t-1)}) | - \varepsilon_{thres}) \\
    \varepsilon_{hp}^{(t)} =& \varepsilon_{hp}^{(t-1)} + \mbox{ReLU} (|\varepsilon_e^{(t-1)} - \varepsilon_p^{(t-1)}| - (\varepsilon_{y0} + \frac{k_h}{k} \varepsilon_{hp}^{(t-1)})) \\
    \varepsilon_p^{(t)} =& \varepsilon_p^{(t-1)} + sgn(\varepsilon_e^{(t-1)} - \varepsilon_p^{(t-1)}) \notag \\ &\frac{k}{k_h + k} \mbox{ReLU} (|\varepsilon_e^{(t-1)} - \varepsilon_p^{(t-1)}| - (\varepsilon_{y0} + \frac{k_h}{k} \varepsilon_{hp}^{(t-1)})) 
\end{align}
In this way, the $\frac{\partial \mathcal{J}}{\partial \varepsilon_{y0}}$, $\frac{\partial \mathcal{J}}{\partial k_h}$, and $\frac{\partial \mathcal{J}}{\partial \varepsilon_{thres}}$ can be calculated by the chain rule:
\begin{gather}
    \frac{\partial \mathcal{J}}{\partial \varepsilon_{thres}} = \frac{\partial \mathcal{J}}{\partial \sigma_{fric}^{(t+1)}} \frac{\partial \sigma_{fric}^{(t+1)}}{\partial \bar{\varepsilon}^{(t)}} \frac{\partial \bar{\varepsilon}^{(t)}}{\partial \varepsilon_{thres}} \\
    \frac{\partial \mathcal{J}}{\partial \varepsilon_{y0}} = \frac{\partial \mathcal{J}}{\partial \varepsilon_{p}^{(t)}} \frac{\partial \varepsilon_{p}^{(t)}}{\partial \varepsilon_{y0}} + \frac{\partial \mathcal{J}}{\partial \varepsilon_{hp}^{(t)}}\frac{\partial \varepsilon_{hp}^{(t)}}{\partial \varepsilon_{y0}} \\
    \frac{\partial \mathcal{J}}{\partial k_h} = \frac{\partial \mathcal{J}}{\partial \varepsilon_{p}^{(t)}} \frac{\partial \varepsilon_{p}^{(t)}}{\partial k_h} + \frac{\partial \mathcal{J}}{\partial \varepsilon_{hp}^{(t)}}\frac{\partial \varepsilon_{hp}^{(t)}}{\partial k_h}
\end{gather} and according to \Cref{eq:anchor_update}, \Cref{eq:hardening_update}, \Cref{eq:plastic_update}, we have
\begin{gather}
    \frac{\partial \bar{\varepsilon}^{(t)}}{\partial \varepsilon_{thres}} = sgn(\varepsilon^{(t-1)}-\bar{\varepsilon}^{(t-1)}) \frac{\partial \mbox{ReLU} (|\varepsilon^{(t-1)}-\bar{\varepsilon}^{(t-1)}) | - \varepsilon_{thres})}{\partial \varepsilon_{thres}} \notag \\
    \frac{\partial \varepsilon_{p}^{(t)}}{\partial \varepsilon_{y0}} = sgn(\varepsilon_e^{(t-1)}-\varepsilon_p^{(t-1)}) \frac{\partial \mbox{ReLU} (|\varepsilon_e^{(t-1)}-\varepsilon_p^{(t-1)}) | - \varepsilon_y)}{\partial \varepsilon_{y0}} \notag \\
    \frac{\partial \varepsilon_{p}^{(t)}}{\partial k_h} = sgn(\varepsilon_e^{(t-1)}-\varepsilon_p^{(t-1)}) \frac{\partial \mbox{ReLU} (|\varepsilon_e^{(t-1)}-\varepsilon_p^{(t-1)}) | - \varepsilon_y)}{\partial k_h} \notag \\
    \frac{\partial \varepsilon_{hp}^{(t)}}{\partial \varepsilon_{y0}} = \frac{\partial \mbox{ReLU} (|\varepsilon_e^{(t-1)}-\varepsilon_p^{(t-1)}) | - \varepsilon_y)}{\partial \varepsilon_{y0}} \notag \\
    \frac{\partial \varepsilon_{hp}^{(t)}}{\partial k_h} = \frac{\partial \mbox{ReLU} (|\varepsilon_e^{(t-1)}-\varepsilon_p^{(t-1)}) | - \varepsilon_y)}{\partial k_h} \notag
\end{gather}

\pagebreak
\clearpage

\begin{table*}[ht]\tiny
\centering
\caption{Fabric Specification Table: Part One}
\label{tab:fabric1}
\resizebox{\textwidth}{!}{%
\begin{tabular}{lllcc}
\toprule
\textbf{No.} & \textbf{Name} & \textbf{Composition} & \textbf{Thickness (mm)} & \textbf{Density (gsm)} \\
\midrule
C01 & Cotton Plain & 100\% Cotton & 0.7 & 134.85 \\
C02 & Cotton Poplin & 100\% Cotton & 0.2 & 108.07 \\
C03 & High-Density Cotton Muslin & 100\% Cotton & 0.3 & 76.46 \\
C04 & Cotton Oxford & 100\% Cotton & 0.5 & 156.93 \\
C05 & Cotton Chambray & 100\% Cotton & 0.7 & 144.69 \\
C06 & Cotton Twill & 100\% Cotton & 0.6 & 115.72 \\
C07 & Cotton Khaki & 100\% Cotton & 0.6 & 228.48 \\
C08 & Cotton Canvas & 100\% Cotton & 0.6 & 238.5 \\
C09 & Cotton Corduroy & 100\% Cotton & 1.5 & 297.16 \\
C10 & Cotton Velveteen & 100\% Cotton & 1.5 & 229.16 \\
C11 & Classic Denim & 100\% Cotton & 1.0 & 293.89 \\
C12 & Stretch Denim & 98\% Cotton / 2\% Spandex & 0.6 & 308.48 \\
C13 & Lightweight Denim & 100\% Cotton & 0.7 & 249.68 \\
\textbf{C14} & \textbf{Thinner Plain Cotton} & 100\% Cotton & 0.4 & 172 \\
KN01 & Cotton Jersey & 100\% Cotton & 0.7 & 174.34 \\
KN02 & Cotton Rib & 100\% Cotton & 0.9 & 309.96 \\
KN03 & Cotton French Terry & 100\% Cotton & 1.4 & 239.61 \\
E-KN01 & Combed Cotton Jersey & 100\% Cotton & 0.7 & 108.77 \\
E-KN03 & Cotton Interlock & 100\% Cotton & 1.6 & 323.11 \\
E-KN04 & Cotton Air Layer & 100\% Cotton & 1.4 & 291.32 \\
E-KN05 & Double-sided French Terry & 100\% Cotton & 3 & 273.11 \\
E-KN06 & Double-sided Terry Towel Fabric & 100\% Cotton & 2.9 & 172.13 \\
E-C01 & High-count Cotton Poplin & 100\% Cotton & 0.4 & 84.26 \\
E-C02 & Low-count Cotton Canvas & 100\% Cotton & 2.3 & 576.45 \\
E-C03 & High-density Cotton Plain & 100\% Cotton & 0.2 & 101.89 \\
E-C04 & Stretch Cotton Khaki & 97\% Cotton / 3\% Spandex & 0.6 & 246.56 \\
E-C05 & Polyester/Cotton Plain & 65\% Polyester / 35\% Cotton & 0.4 & 102.97 \\
E-C06 & Polyester/Cotton Twill & 80\% Polyester / 20\% Cotton & 0.4 & 150.29 \\
E-C07 & Mercerized Cotton Plain & 100\% Cotton & 0.4 & 123.38 \\
E-C08 & Wrinkle-free Cotton Poplin & 100\% Cotton & 0.3 & 109.51 \\
E-C09 & Brushed Cotton Twill & 100\% Cotton & 0.8 & 219.76 \\
E-C10 & Cotton Herringbone Twill & 100\% Cotton & 1 & 246.47 \\
L01 & Linen Plain & 100\% Linen & 0.4 & 117.38 \\
L02 & Dew Retting Linen & 100\% Linen & 0.5 & 136.97 \\
L03 & Ramie Plain & 100\% Ramie & 0.3 & 65.77 \\
L04 & Heavier Linen & 100\% Linen & 0.3 & 151 \\

\bottomrule
\end{tabular}
}
\end{table*}

\begin{table*}[ht]\tiny
\centering
\caption{Fabric Specification Table: Part Two}
\label{tab:fabric2}
\resizebox{\textwidth}{!}{%
\begin{tabular}{lllcc}
\toprule
\textbf{No.} & \textbf{Name} & \textbf{Composition} & \textbf{Thickness (mm)} & \textbf{Density (gsm)} \\
\midrule
E-W01 & Wool/Polyester Blend & 70\% Wool / 30\% Polyester & 0.4 & 176.54 \\
E-W02 & Wool/Cashmere Blend & 90\% Wool / 10\% Cashmere & 0.4 & 143.68 \\
E-W03 & Hard-twist Worsted Valitin & 100\% Wool & 0.4 & 184.13 \\
E-W04 & Machine-washable Worsted & 100\% Wool & 1.8 & 338.18 \\
S01 & Silk Satin & 100\% Mulberry Silk & 0.2 & 72.67 \\
S02 & Silk Crepe de Chine & 100\% Mulberry Silk & 0.3 & 92.03 \\
S03 & Silk Habotai & 100\% Mulberry Silk & 0.1 & 52.44 \\
S04 & Silk Georgette & 100\% Mulberry Silk & 0.2 & 45.11 \\
S05 & Silk Chiffon & 100\% Mulberry Silk & 0.1 & 25.39 \\
S06 & Silk Taffeta & 100\% Mulberry Silk & 0.1 & 67.65 \\
S07 & Crêpe de Chine & 100\% Mulberry Silk & 0.5 & 85.56 \\
S08 & Sanma Song Brocade & 100\% Mulberry Silk & 0.3 & 140.25 \\
S09 & Song Brocade & 100\% Mulberry Silk & 0.5 & 217.86 \\
S10 & Habotai-base Gambiered Guangdong Gauze & 100\% Mulberry Silk & 0.4 & 103.62 \\
S11 & Leno-base Gambiered Guangdong Gauze & 100\% Mulberry Silk & 0.4 & 130.13 \\
\textbf{S12} & \textbf{Heavy Silk Crepe (Light)} & 100\% Mulberry Silk & 0.3 & 126 \\
E-S01 & Heavy Silk Crepe & 100\% Mulberry Silk & 0.4 & 142.81 \\
E-S02 & Silk Crepe Georgette & 100\% Mulberry Silk & 0.2 & 19.03 \\
E-S03 & Silk Douppioni & 100\% Mulberry Silk & 0.3 & 86.94 \\
E-S04 & Silk Burn-out & 100\% Mulberry Silk & 0.3 & 53.13 \\
R01 & Viscose Plain & 100\% Viscose & 0.3 & 63.57 \\
R02 & Viscose Twill & 100\% Viscose & 0.3 & 191.9 \\
R03 & Viscose Satin & 100\% Viscose & 0.3 & 72.67 \\
\textbf{R04} & \textbf{Acetate Lining} & 100\% Diacetate & 0.1 & 100 \\
CU01 & Cupro Plain & 100\% Cupro & 0.1 & 68.12 \\
P01 & Polyester Taffeta & 100\% Polyester & 0.1 & 83.61 \\
P02 & Polyester Chiffon & 100\% Polyester & 0.4 & 97.74 \\
N01 & Nylon Taffeta & 100\% Nylon & 0.2 & 81.74 \\
N02 & Nylon Twill & 100\% Nylon & 0.4 & 178.13 \\
A01 & Acrylic Wool-like & 100\% Acrylic & 0.5 & 239.5 \\
E-R02 & Viscose/Linen Plain & 70\% Viscose / 30\% Linen & 0.6 & 206.81 \\
E-P02 & Cationic Polyester & 100\% Cationic Polyester & 0.3 & 121.9 \\
E-P03 & Polyester Peach Skin & 100\% Polyester (Microfiber) & 0.4 & 148.37 \\
E-P04 & Polyester Suede & 100\% Polyester (Sea-island) & 0.5 & 185.45 \\
E-N01 & Nylon Taslon & 100\% Nylon (ATY) & 0.4 & 131.14 \\
E-N02 & Nylon Oxford & 100\% Nylon & 0.2 & 122.73 \\
E-KN07 & Polyester Polar Fleece & 100\% Polyester & 1.2 & 231.71 \\
E-F01 & Coated Nylon & Nylon + PU Coating & 0.1 & 41.9 \\
E-F02 & Waterproof Breathable & Polyester + TPU Membrane & 0.2 & 126.12 \\
E-F05 & Antistatic Polyester & Polyester & 0.6 & 318.07 \\
\bottomrule
\end{tabular}
}
\end{table*}

\begin{table*}[ht]\tiny
\centering
\caption{Fabric Specification Table: Part Three}
\label{tab:fabric3}
\resizebox{\textwidth}{!}{%
\begin{tabular}{lllcc}
\toprule
\textbf{No.} & \textbf{Name} & \textbf{Composition} & \textbf{Thickness (mm)} & \textbf{Density (gsm)} \\
\midrule
EX-01 & Lyocell (Tencel™) Plain & 100\% Lyocell & 0.5 & 122.78 \\
EX-02 & Lyocell Twill & 100\% Lyocell & 0.3 & 165.46 \\
EX-03 & Modal Plain & 100\% Modal & 0.6 & 157.93 \\
EX-04 & Modal/Cotton Jersey & 50\% Modal / 50\% Cotton & 0.6 & 230.77 \\
EX-05 & Bamboo Rayon Plain & 100\% Bamboo Viscose & 0.6 & 251.12 \\
EX-08 & Coolmax Jersey & 100\% Profiled Polyester & 0.6 & 128.73 \\
EX-11 & Reflective Fabric & Polyester + Glass Beads/Prisms & 0.1 & 129.88 \\
EX-15 & Recycled Polyester Plain & 100\% rPET & 0.6 & 185.28 \\
EX-17 & Tencel™/Linen Blend & 70\% Lyocell / 30\% Linen & 0.4 & 138.15 \\
EX-18 & Modal Plain & 100\% Modal & 0.6 & 153.83 \\
DP 1 & Low-Density Cotton Muslin & 100\% Cotton & 0.3 & 60.9 \\
DP 2 & Cotton Chambray & 100\% Cotton & 0.7 & 192.86 \\
DP 3 & White Viscose & 95\% Viscose / 5\% Elastane & 0.7 & 225.44 \\
DP 4 & Gingham-Pink White Cotton & 100\% Cotton & 0.5 & 114.62 \\
DP 5 & Tartan Wool Fabrics & 100\% Wool & 0.8 & 114.55 \\
DP 6 & Duchess Satin & Polyester & 0.3 & 186.8 \\
DP 7 & Plain PolyCotton & 65\% Polyester / 35\% Cotton & 0.4 & 105.39 \\
DP 8 & Washed Linen Aqua & 100\% Linen & 0.8 & 240.26 \\
DP 9 & Cotton Canvas & 100\% Cotton & 0.7 & 249.75 \\
DP 10 & Pale Gold Metallic Viscose & 70\% Viscose / 30\% Polyester & 0.7 & 204.47 \\
DP 11 & Brushed Cotton Check & 65\% Polyester / 35\% Cotton & 0.9 & 162.34 \\
DP 12 & Polyester Plain Satin & Polyester & 0.7 & 171.58 \\
Paper 1 & Printing Paper & Paper & 0.1 & 83.15 \\
Paper 2 & Folding Paper & Paper & 0.2 & 81.33 \\
Paper 3 & Thick Folding Paper & Paper & 0.2 & 173.71 \\
\textbf{Paper 4} & \textbf{Thinner Printing Paper} & Paper & 0.08 & 79.43 \\
\bottomrule
\end{tabular}
}
\end{table*}

\pagebreak
\clearpage

\begin{figure*}[htb]
    \centering
    \includegraphics[width=\textwidth]{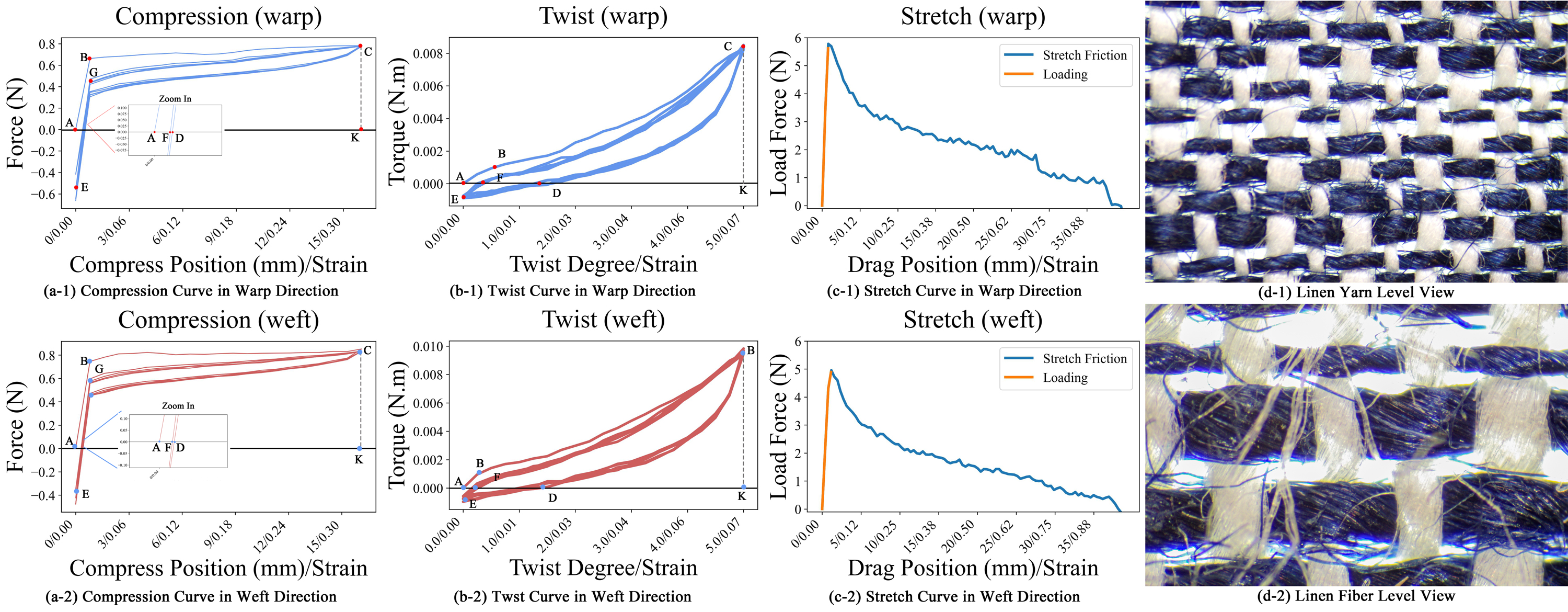}
    \caption{The load-deformation curve of linen used in our experiments (a,b,c-1,2). The yarn-level (d-1) and fiber-level (d-2) view of the linen fabric.}
    \label{fig:linen_curve}
\end{figure*}

\begin{figure*}[htb]
    \centering
    \includegraphics[width=\textwidth]{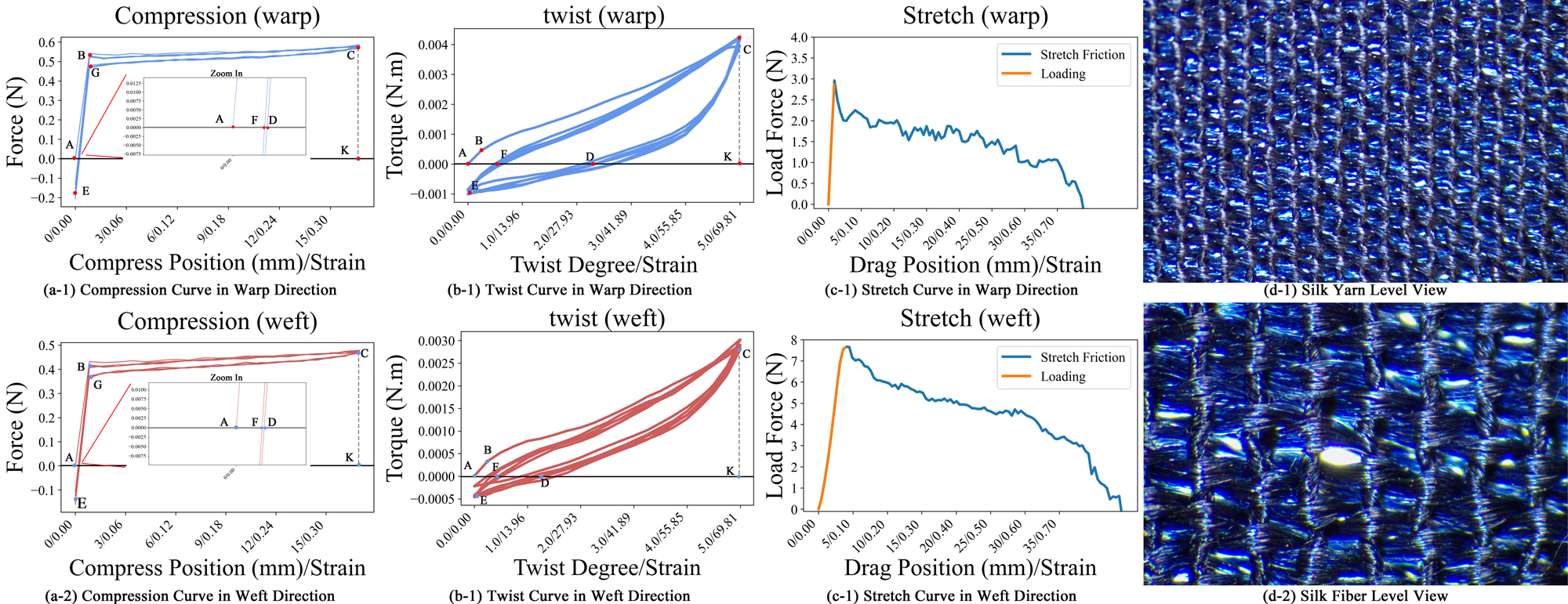}
    \caption{The load-deformation curve of silk used in our experiments (a,b,c-1,2). The yarn-level (d-1) and fiber-level (d-2) view of the silk fabric.}
    \label{fig:silk_curve}
\end{figure*}

\begin{figure*}[htb]
    \centering
    \includegraphics[width=\textwidth]{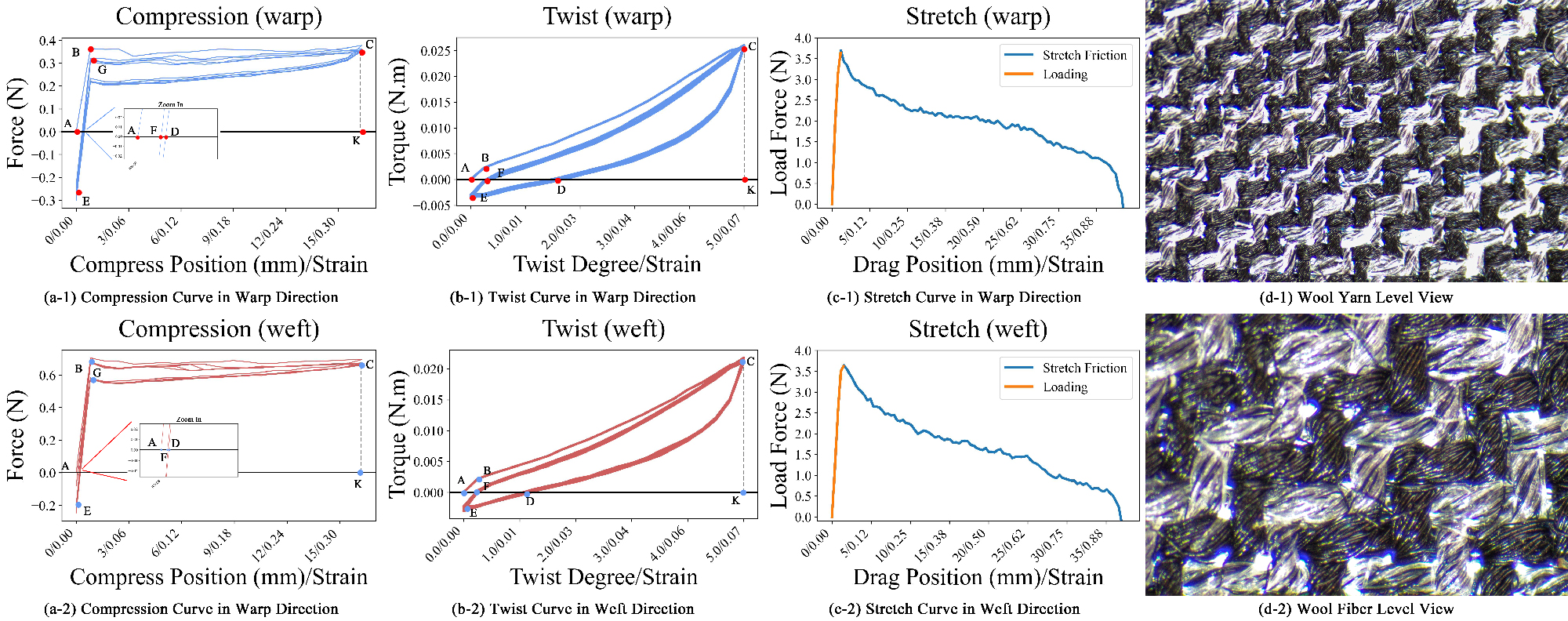}
    \caption{The load-deformation curve of wool used in our experiments (a,b,c-1,2). The yarn-level (d-1) and fiber-level (d-2) view of the wool fabric.}
    \label{fig:wool_curve}
\end{figure*}

\begin{figure*}[htb]
    \centering
    \includegraphics[width=\textwidth]{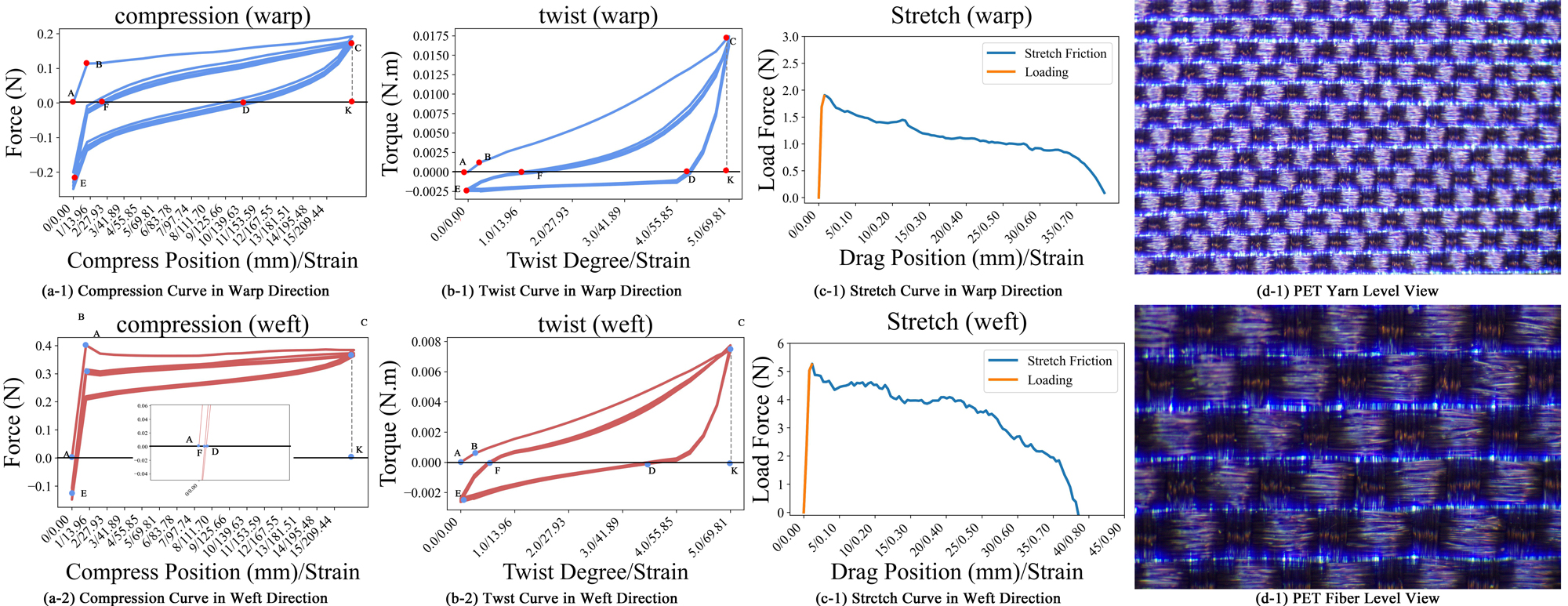}
    \caption{The load-deformation curve of PET used in our experiments (a,b,c-1,2). The yarn-level (d-1) and fiber-level (d-2) view of the PET fabric.}
    \label{fig:PET_curve}
\end{figure*}